\documentclass[11pt,letterpaper]{article}

\usepackage{amsmath}
\usepackage{amsfonts}
\usepackage{amssymb}
\usepackage{graphicx}
\usepackage{graphics,epsfig}
\usepackage{natbib}

\DeclareRobustCommand{\SkipTocEntry}[4]{}

\newcommand{\be}{\begin{equation}}
\newcommand{\ee}{\end{equation}}
\newcommand{\ba}{\begin{eqnarray}}
\newcommand{\ea}{\end{eqnarray}}

\newcommand{\mb}{\mathbf}

\newcommand{\bg}[1]{\mbox{\boldmath$#1$}}
\newcommand{\bgb}{\bg{\beta}}
\newcommand{\bga}{\bg{\alpha}}
\newcommand{\nside}     {\mbox{$N_{\rm side}$}} 
\newcommand{\kpc}{\rm{kpc}}
\def\ltsima{$\; \buildrel < \over \sim \;$}
\def\ltsim{\lower.5ex\hbox{\ltsima}}

\def\gtsima{$\; \buildrel > \over \sim \;$}
\def\gtsim{\lower.5ex\hbox{\gtsima}}

\newcommand{\apgt}{\ {\raise-.5ex\hbox{$\buildrel>\over\sim$}}\ }
\newcommand{\aplt}{\ {\raise-.5ex\hbox{$\buildrel<\over\sim$}}\ }

\newcommand{\bfig}{\begin{figure}}
\newcommand{\efig}{\end{figure}}
\newcommand{\bcn}{\begin{center}}
\newcommand{\ecn}{\end{center}}
\newcommand{\ben}{\begin{enumerate}}
\newcommand{\een}{\end{enumerate}}
\newcommand{\btab}{\begin{tabular}}
\newcommand{\etab}{\end{tabular}}
\newcommand{\bt}{\begin{table}}
\newcommand{\et}{\end{table}}

\newcommand{\apjs}{Astrophys. J., Suppl. Ser.}
\newcommand{\apj}{Astrophys. J.}
\newcommand{\aap}{Astron. Astrophys.}
\newcommand{\mnras}{Mon. Not. R. Astron. Soc.}
\newcommand{\aj}{Astron. J.}
\newcommand{\apjl}{Astrophys. J., Lett.}
\newcommand{\prd}{Phys. Rev. A}

\newcommand{\nat}{Nature}
\newcommand{\apss}{APSS}
\newcommand{\araa}{ARAA}

\newcommand{\iras}{{\sl IRAS}}
\newcommand{\firas}{{\sl FIRAS}}
\newcommand{\cobedirbe}{{\sl COBE/DIRBE}}
\newcommand{\cobedmr}{{\sl COBE/DMR}}
\newcommand{\cobefiras}{{\sl COBE/FIRAS}}
\newcommand{\wmap}    {{\sl WMAP}}
\newcommand{\cobe}   {{\sl COBE}}
\newcommand{\beast}   {{\sl BEAST}}

\bibpunct{(}{)}{;}{a}{}{,}

\begin{document}

\vspace{5mm}
\vspace{0.5cm}

\begin{center}

% TITLE
{\large CMBPol Mission Concept Study} \\
\vskip 15pt {\Large Prospects for polarized foreground removal}
\\[1.0cm]

% AUTHORS
{J.~Dunkley$^{\rm 1,2,3}$, A.~Amblard$^{\rm 4}$, C.~Baccigalupi$^{\rm 5}$, M.~Betoule$^{\rm 6}$, D.~Chuss$^{\rm 7}$, A.~Cooray$^{\rm 4}$, J.~Delabrouille$^{\rm 6}$, C.~Dickinson$^{\rm 8,9}$, G.~Dobler$^{\rm 10}$, J.~Dotson$^{\rm 11}$, H.~K.~Eriksen$^{\rm 12}$, D.~Finkbeiner$^{\rm 10}$, D.~Fixsen$^{\rm 7}$, P.~Fosalba$^{\rm 13}$, A.~Fraisse$^{\rm 3}$, C.~Hirata$^{\rm 14}$, A.~Kogut$^{\rm 7}$, J.~Kristiansen$^{\rm 12}$, C.~Lawrence $^{\rm 9}$, A.~M.~Magalh\~aes$^{\rm 15}$, M.~A.~Miville-Deschenes$^{\rm 16}$, S.~Meyer$^{\rm 17}$, A.~Miller$^{\rm 18}$, S.~K.~Naess$^{\rm 12}$, L.~Page$^{\rm 2}$, H.~V.~Peiris$^{\rm 19}$, N.~Phillips$^{\rm 7}$, E.~Pierpaoli$^{\rm 20}$, G.~Rocha$^{\rm 8}$,  J.~E.~Vaillancourt$^{\rm 21}$, L.~Verde$^{\rm 22}$}
\\[0.5cm]

\end{center}

\vspace{.8cm}

% ABSTRACT
\hrule \vspace{0.3cm}
{\small  \noindent \textbf{Abstract} \\[0.3cm]
\noindent

In this report we discuss the impact of polarized foregrounds on a future
CMBPol satellite mission. We review our current knowledge
of Galactic polarized emission at microwave frequencies, including synchrotron and thermal
dust emission. We use existing data and our
understanding of the physical behavior of the sources of foreground
emission to generate sky templates, 
and start to assess how well primordial gravitational wave signals
can be separated from foreground contaminants for a CMBPol mission.
At the estimated foreground minimum of $\sim 100$~GHz, the polarized foregrounds 
are expected to be lower than a primordial
polarization signal with tensor-to-scalar ratio 
$r=0.01$, in a small patch ($\sim 1\%$) of the sky known to have low Galactic emission.
Over 75\% of the sky we expect the foreground
amplitude to exceed the primordial signal by about a
factor of eight at the foreground minimum and on scales of two degrees.
Only on the largest scales does the
polarized foreground amplitude exceed the primordial signal 
by a larger factor of about 20. 
The prospects for detecting an $r=0.01$ signal 
including degree-scale measurements appear promising,
with $5\sigma_r \sim 0.003$ forecast from multiple methods. 
A mission that observes a range of scales offers better prospects 
from the foregrounds perspective than one targeting only the lowest few multipoles.
We begin to explore how optimizing the composition of frequency channels in 
the focal plane can maximize our ability to perform component separation, with a 
range of typically $40\ltsim \nu \ltsim 300$ GHz preferred for ten channels. 
Foreground cleaning methods are already in place to tackle
a CMBPol mission data set, and further investigation of the optimization and 
detectability of the primordial signal will be useful for mission design.
\vspace{0.5cm}  \hrule
\def\thefootnote{\arabic{footnote}}
\setcounter{footnote}{0}

\vspace{1.0cm}

\vfill \noindent
$^\dagger$ {\footnotesize {\tt j.dunkley@physics.ox.ac.uk}}\\

\begin{center}

% AFFILIATIONS

{\small \textit{$^{\rm 1}$ Astrophysics, University of Oxford, Keble Road, Oxford, OX1 3RH, UK}}

{\small \textit{$^{\rm 2}$ Department of Physics, Princeton University, Princeton, NJ 08540, USA}}

{\small \textit{$^{\rm 3}$ Department of Astrophysical Sciences, Princeton, NJ 08540, USA}}

{\small \textit{$^{\rm 4}$ Center for Cosmology, University of California, Irvine, CA 92697, USA}}

{\small \textit{$^{\rm 5}$ SISSA, Via Beirut, Trieste 34014, Italy}}

{\small \textit{$^{\rm 6}$ Laboratoire AstroParticule et Cosmologie, UMR 7164, CNRS and Observatoire de Paris, 10, 75205 Paris Cedex 13, France}}

{\small \textit{$^{\rm 7}$ Code 665, NASA/Goddard Space Flight Center, Greenbelt, MD 20771, USA}}

{\small \textit{$^{\rm 8}$ Infrared Processing \& Analysis Center, California Institute of Technology, MC 220-6, Pasadena, CA 91125, USA}}

{\small \textit{$^{\rm 9}$ NASA Jet Propulsion Laboratory, 4800 Oak Grove Drive, Pasadena, CA 91109, USA}}

{\small \textit{$^{\rm 10}$ Harvard-Smithsonian Center for Astrophysics, 60 Garden St. - MS 51,
Cambridge, MA 02138, USA}}

{\small \textit{$^{\rm 11}$ NASA Ames Research Center, MS 240-2, Moffett Field, CA  94035  USA}}

{\small \textit{$^{\rm 12}$ Institute of Theoretical Astrophysics, University of
Oslo, P.O.\ Box 1029 Blindern, N-0315 Oslo, Norway}}

{\small \textit{$^{\rm 13}$ Institut de Ciencies de l'Espai, IEEC-CSIC, F. de Ciencies, Torre C5 par-2, Barcelona 08193, Spain}}

{\small \textit{$^{\rm 14}$ Division of Physics, Mathematics, \& Astronomy, California Institute of Technology, MC 130-33, Pasadena, CA 91125, USA}}

{\small \textit{$^{\rm 15}$ IAG, Universidade de S\~ao Paulo, Rua do Mat\~ao 1226, S\~ao Paulo 05508-900, Brazil}}

{\small \textit{$^{\rm 16}$  Institut d'Astrophysique Spatiale, Universite Paris XI, Orsay, 91405, France}}

{\small \textit{$^{\rm 17}$ Kavli Institute for Cosmological Physics, University of Chicago, Chicago, IL 60637, USA}}

{\small \textit{$^{\rm 18}$ Physics Department, Columbia University, New York, NY 10027, USA}}

{\small \textit{$^{\rm 19}$ Institute of Astronomy, University of Cambridge, Cambridge, CB3 0HA, UK}}

{\small \textit{$^{\rm 20}$ University of Southern California, Los Angeles, CA, 90089-0484, USA}}

{\small \textit{$^{\rm 21}$ Division of Physics, Mathematics, \& Astronomy, California Institute of Technology, MC 320-47, Pasadena, CA 91125, USA}}

{\small \textit{$^{\rm 22}$ ICREA \& Institute of Space Sciences (IEEC-CSIC), Barcelona, Spain}}

\end{center}

\newpage
\tableofcontents

%-----------------------------------------------------------------------------------------------
% MAIN BODY

\newpage
\section{Introduction}
\label{sec:intro}

Measurements of the Cosmic Microwave Background (CMB) temperature anisotropy 
have led to the establishment of a standard $\Lambda$CDM cosmological model for the universe \citep{miller/etal:1999,lee/etal:2001,netterfield/etal:2002,halverson/etal:2002,pearson/etal:2003,scott/etal:2003,spergel/etal:2003,spergel/etal:2007,dunkley/etal:prep}. The polarization 
anisotropy, two orders of magnitude smaller, was measured for the 
first time in 2002 by DASI \citep{kovac/etal:2002,leitch/etal:2002} and confirmed by the ground and balloon-based experiments CBI, CAPMAP, Boomerang, and 
QuAD \citep{readhead/etal:2004,barkats/etal:2005,montroy/etal:2006,ade/etal:2007,pryke/etal:prep}. The first all-sky observations by the \wmap\ satellite \citep{page/etal:2007,hinshaw/etal:prep}, with a measurement of the power in the 
curl-free polarized `E-mode', have provided a 
cross-check of the cosmological model, led to improved constraints on 
cosmological parameters, and a measurement of the optical depth to the
reionization of the universe \citep{page/etal:2007,hinshaw/etal:prep}. 

The next challenge of CMB observation is to test the inflationary scenario for
the early universe, by looking for the signature of 
primordial gravitational waves. They are predicted to leave a distinct 
divergence-free, or `B-mode', pattern in the large-scale 
CMB polarization anisotropy that may be observable with a 
future experiment (see \citet{baumann/etal:prepa} for details). 
This signal is already constrained to be 
over an order of magnitude smaller than the observed E-mode signal 
arising from scalar fluctuations, from ground and space-based observations 
\citep{page/etal:2007,hinshaw/etal:prep,pryke/etal:prep}. 
The Planck satellite, due for launch 2009, is projected to 
reach limits on the ratio of power in primordial tensor 
to scalar fluctuations $r$, of $r\sim0.1$ \citep{planck:2006},
while ground and 
balloon-based experiments currently observing or being commissioned 
are designed to reach below $r=0.1$. These include EBEX, 
Clover, QUIET, and SPIDER 
(see e.g., \citet{oxley/etal:2004,taylor:2006,samtleben:2008,crill/etal:2008} for descriptions). 

This document is part of a study focusing on the capabilities of 
a future satellite mission targeting limits of $r=0.01$ or lower, known as 
CMBPol. Here we address the issue of polarized foreground emission 
from the Galaxy, which on average dominates the primordial B-mode 
over the whole sky.
Establishing how well we can realistically hope to extract the 
primordial signal, and how this influences mission design, has been 
addressed in a number of previous studies 
(e.g., \citet{amarie/hirata/seljak:2005,verde/peiris/jimenez:2006,bock/etal:2006,bock/etal:prep}). 
In companion documents the theoretical case for inflation and its predictions 
are presented \citep{baumann/etal:prepa}, with weak lensing effects studied in 
\citet{smith/etal:prep}, 
and reionization prospects in \citet{zaldarriaga/etal:prep}. 
The Galactic science case is presented in \citet{fraisse/etal:prep}, 
and \citet{baumann/etal:prepb} provide a summary.

The document is structured as follows. In \S\ref{sec:current} we review our 
current understanding of polarized foreground emission, and discuss 
predictions and models for the emission, with 
estimates of large-scale polarization maps and power spectra. 
In \S\ref{sec:methods} we 
describe various methods that are used to estimate polarized CMB maps, and in 
\S\ref{sec:forecast} apply these and Fisher matrix methods to 
begin to forecast limits for a specific template mission, and test 
optimal allocation of channels among frequencies. We also 
describe future tests, and summarize in \S\ref{sec:summary}.

\section{Knowledge of polarized foregrounds}
\label{sec:current}

In this section we focus on observations of polarized foregrounds
relevant to the microwave regime, concentrating on
large-scale diffuse emission. The discussion is organized by physical
emission mechanism, and outlines how our 
understanding of the physics of the interstellar medium is used to construct
two-dimensional emission templates on the sky. This is an inherently simpler
problem than a three-dimensional model. Issues that remain open
for investigation include: possible spatial and frequency variation of
spectral indices, the number of parameters required for a sufficiently
sophisticated model to describe and subtract foregrounds at the level
required if $r<0.01$, and methods for assessing the
validity of the chosen model.

\subsection{Synchrotron emission}

Synchrotron emission results from the acceleration of 
cosmic-ray electrons
in the magnetic field of the Galaxy.
For a power-law distribution of electron energies,
$N(E) \propto E^{-p}$,
propagating in a uniform magnetic field,
the resulting emission
is partially polarized
with fractional linear polarization
\begin{equation}
f_s = \frac{p+1}{p + 7/3}
\label{fs_vs_p_eq}
\end{equation}
aligned perpendicular to the magnetic field
\citep{rybicki/lightman:1979}.
The frequency dependence of synchrotron emission
is also related to the electron energy distribution,
$T(\nu) \propto \nu^\beta$,
with spectral index
\begin{equation}
\beta = -\frac{p+3}{2}
\label{beta_vs_p_eq}
\end{equation}
where $T$ is in units of antenna temperature.
For spectral index $\beta \approx -3$
observed at microwave frequencies,
synchrotron emission
could have fractional polarization as high as $f_s \sim 0.75$,
although in practice this is almost never observed.
Line-of-sight and beam averaging effects
tend to reduce the observed polarization
by averaging over regions with different 
electron energy distribution
or magnetic field orientation.
At frequencies below a few GHz,
Faraday rotation will induce additional depolarization.

%--------------------------------------------------------
% Synchrotron fig 1: \wmap\ 5-yr K-band polarization
%--------------------------------------------------------
\begin{figure}[t]
\center{
\includegraphics[angle=270,width=0.8\hsize]{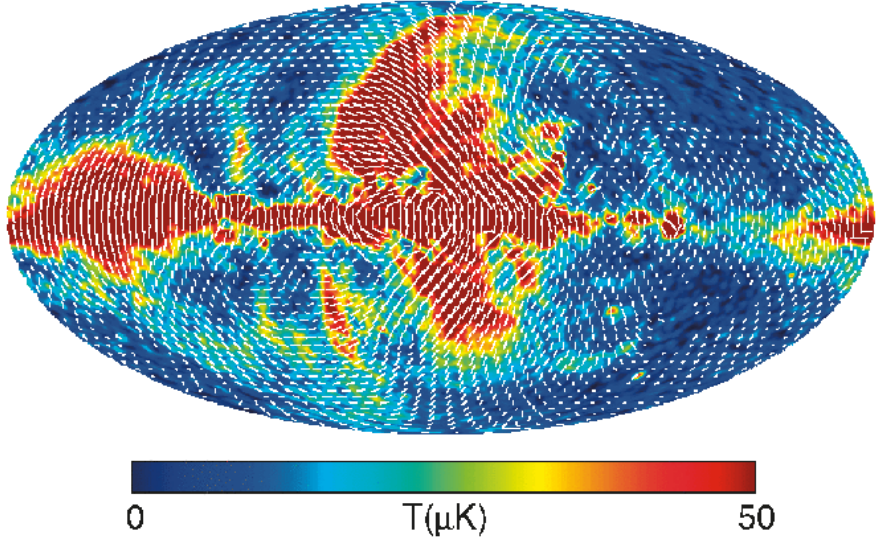}
}
\caption{
The \wmap\ measurements of polarization at 23 GHz
map the large-scale distribution of polarized synchrotron emission
(Mollweide projection in Galactic coordinates).
White lines indicate the polarization angle $\gamma_d$ and amplitude,
and are oriented perpendicular to the magnetic field.
\label{fig:wmap_5yr_kband} 
}
\end{figure}

%--------------------------------------------------------
% Synchrotron  fig 2: \wmap\ foreground power spectra
%--------------------------------------------------------
\begin{figure}[t]
\center{
\includegraphics[width=0.6\hsize]{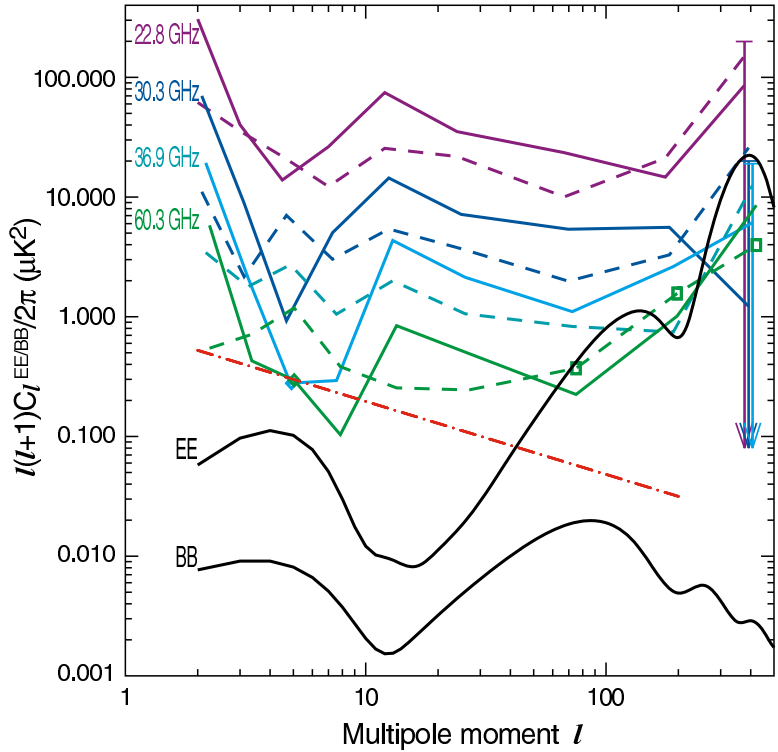}}
\caption{Power spectra for the \wmap\ 3-year polarization maps
are dominated by foreground emission.
Solid lines show EE spectra and dashed lines show BB spectra outside the Galactic plane;
the color indicates frequency band. The rise in power for $\ell > 100$ is an artifact of the instrument noise. The red dot-dashed line shows the estimated
BB foreground power at 60 GHz, using a parameterized model fit to the multi-frequency 
data.  Diamonds (EE) and boxes (BB) indicate points with negative values.
The cosmological model shown has $r=0.3$ and $\tau=0.09$.
From \citet{page/etal:2007}.
\label{fig:pol_fg_cl_fig}}
\end{figure}
%---------------------------------------------------------
%---------------------------------------------------------

The polarized synchrotron emission may be parameterized as
\begin{eqnarray}
Q(\hat{n}) &=& T_s^P(\hat{n}) \cos(2 \gamma(\hat{n})) 
   \left( \frac{\nu}{\nu_0} \right)^{\beta_s(\hat{n}) + C(\hat{n}) \log(\nu / \nu_0)}	\nonumber \\
U(\hat{n}) &=& T_s^P(\hat{n}) \sin(2 \gamma(\hat{n})) 
   \left( \frac{\nu}{\nu_0} \right)^{\beta_s(\hat{n}) + C(\hat{n}) \log(\nu / \nu_0)}	
\label{synch_param_eq}
\end{eqnarray}
where
$Q(\hat{n})$ and $U(\hat{n})$ are the Stokes parameters 
in direction $\hat{n}$,
$T_s^P(\hat{n})$ is the (polarized) amplitude $T^P = \sqrt{ Q^2 + U^2}$,
$\gamma(\hat{n})$ is the polarization angle,
$\beta_s(\hat{n})$ is the spectral index,
and
$C(\hat{n})$ parameterizes any spectral curvature 
(flattening or steepening)
relative to a pure power law.

The Wilkinson Microwave Anisotropy Probe (\wmap)
provides our best estimate of the synchrotron morphology
on angular scales of a few degrees or larger.
Figure \ref{fig:wmap_5yr_kband}
shows the full-sky \wmap\ observations
at 23 GHz
\citep{page/etal:2007,hinshaw/etal:prep}.
At 4$^\circ$ pixelization,
the signal to noise ratio for polarized synchrotron emission
is better than three over 90\% of the sky.
The polarization angle has coherent structure
over large swaths of the sky,
creating significant emission at low multipole moments $\ell$.
Figure \ref{fig:pol_fg_cl_fig} shows the power spectra
of the \wmap\ polarization data from 23 to 94 GHz
\citep{page/etal:2007}.
The power spectra are dominated by foreground emission
with roughly equal power in E- and B-modes,
which, averaged over the high-latitude sky,
are brighter than the CMB polarization
even at the minimum near 70-100 GHz.
A simple parameterization of the foreground emission in \citet{page/etal:2007}
gives a power spectrum with $\ell(\ell+1)C_\ell/{2\pi} \propto \ell^{-0.6}$, 
shown in red in Figure \ref{fig:pol_fg_cl_fig}.
This is consistent with the power spectrum for synchrotron emission
observed from radio maps (see e.g. \citet{laporta/etal:2008}). 

We can estimate the fractional polarization of synchrotron emission
by comparing the \wmap\ 23 GHz map
to an estimate of the unpolarized synchrotron emission.
The synchrotron intensity
suffers from confusion
between synchrotron, free-free,
and other microwave components, so extracting this signal from observed sky 
maps depends on modeling assumptions.
\citet{kogut/etal:2007} use the 
\wmap\ three-year maximum-entropy map of synchrotron emission as a tracer of
the intensity \citep{hinshaw/etal:2007}, finding the Galactic plane region
($|b| < 5^\circ$)
to be largely depolarized,
with mean fractional polarization
$f_s = P/I = 0.05$. 
The North Galactic spur region at mid-latitudes
has fractional polarization $f_s \approx 0.3$
while the high latitude sky
outside the P06 mask 
has a very broad distribution 
with mean $f_s = 0.15$ 
\citep{kogut/etal:2007}. 
The synchrotron polarization
is consistent with the magnetic field
having rough equipartition
between a large-scale smooth component
and a smaller scale turbulent component
\citep{page/etal:2007}.
\citet{miville-deschenes/etal:prep} estimate the polarization 
fraction at 23~GHz for a set of synchrotron intensity maps.
One map is derived from \wmap\ data assuming that no 
anomalous dust emission contributes at 23 GHz, and gives a polarization
fraction, shown in Figure \ref{fig:polar_fraction}, similar to the one 
studied in \cite{kogut/etal:2007}. An alternative intensity map 
is derived from \wmap\ data allowing for an anomalous 
unpolarized emission component to contribute to the 
total observed intensity. This results in a modified estimate 
for the synchrotron intensity, and an increase in polarization 
fraction over much of the sky, also shown in Figure \ref{fig:polar_fraction}. 
In this second map, there is a larger polarization fraction in the 
anti-center region compared to the inner galaxy. These estimates for the 
polarization fraction would be 
modified if the anomalous component(s) had a non-negligible polarization 
fraction.
We note however that despite 
these uncertainties in the polarization fraction, our 
estimate of the large-scale polarization of synchrotron emission, $P=\sqrt{Q^2+U^2}$, is robust at 23 GHz.

\begin{figure}
\center{
\includegraphics[width=0.5\hsize, draft=false]{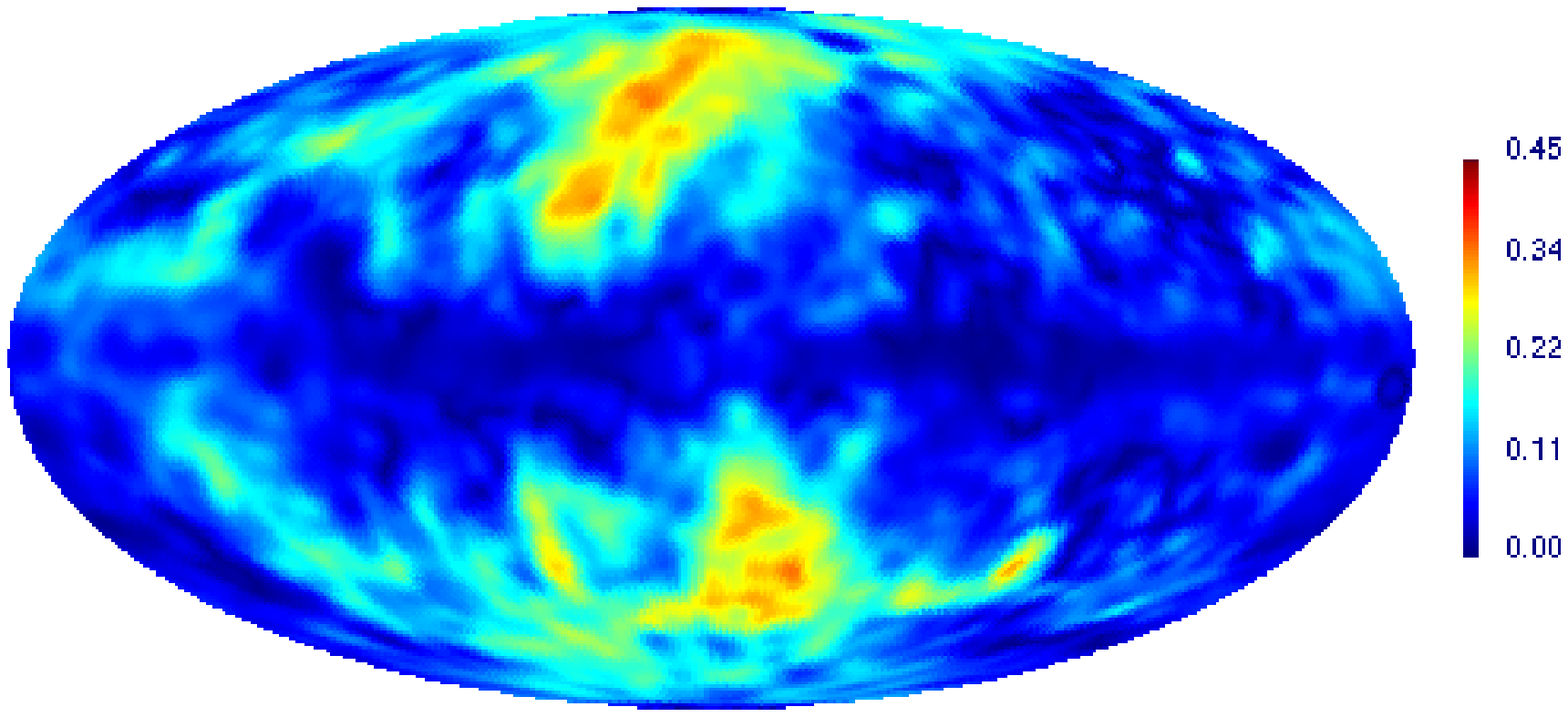}
\includegraphics[width=0.5\hsize, draft=false]{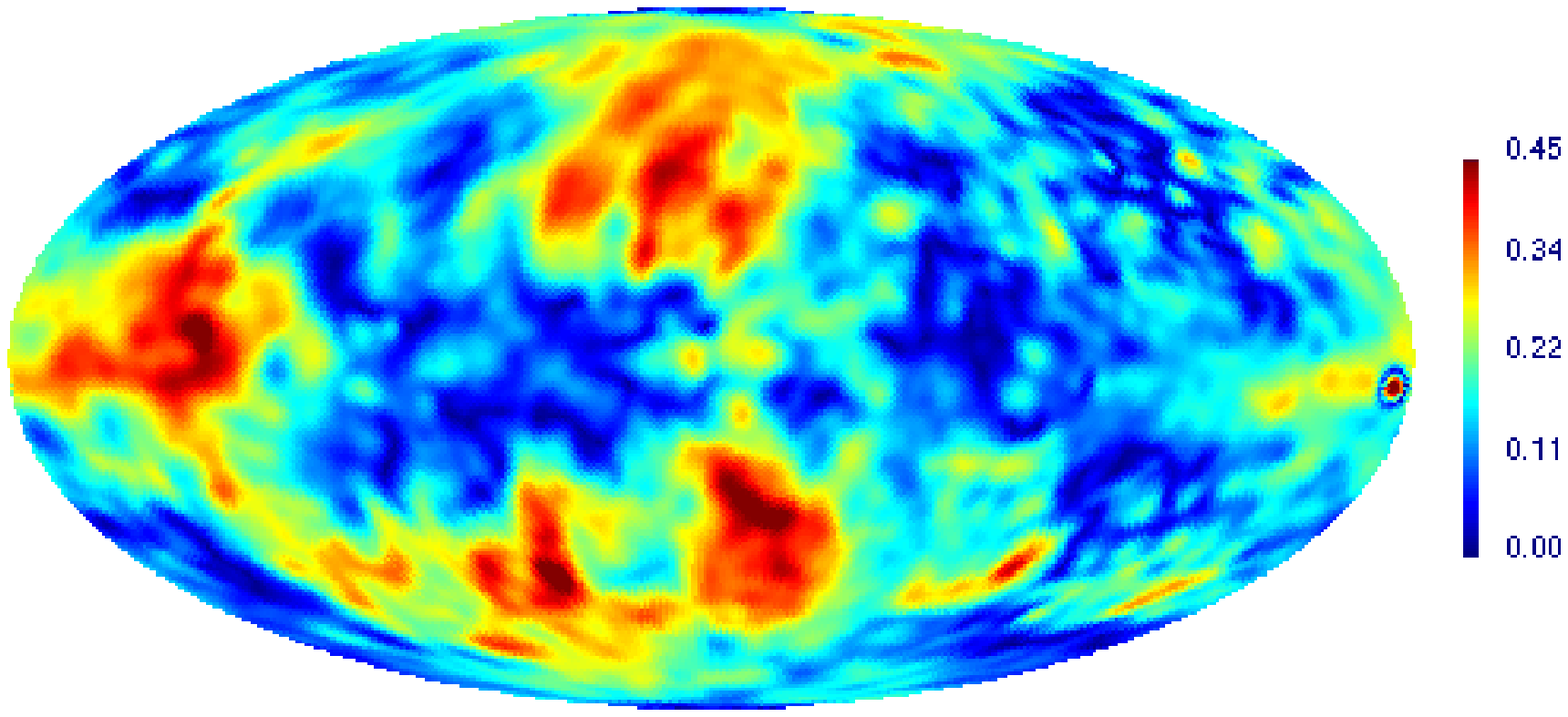}
}
\caption{
\label{fig:polar_fraction} 
Estimated polarization fraction $f=P/I$ at $5^\circ$ resolution.
The polarization $P=\sqrt{Q^2+U^2}$ 
is well estimated from the \wmap\ K band data \citep{page/etal:2007} 
and the intensity $I$, which is less certain, is modeled excluding (top), and including 
(bottom), an unpolarized `anomalous' dust 
component (from \citet{miville-deschenes/etal:prep}).}
\end{figure}

The \wmap\ data provide a reasonable signal to noise ratio
for the polarized synchrotron morphology
on angular scales of a few degrees or larger.
The spectral index $\beta_s(\hat{n})$ is less well constrained.
Estimates of the spectral index for {\it unpolarized} synchrotron emission
suffer from confusion with other emission components,
while estimates based on polarization data
(where confusion is less severe)
suffer from the weaker synchrotron signal at frequencies above 23 GHz.
Figure \ref{fig:pol_variance_vs_freq}
shows the observed polarization variance $\langle P^2 \rangle$ 
of the \wmap\ data
as a function of frequency
for $4^\circ$ pixelization.
The spectra are consistent with a superposition
of two power-law components,
with synchrotron $\beta_s \approx -3$
and dust $\beta_d \approx 2$. The CMB contributes lower power relative
to the foreground components.
\citet{kogut/etal:2007} find that the synchrotron 
spectral index steepens off the plane
by a modest amount $\Delta \beta_s \sim 0.2$.
\citet{gold/etal:prep} map the synchrotron spectral index
using a combined analysis of the temperature
and polarization data, finding a trend towards a steepening of the 
spectral index
off the Galactic plane. A single index of $\beta_s=-3.3$ currently works 
sufficiently well outside the Galactic plane for cleaning the 
\wmap\ maps \citep{page/etal:2007}.

Similar problems affect estimates of the synchrotron spectral curvature.
The spectral index depends on the energy spectrum 
of cosmic ray electrons (Eqn.~\ref{beta_vs_p_eq}).
Higher energy electrons lose energy more rapidly,
which can lead to a steepening or break in the synchrotron spectrum.
The frequency of the break is not well known, but is 
thought to vary with position on the sky and be in the 
range 10-100 GHz, corresponding to a break in the electron 
spectrum at the 
energy where the characteristic cooling time and the escape time are the 
same.  
To a large extent,
efforts to characterize any spectral steepening
are limited by confusion with thermal dust emission.
The effects of steepening for the synchrotron component
are largest near the foreground minimum near 70-100 GHz,
where the amplitudes of synchrotron and thermal dust emission
are comparable. 
New measurements of polarized dust emission
at frequencies above 100 GHz,
from the Planck satellite and other experiments,
will help break this degeneracy
and provide tighter limits
on both the dust and synchrotron spectra. Low frequency observations 
from experiments at 5-15 GHz, including the C-Band All Sky Survey (C-BASS), 
will 
also help determine the frequency dependence of the 
polarized synchrotron emission.

%--------------------------------------------------------
% Synchrotron  fig 4: Polarized spectra
%--------------------------------------------------------
\begin{figure}[t]
\center{
\includegraphics[width=0.4\hsize]{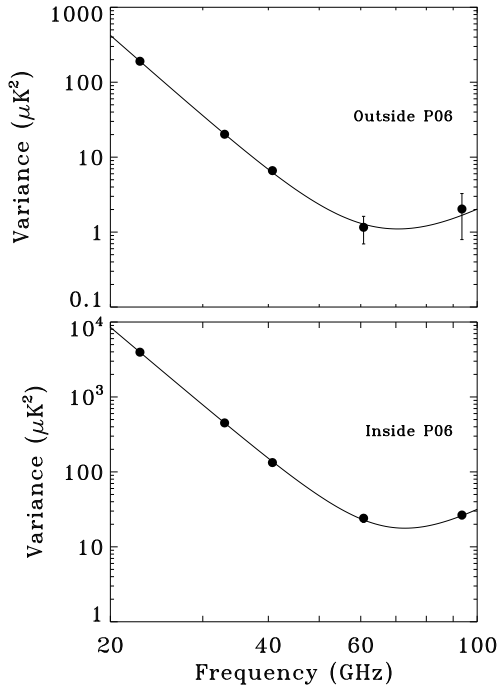}}
\caption{
Spectrum of the variance in the \wmap\ polarization data,
computed at each frequency from the cross-spectra of one-year maps,
to avoid a bias from instrument noise.
The spectra are consistent with power-law
synchrotron and dust emission,
with synchrotron $\beta_s \approx -3$.  
The P06 cut excludes 25\% of the sky in the Galactic plane. 
From \citet{kogut/etal:2007}.
\label{fig:pol_variance_vs_freq} 
}
\end{figure}

%---------------------------------------------------------

Little information is available
for synchrotron polarization
on angular scales below 1$^\circ$.
Existing measurements on small scales
are heavily affected by Faraday rotation at low frequencies.
Figure \ref{fig:drao_map} shows a polarization map
with angular resolution 36$^\prime$
from the Dominion Radio Astrophysical Observatory
at 1.4 GHz
\citep{wolleben/etal:2006}.
Although some features
such as the North Galactic spur at high latitude
and the Fan region near longitude 150$^\circ$
are recognizable from the polarization map at 23 GHz
(Fig \ref{fig:wmap_5yr_kband}),
the 1.4 GHz map is dominated by
Faraday depolarization, whereas the 23 GHz emission has almost 
no Faraday rotation.
Similar depolarization affects interferometric maps
at finer angular resolution
\citep{uyaniker/etal:2003,
taylor/etal:2003},
limiting their utility for foreground modeling
at millimeter wavelengths.

An additional component of synchrotron emission that may be 
constrained with future polarization data 
is anomalously hard synchrotron emission towards the Galactic 
center. This is known as the \wmap\ `Haze', found by
\citet{finkbeiner:2004,dobler/finkbeiner:2008a,bottino/banday/maino:2008} 
in analyses of the \wmap\ data. This does not have a conclusively 
confirmed origin, but as it is confined to the Galactic 
center, it is discussed in more detail in 
the companion CMBPol document on foreground science \citep{fraisse/etal:prep}.

%--------------------------------------------------------
% Synchrotron fig 6: DRAO polarization map at 1.4 GHz 
%--------------------------------------------------------
\begin{figure}[t]
\center{
\includegraphics[width=0.7\hsize]{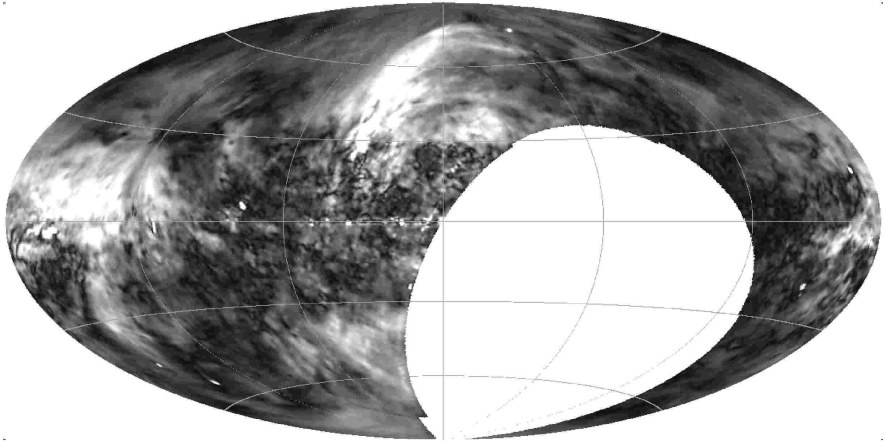}}
\caption{
Polarization intensity measured at 1.4 GHz.
The prominent Faraday depolarization
limits use of low-frequency maps
for foreground modeling,
despite the finer angular resolution.
From \citet{wolleben/etal:2006}. The gray-scale runs from 
0 to 450 mK in brightness temperature.
\label{fig:drao_map} 
}
\end{figure}
%---------------------------------------------------------

\subsubsection{Modeling considerations}

Given the spatial variation in the observed synchrotron polarization fraction, 
it is unrealistic to model the polarized signal as a 
synchrotron intensity scaled by a single global polarization fraction, as 
illustrated in Figure \ref{fig:polar_fraction}. This implies that for a given
frequency, the emission in every pixel should be modeled with at 
least two parameters for the synchrotron Q and U (or P and $\gamma$).
In terms of the spectral behavior, the power law approximation 
seemed to be sufficient for cosmological analyses well outside the plane for 
\wmap\ \citep{page/etal:2007,gold/etal:prep}. However, 
a power law fit in smaller pixels, and a curvature fit either globally or
in pixels, are likely necessary when pushing the foregrounds down 
another factor of 10-100. 
Values for the synchrotron index in the range $-3.5 \ltsim \beta_s \ltsim -2.5$ 
are 
physically reasonable, with index curvature expected to lie 
in the range $-0.5 \ltsim C \ltsim 0.5$, consistent with current observations. 
Additional parameters beyond the curvature 
may be required to quantify the frequency 
dependence, although they are not necessary to fit current data.
Diffusion of electrons limits the possible spatial variation of 
the spectral index at small angular scales (see e.g., 
\citet{strong/moskalenko/ptuskin:2007}),
so models where the index is fit in larger 
regions may be considered. 
In principle the spectral indices of the 
observed synchrotron intensity and polarization 
should be modeled with independent parameters, due to 
possible depolarization effects from the Galactic magnetic field, 
although a difference between the two 
has not yet been conclusively detected. This is due 
to both uncertainty in the spectral 
index of the synchrotron intensity, and the instrumental 
noise levels of the polarization measurements.

\subsection{Dust emission}
\label{subsec:dust}

Galactic emission in the 100 -- 6000 GHz frequency range ($\lambda =
50$ -- 3000 $\mu$m) is dominated by thermal emission from warm ($\sim
10$ -- 100 K) interstellar dust grains.  Here `thermal' refers to
emission from thermal fluctuations in the electric dipole moments of grains. 
The spectral shape of this
emission is generally modeled with one or more thermal components
modified by a frequency-dependent emission. 
That is, for a single
temperature, the intensity is proportional to $I(\nu) \sim \nu^\beta
\, B_\nu(T)$, where $B_\nu(T)$ is the Planck function at frequency
$\nu$ and temperature $T$\@. Multiple temperatures and spectral indices
($\beta$) are often needed to model the intensity spectrum at any
single point on the sky.  Using data from the all-sky surveys of
\iras\  at 100\,$\mu$m \citep{neugebauer/etal:1984}, and \cobedirbe\
at 100 and 240 $\mu$m \citep{fixsen/etal:1998}, \citet{finkbeiner/davis/schlegel:1999}
have modeled this Galactic dust emission using two temperature
components, with $\langle T_{1,2} \rangle = 9.5$ and 16 K; $\beta_{1,2} =
1.7$ and 2.7.  Both components are in equilibrium with 
the interstellar radiation field. 
They then used this model to predict thermal dust
emission at microwave frequencies.  The resulting model is 
calibrated from \cobedirbe\ and has the angular
resolution of \iras\ ($\approx 6^\prime$).  The predicted signal, shown in 
Figure \ref{fig:fds},
has been found consistent with \wmap\ total intensity data
\citep{bennett/etal:2003}, especially in the 61 and 94 GHz (V and W)
bands where the dust contribution is greatest\@.  
This model has become the standard template for
removing the contribution of Galactic dust from microwave
observations.

To date there are no similar models that can be used to remove the
polarized emission from these same interstellar dust grains.  \wmap\
polarization observations
\citep{page/etal:2007,kogut/etal:2007,dunkley/etal:prepb} at V and W-band
show evidence of a dust polarization fraction ranging from $\sim 1\%$
in the direction of the Galactic center to a few percent at
intermediate latitudes, consistent with the observations of Archeops
at 353 GHz \citep{ponthieu/etal:2005}.  However, due to the
high noise level of the V and W-band polarization measurements, an
estimate of the polarization angle is possible only on very large
angular scales ($\gtrsim 4^\circ$, $\ell\sim50$), and does not have
high signal-to-noise.  Therefore, additional information is needed in
order to remove polarized dust foregrounds on both large and smaller
angular scales.
However, this is not a limitation to initiating a CMBPol mission, as a 
dust model can be generated from the data if the 
mission has a rich enough data set in terms of frequency coverage 
and number of channels.
This scenario occurred for both \iras\ and \wmap, where data
was used both to construct foreground templates and to measure the CMB spectrum.

\begin{figure}[t]
\center{
\epsfig{file=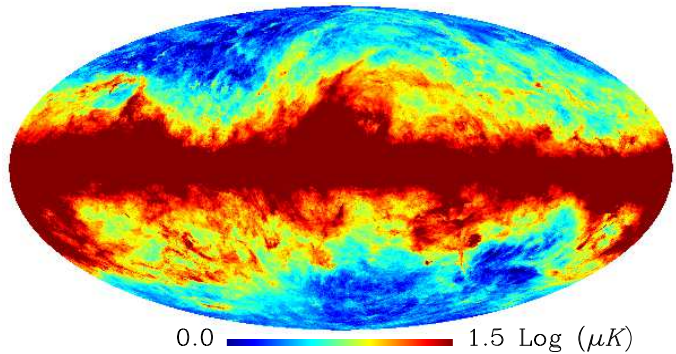,angle=90.,width=9cm,height=5.5cm}
\caption{Thermal dust intensity observed by \iras\ 
and extrapolated to 94 GHz by \citet{finkbeiner/davis/schlegel:1999}. This 
map is commonly used to identify regions of expected low dust 
emission, with the minimum at  Galactic coordinates $(l,b)\sim(240,-70)$. 
Although the dust polarization fraction is expected to vary spatially, this
map is also useful for identifying regions of potential low dust polarization. 
\label{fig:fds}}
}
\end{figure}
 
\subsubsection{Dust Alignment}  \label{sec-alignment}

Thermal dust polarization results from aspherical dust grains that
have been aligned by interstellar magnetic fields.  The exact grain
alignment mechanism is complex so we provide only a brief description
here. Basic alignment requires the grain's small axis, with the
largest moment of inertia, to align with the spin 
axis, followed by alignment of the the spin axis with the local
magnetic field (e.g.,
\citealt{davis/greenstein:1951,whittet:2003,lazarian:2003,lazarian:2007,roberge:2004}).
While other alignment mechanisms (e.g., mechanical alignment;
\citealt{gold:1952}) may dominate in some select environments, the
above mechanism is favored in conditions prevalent throughout most of
the ISM\@.  Modern grain alignment theory favors grain-photon
interactions, or radiative torques, as the mechanism by which grains
achieve high-angular velocities and subsequent alignment (e.g.,
\citealt{dolginov:1972,dolginov/mytrophanov:1976,draine/weingartner:1996,draine/weingartner:1997,lazarian/hoang:2007,Hoang/Lazarian:2008}). The
result of these mechanisms is to align the longest grain axis
perpendicular to the local magnetic field.
All grains will emit, and absorb, most efficiently along the long
grain axis. Therefore, polarization is oberved perpendicular to the
aligning field in emission, but parallel to the field in absorption or
extinction.  

Polarized thermal emission at $\lambda = 60$ -- 1000
$\mu$m has been observed in a wide variety of dusty Galactic objects
and typically has a polarization fraction of 0.5 -- 10 \% (e.g.,
\citealt{dotson/etal:2000,curran/chrysostomou:2007,matthews/etal:2008,dotson/etal:2009}).
To date, polarization from the extinction of background starlight at
near-visible wavelengths, believed to be due to differential absorption, has been observed for more than 10,000 stars,
typically with values $\sim 1$ -- 4\% (e.g.,
\citealt{heiles:2000,berdyugin/etal:2004}).  Both types of
observations have been used to study magnetic field structure
throughout the Galaxy (e.g.,
\citealt{chuss/etal:2003,schleuning:1998,matthews/etal:2001,fosalba/etal:2002,zweibel/heiles:1997,pereyra/magalhaes:2004,pereyra/magalhaes:2007})
as well as to investigate the alignment mechanism and physical dust
characteristics (e.g.,
\citealt{hildebrand/dragovan:1995,aitken:1996,hildebrand/etal:1999,whittet/etal:2001}). For further discussion on grain alignment see \citet{fraisse/etal:prep}.

\subsubsection{Polarized Dust Emission} \label{sec:poldust}

Current instrument sensitivity limits preclude a comprehensive
characterization of the properties of polarized dust emission at
millimeter wavelengths.  Existing measurements of diffuse regions are
limited to large beams (intrinsic or binned) usually greater than
$\sim 2^\circ$ (e.g.,
\citealt{ponthieu/etal:2005,page/etal:2007,takahashi/etal:2008}).
High-sensitivity measurements of polarized dust have thus far been
limited to
dense regions of the ISM\@.  These are generally regions in the
Galactic plane (with optical extinctions exceeding $A_V \sim 20$
magnitudes) which are avoided for CMB analysis. Because dense dusty
regions involve multiple populations of dust grains in a complex
magnetic field structure, simple parameterizations describing dust
behavior are likely to be significantly different in dense Galactic
regions than in diffuse regions targeted for CMB observations, so
care must be taken when using these measurements to predict emission
characteristics in diffuse regions. However, useful information can
still be gleaned.

Multi-wavelength observations of dense clouds in the Galactic plane
at far-infrared through millimeter wavelengths
show clear evidence of a wavelength-dependent
polarization in terms of both the fractional polarization
\citep{hildebrand/etal:1999,vaillancourt:2002,vaillancourt:2007} and
position angle (e.g., \citealt{schleuning/etal:2000}).  In a sample of
the most massive clouds, the spectrum has a minimum at approximately
$350\,\mu$m (860~GHz) and can vary by up to a factor of 4 in the 60\,$\mu$m --
1\,mm wavelength range.  These changes in polarization amplitude and
position angle are attributed to multiple populations of dust grains
with different temperatures, emissivities, and polarization
efficiencies, as well as differences in the aligning magnetic field
direction.
\citet{draine/fraisse:prep} demonstrate the expected
variation of the polarization fraction with wavelength due to a
mixture of silicates and carbonaceous grains.

The structure in the 50--1000 $\mu$m (300-6000 GHz) polarization
spectrum of dense clouds results from the existence of multiple dust
temperature components at 20--60 K \citep{vaillancourt:2002}.  On the
other hand,
lower frequency emission from the diffuse IR cirrus ($\nu<300$~GHz)
is expected to be dominated by a single cold dust component
($T\sim10$~K), in thermal equilibrium with the interstellar radiation
field \citep{finkbeiner/davis/schlegel:1999}. Therefore, although
multiple polarization domains are also likely to exist along most
lines of sight in low-extinction regions (preferred for CMB analysis) 
far outside the Galactic plane, %if this prediction is accurate
this indicates that the polarization fraction will be nearly 
frequency-independent in diffuse
clouds at $\lambda > 1$\,mm, $\nu < 300$\,GHz \citep{hildebrand/kirby:2004}.

\subsubsection{Starlight Polarization} \label{sec-starlight}

Background-starlight polarization is only possible in regions of
fairly low extinction ($A_V$ less than a few magnitudes for
near-infrared observations) where near-visible photons have sufficient
mean-free-path to traverse the ISM\@. This makes it a feasible tool 
for inferring the Galactic magnetic field in regions of interest for
CMB analysis. However, the extinction places a practical limit on the most
distant stars for which polarization can be observed.
The vast majority of observed high-latitude stars are within 1 kpc of the
Sun, however at low latitude some luminous stars as far away as 2 kpc
contribute to the measured polarization signal 
(\citealt{heiles:2000,fosalba/etal:2002}).
Despite this limitation, recent analyses of such measurements suggest that
they do contain information about the uniform and random components of
the magnetic field on large scales \citep{fosalba/etal:2002}. In 
the foreground polarization model inferred from \wmap\ data,
\citet{page/etal:2007} and 
\citet{kogut/etal:2007} showed that the polarization angle derived
from starlight polarization provides a slightly better fit to the
data than the synchrotron polarization angle.  Measurements 
of polarization for stars of different distances can
also reveal the 3-D distribution of magnetic field orientations averaged
along the line of sight.

\begin{table}[!tb]
\begin{center}
\begin{tabular}{ccccc}
\hline
{Latitude} & {Distance} & {No. Stars (\%)} & {$f=P/I$($\%$)} & {E(B-V)}\\ 
\hline
      			 & Total & 4114 (75) & $1.69$  & $0.49$ \\
Low Latitude		 & Nearby & 1451 (26) & $0.94$ & $0.29$ \\
($|b| < 10^\circ$)       & Distant & 2663 (48) & $2.09$ & $0.60$ \\
\hline
  			& Total  & 1399 (25) & $0.45$ & $0.15$ \\
High Latitude	        & Nearby & 1315 (24) & $0.42$ & $0.14$ \\
 ($|b| > 10^\circ$)     & Distant & \phantom{00}84 (\phantom{0}1) & $0.89$ & $0.26$ \\
\hline
\end{tabular}
\end{center}
\caption{Mean stellar parameters by distance and Galactic latitude
  ($b$) from \citet{fosalba/etal:2002}. `Nearby' and `distant'
  denote stars within and beyond 1\,\kpc, respectively.  The
  quantities in parentheses denote the fraction, in percent, of all
  stars in the sample.}
\label{tab-pstars}
\end{table}

Tables \ref{tab-pstars} and Figure \ref{fig:star_polmap} show an analysis
from \citet{fosalba/etal:2002} using the catalog
of starlight polarization data from \citet{heiles:2000}. This analysis includes
only stars with reliable distance and extinction estimates, and small
polarization measurement uncertainties (5513 stars, 60\% of 
the \citeauthor{heiles:2000} catalog).  
Low-latitude stars have
large polarization fractions and extinctions ($f_d=P/I \approx
1.7$\%, $E(B-V) \approx 0.5$\,mag), while high-latitude sources
exhibit significantly lower values ($f_d \approx 0.5$\%, $E(B-V) \approx
0.15$\,mag).
As shown in Figure \ref{fig:star_polmap} there is a strong net
alignment of starlight polarization vectors with the Galactic plane
(see lower panel) as well as a clear alignment with the spherical
shell of Loop 1 as seen from the polarization vectors in local clouds
(upper panel).  The amplitude of the polarization vectors are also
found to vary nearly sinusoidally with Galactic longitude \citep{fosalba/etal:2002}.
The plane-of-sky polarizations are
near-minimum towards lines-of-sight tangent to Galactic spiral arms
and near-maximum towards lines-of-sight perpendicular to spiral arms.
This observation is consistent with a model in which the Galactic
magnetic field follows the spiral arm structure
(e.g. \citealt{heiles:1996,beck:2007}).  Further reconstruction of the
three-dimensional Galactic magnetic field structure requires
complementary observations including radio synchrotron
polarization, IR--submillimeter dust polarization and Faraday
rotational measures from distant pulsars and quasars (e.g.,
\citealt{han/etal:2006,han/zhang:2007,han:2007}). 
New stellar polarization data from an optical survey 
described in \citet{magalhaes/etal:2005} is expected in the next 
few years, and proposals for new observations have been submitted. 
See \citet{fraisse/etal:prep} for further discussion.

\begin{figure}
\center{
  \includegraphics[angle=90, width=0.7\hsize, height=0.5\hsize]{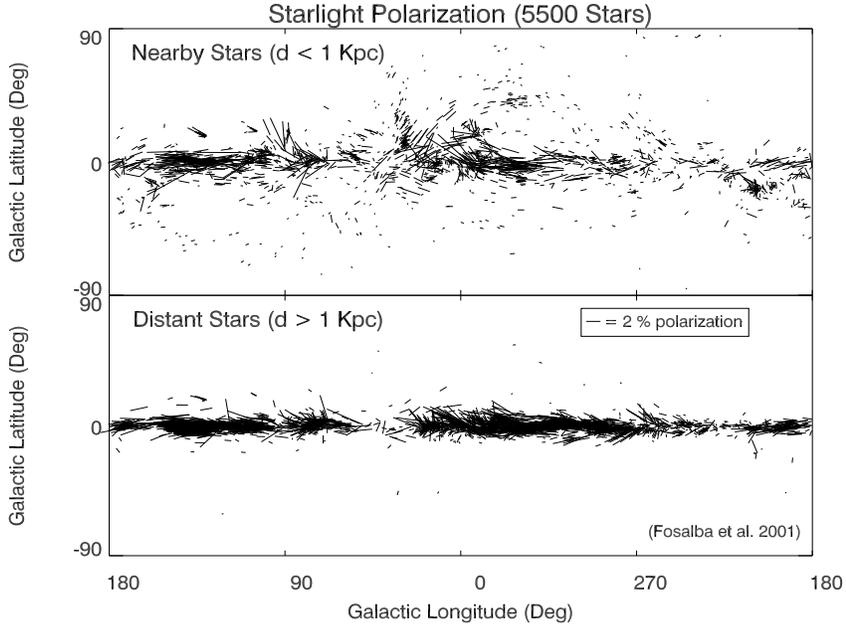}
  \caption{\label{fig:star_polmap}
Starlight polarization vectors in Galactic coordinates for
a sample of 5513 stars. The upper panel shows 
polarization vectors in local clouds, while the lower panel displays 
polarization averaged over many clouds in the Galactic plane.
The length of the vectors is proportional to the polarization degree
and the scale used is shown in the lower panel. For starlight the direction of 
polarization is parallel to the magnetic field. From \citet{fosalba/etal:2002}.}
}
\end{figure}

\subsubsection{Modeling considerations}

The expectation of changing polarization with position indicates that
at a given frequency, two parameters are required to quantify the Q and U 
Stokes parameters (or P, $\gamma$) in each pixel. The fractional polarization
$f=P/I$ is expected to be everywhere less than $\sim$15\% 
\citep{draine/fraisse:prep}, but may range from less than 1\% in some regions 
to 10-15\% in highly polarized regions.
The dust intensity can be used to provide additional constraints and 
will be better understood with Planck observations in hand. 
Possible models for the polarization fraction include inverse variation 
with dust column density, or with some measure of the interstellar
radiation field intensity.
The number of parameters required to model 
the frequency dependence of the polarized intensity in a given direction 
is not yet established, although Sec \ref{sec:poldust} 
indicates that optimistically 
one may be sufficient at frequencies below $300$ GHz, as the dust 
polarization is likely to be dominated by a single 
component. A simple power law index can be modeled with one 
parameter in each pixel, and could take values in the range 
$1 \ltsim \beta_d \ltsim 3$, with 
$\beta_d=1.7$ fitting current temperature and polarization 
data at $\nu<100$~GHz. However, since 
the dust consists of at least two
components with different indices \citep{finkbeiner/davis/schlegel:1999},
additional parameters may be required to quantify their frequency dependence.
Extrapolating from frequencies above $\sim 300$~GHz 
is likely to lead to modeling 
errors, and therefore bias the estimated signal if power law behavior is 
assumed. 

\subsection{Free-free emission}
\label{subsec:free}

Free-free (bremsstrahlung) emission is due to electron-electron 
scattering from warm ($T_{e}\approx 10^4$~K) ionized gas in the ISM. In the 
optically thin regime, and at radio frequencies, the brightness temperature 
is given by 
$T_b=8.235\times10^{-2}a T_e^{-0.35}\nu_{\rm GHz}^{-2.1}(1+0.08)({{\rm EM})_{{{\rm cm}^{-6}}{\rm pc}}}$ where $T_e$ is the electron temperature, EM is the 
Emission Measure and $a$ is a factor close to unity 
\citep{rybicki/lightman:1979,dickinson/etal:2003}. The electron temperature 
in the warm ionized medium ranges from $\sim 3000$ to $20000$~K but 
is $\sim 8000$~K at the Solar Galactocentric distance. The spectrum follows 
a well-defined power-law with a spectral index of $-2.1$ showing only a small 
variation with temperature. At \wmap\ frequencies, the effective free-free 
spectral index is $\sim -2.14$ for $T_e\sim 8000$K.
For modeling purposes, a reasonable prediction for the intensity signal outside
the Galactic plane is the H$\alpha$ map compiled by 
\cite{finkbeiner:2003} with a correction for dust extinction 
\citep{bennett/etal:2003}, or the MEM map obtained from 
the \wmap\ analysis \citep{gold/etal:prep}.

Free-free emission is intrinsically unpolarized because the scattering 
directions are random. However, a secondary polarization signature can 
occur at the edges of bright free-free features (i.e. HII regions) from 
Thomson scattering \citep{rybicki/lightman:1979,keating/etal:1998}. This 
could cause significant polarization ($\sim 10\%$) in the Galactic plane, 
particularly when observing at high angular resolution. However, at high 
Galactic latitudes, and with a relatively low resolution, we expect the 
residual polarization to be $<1\%$. Therefore, free-free emission is 
unlikely to be a major foreground contaminant for CMB polarization 
measurements in clean regions of sky.

\subsection{Additional polarized components}
\label{subsec:anom} 

There are at least two additional dust emission 
mechanisms that have been suggested in the literature that 
could produce a low level of polarized emission.
A number of high Galactic latitude studies
\citep{kogut/etal:1996a,deoliveira-costa/etal:1997,leitch/etal:1997,hildebrandt/etal:2007},
as well as studies of individual Galactic clouds
\citep{finkbeiner/langston/minter:2004,watson/etal:2005}, have
observed emission in excess of that expected from the
three foreground components discussed above, namely synchrotron,
thermal dust, and free-free emission.  This emission has been termed
\emph{anomalous} for the simple reason that its provenance is not
completely understood at this time.  However, it is clear that this
anomalous emission is highly correlated with large-scale maps of far infrared
emission from thermally emitting dust grains (e.g.,
\citealt{finkbeiner:2004,davies/etal:2006, dobler/finkbeiner:2008a,dobler/draine/finkbeiner:prep}). \citet{bennett/etal:2003}
suggested that this correlation with the \wmap\ data could be explained by
synchrotron emission from dusty star-forming regions. Alternatively,
\citet{draine/lazarian:1998a,draine/lazarian:1998b,draine/lazarian:1999}
invoked microwave emission from the dust itself to explain the excess.  
The hypothesis that this
emission is dust-correlated synchrotron is disfavored by \wmap\
polarization maps since they resemble the 408\,MHz synchrotron map at
23\,GHz, whereas the total intensity map at that frequency much more
closely resembles the 100\,$\mu$m dust map.  Dust-correlated
synchrotron would be completely ruled out by upcoming surveys like 
C-BASS at 5\,GHz.  Such low frequency surveys
will be essential for further tests of these hypotheses as synchrotron
and microwave dust emission have divergent spectra below the lowest
\wmap\ frequency at 23\,GHz.

The correlation of anomalous emission and thermal dust emission has
led to two hypotheses in which interstellar dust grains themselves are
the source of this emission (see review by
\citealt{lazarian/finkbeiner:2003}).  In the first mechanism, small
(radii $\sim 0.001$\,$\mu$m), rapidly rotating grains emit electric
dipole radiation at microwave frequencies
\citep{draine/lazarian:1998a,draine/lazarian:1998b,ali-haimoud/hirata/dickinson:prep}.  In the second
mechanism, larger (radii $\gtrsim 0.1$\,$\mu$m), thermally vibrating
grains undergo fluctuations in their magnetization, thereby emitting
magnetic dipole radiation, also at microwave frequencies
\citep{draine/lazarian:1999}.

The small `spinning-dust' grains can be stochastically heated to
temperatures approaching 100\,K \citep{draine/li:2001,ali-haimoud/hirata/dickinson:prep} resulting in
significant mid-infrared emission ($\lambda \sim 10$ -- 30 $\mu$m)
whereas the larger `magnetic-dust' grains typically emit in thermal
equilibrium in the far-infrared ($\lambda \sim 50$--200 $\mu$m). 
Observations which show stronger correlations between the excess
emission and the shorter-wavelength (12 and 25 $\mu$m) \iras\ bands,
as compared to the longer-wavelength bands (60 and 100 $\mu$m),
clearly appear to favor spinning-dust over magnetic-dust
\citep{deoliveira-costa/etal:2002,casassus/etal:2006}.   
Furthermore, correlations of the microwave excess with H$\alpha$ emission
peaking at $\sim$40 GHz \citep{dobler/finkbeiner:2008b,dobler/draine/finkbeiner:prep} also 
favor spinning-dust in the warm ionized medium (WIM), as 
both H$\alpha$ and WIM spinning dust are expected to be proportional to the 
density squared of ions. Despite the
strong evidence for spinning dust, emission from magnetic-dust must
exist at some level, as large grains are known to exist from observed 
emission in the far infrared, and contain ferromagnetic material which is depleted from the gas-phase (e.g. \citealt{savage/sembach:1996,sembach/savage:1996}). This is
especially important for polarization observations as magnetic dust is
predicted to be much better aligned than spinning dust.

The small grains responsible for spinning-dust emission are the same
grains responsible for starlight extinction at ultraviolet wavelengths
(e.g., \citealt*{mathis/etal:1977}).  Theoretical models of 
grain alignment predict spinning-dust
polarization could be as high as 7\% at 2\,GHz, falling to $\lesssim
0.5$\% at frequencies above 30\,GHz \citep{lazarian/draine:2000}.
Observationally, the relatively low polarization
observed at UV wavelengths suggests that these small grains are
inefficiently aligned \citep{whittet:2004,martin:2007} and will,
therefore, produce little polarized emission at any frequency.
The larger vibrating magnetic-dust grains are also responsible for the
thermal emission seen at far-infrared and submillimeter wavelengths.
There is clear evidence that these grains are well aligned since their
polarized thermal emission is observed to be as high as 10\%
(\S\ref{sec-alignment}).  Theoretical models suggest that magnetic
dipole emission from these grains can reach levels as high as 40\%
\citep{draine/lazarian:1999}.  Additionally, the polarization angle is
predicted to exhibit $90^\circ$ flips as a function of frequency
within the $\sim 1$ -- 100 GHz range; at high frequencies the position
angle is perpendicular to the aligning magnetic field (parallel to the
angles of polarized spinning-dust, thermal dust, and synchrotron
emission), but parallel to the field at lower frequencies.

Regardless of the exact emission mechanism, the 
polarization will need to be explored and accounted for in 
foreground cleaning, especially
since the morphology and frequency dependence of the emission is not
particularly well known. The 23\,GHz \wmap\ map, as well as targeted
observations of specific Galactic clouds
(\citealt*{mason/robishaw/finkbeiner:2008};
\citealt{battistelli/etal:2006,dickinson/etal:2007}) suggest that the
polarization of the anomalous emission is low, $\aplt$ 5\%, with a 
measurement of $3.4^{+1.5}_{-1.9}$\% by \citet{battistelli/etal:2006}.  This is
consistent with the models of polarized spinning-dust discussed above, but 
cannot rule out significant
contributions from polarized magnetic-dust emission.  
It is worth noting however that the challenges of addressing these anomalous mechanisms 
are not insurmountable, and similar challenges have been addressed successfully with 
previous experiments including {\sl FIRS}, {\sl DMR}, \firas, and \wmap. There are 
many ways to approach and remove foregrounds provided that low noise 
observations are made at a 
sufficient range of frequency channels at 20 -- 1000 GHz. 
Prior to the launch of a CMBPol mission, ground based surveys will cover 
the lower frequency range. In
addition, complimentary targeted searches of specific Galactic clouds
will continue to constrain the polarization properties.

Finally, it is possible that in addition to these dust emission mechanisms, 
there may also be other exciting discoveries awaiting us that at present 
are not in our models. 

\subsection{Simulated foreground maps and spectra}
\label{subsec:sim}

By drawing together current observations and theoretical 
predictions,
we generate simulated maps of the polarized Galactic 
emission with pixels of side 4 and 0.5 degrees\footnote{We use 
HEALPix \nside=16 and 128.}. In anticipation of the Planck satellite mission, 
the `Planck Sky Model' (PSM) has been developed \citep{leach/etal:prep}, 
drawing on observations
including \wmap, \iras, and 408 MHz radio data. 
The Planck Working Group 2 have developed
this model and allowed us to use components of it for this work.
For initial forecasting purposes we generate a set of foreground 
maps with 
synchrotron and dust 
components, each with power-law spectral indices. This physical 
behaviour is over-simplified but useful 
for preliminary tests. The maps with 4 degree pixels are 
smoothed with a 7 degree Gaussian beam, and those with 0.5 degree pixels with a
1.5 degree beam.

\begin{figure}[t]
\center{
\epsfig{file=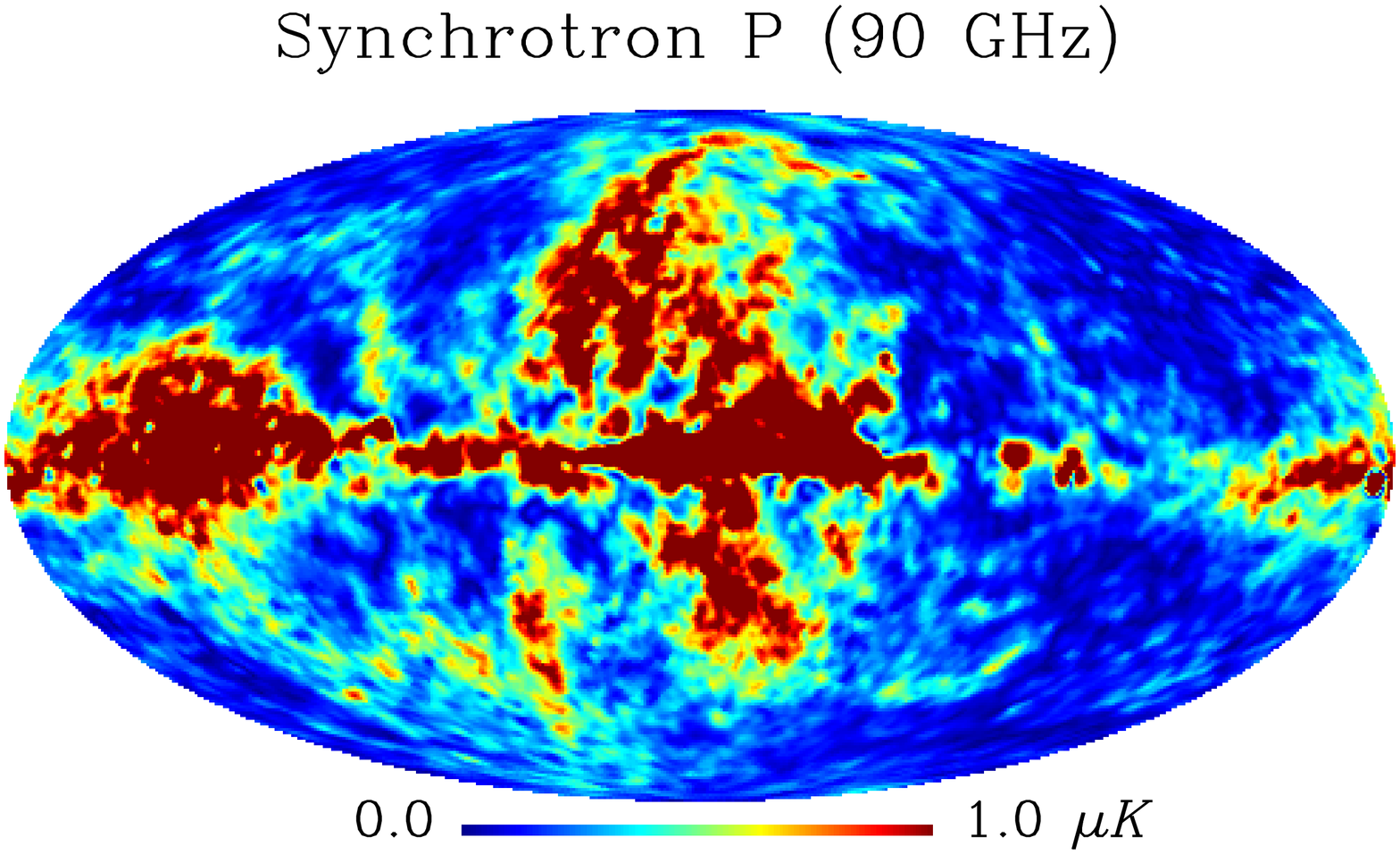,angle=90.,width=7cm,height=4cm}
\epsfig{file=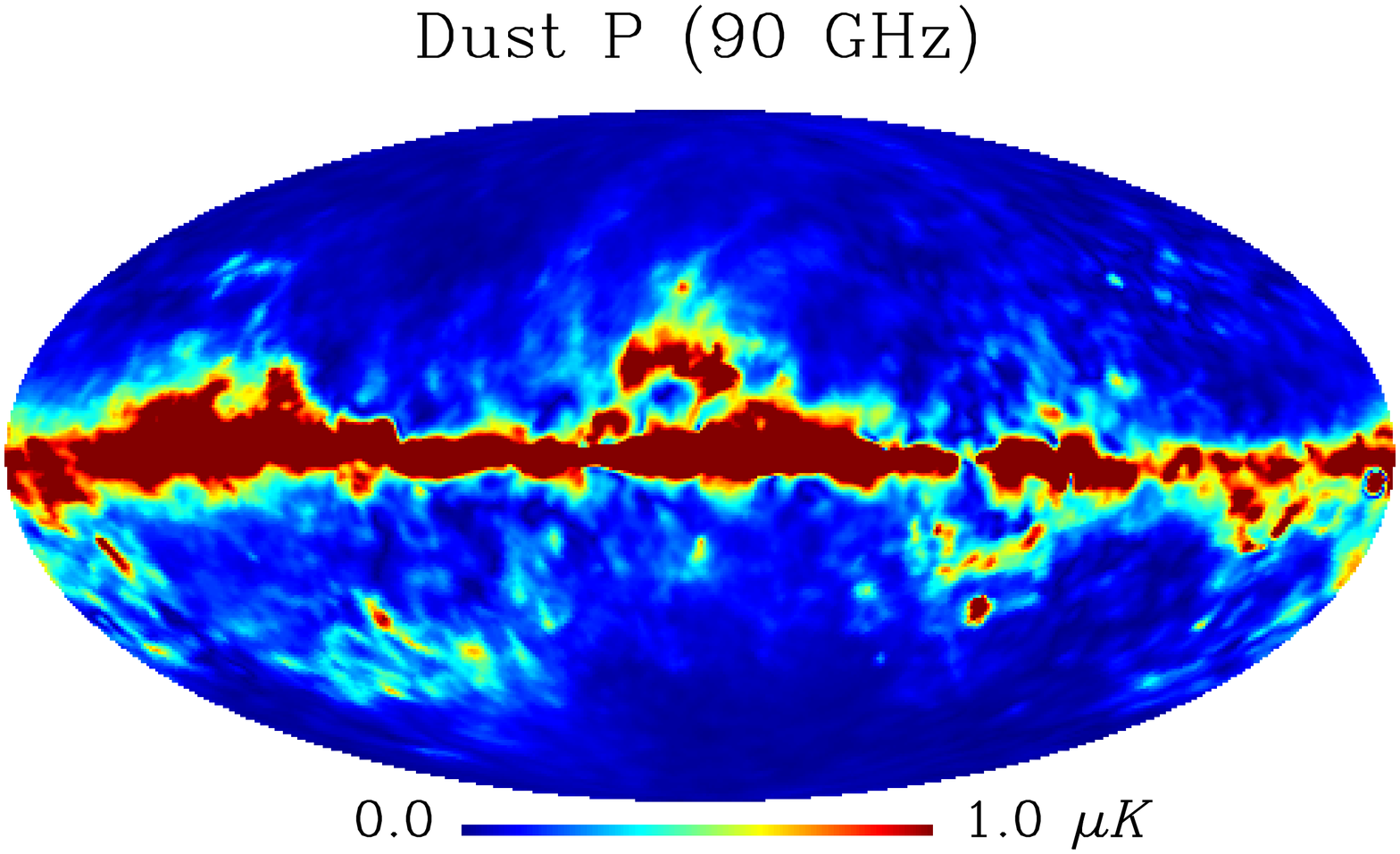,angle=90.,width=7cm,height=4cm}
\epsfig{file=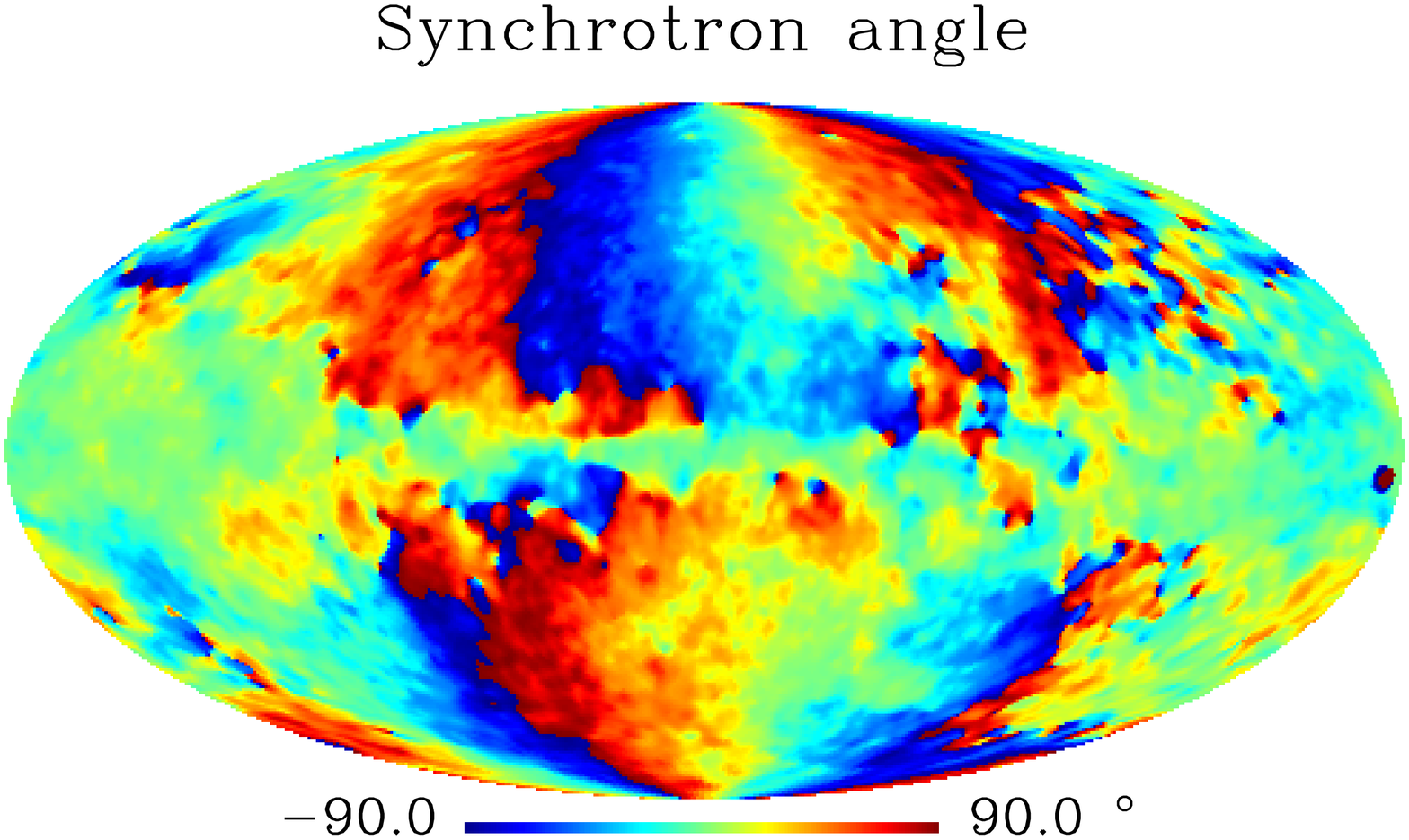,angle=90.,width=7cm,height=4cm}
\epsfig{file=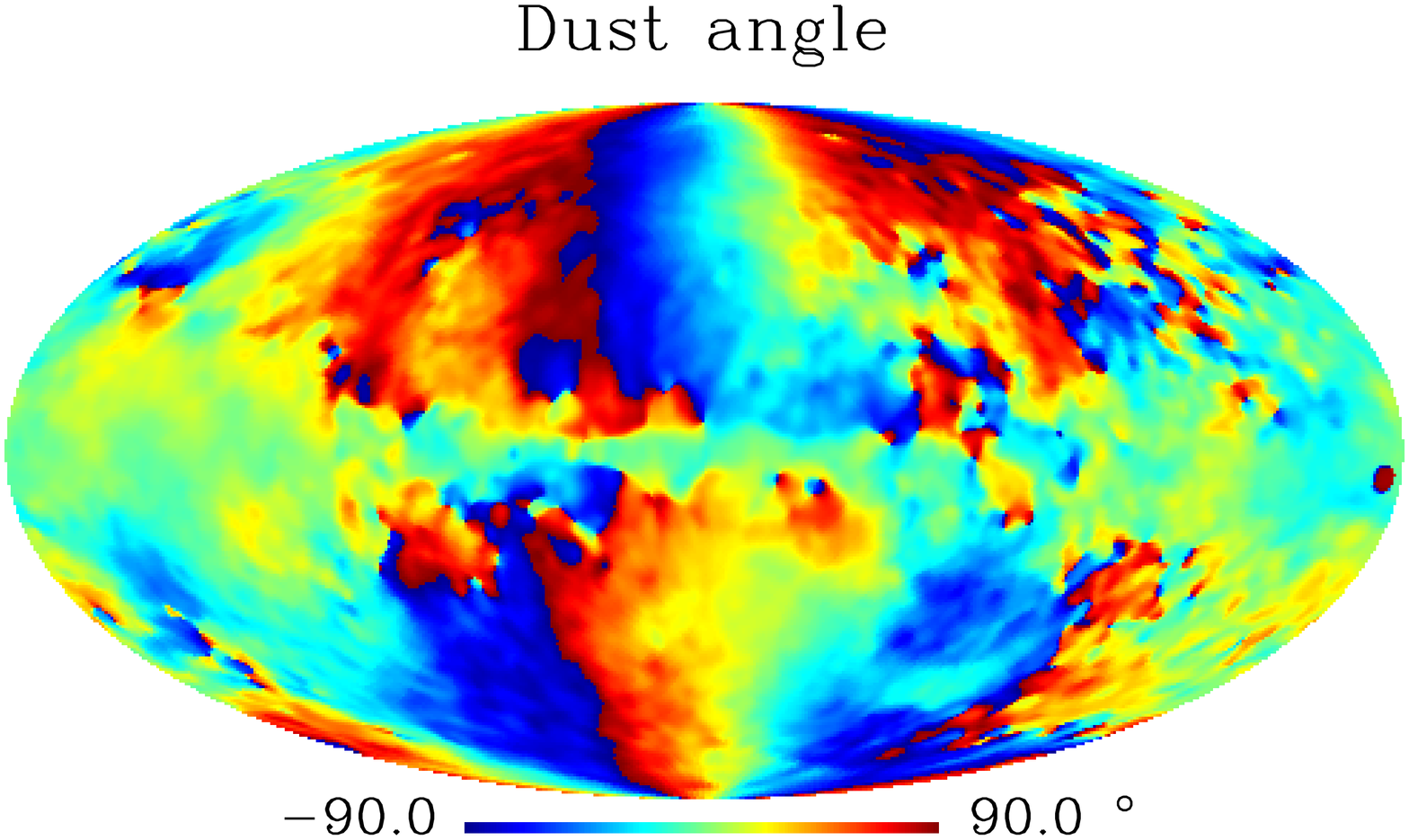,angle=90.,width=7cm,height=4cm}
\epsfig{file=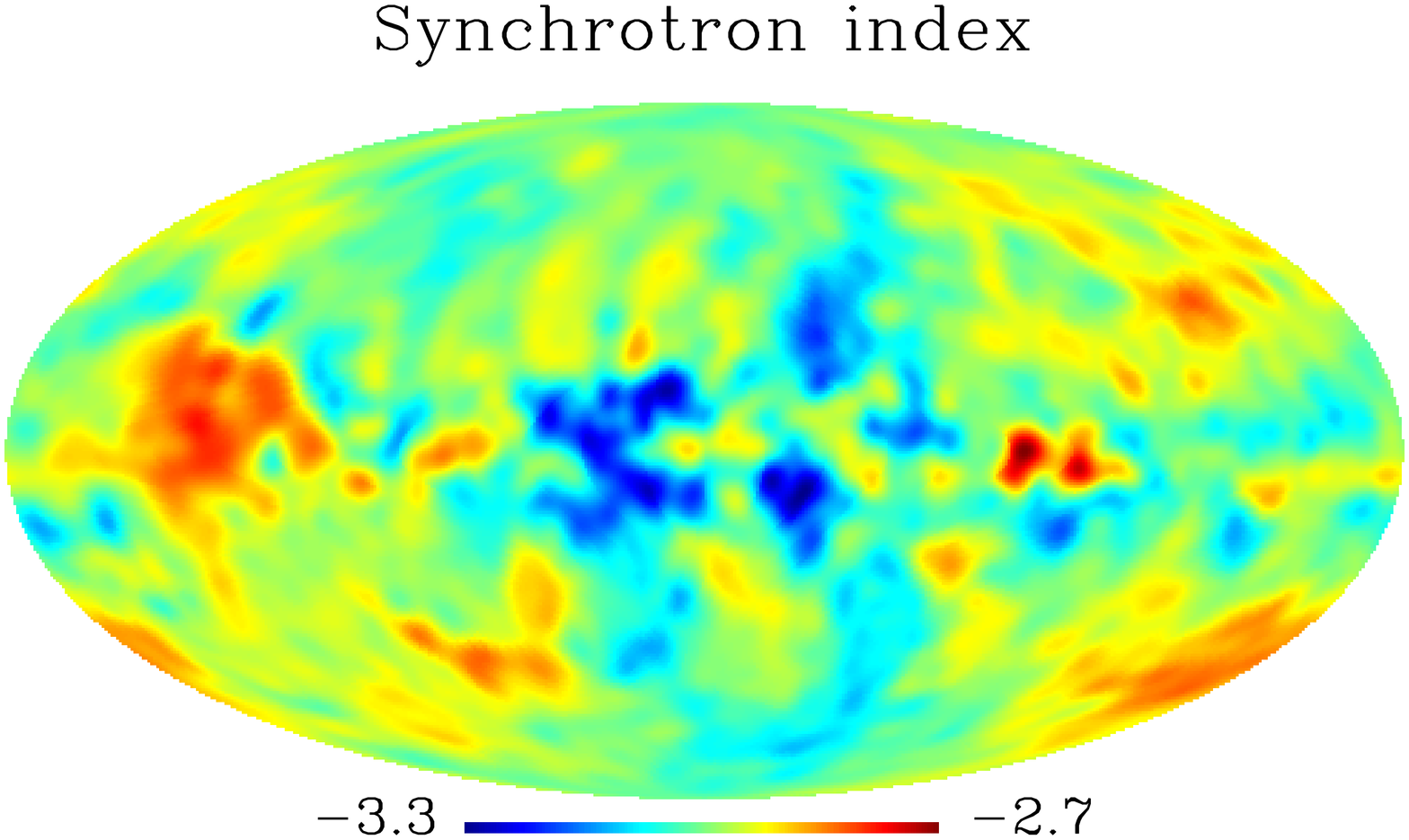,angle=90.,width=7cm,height=4cm}
\epsfig{file=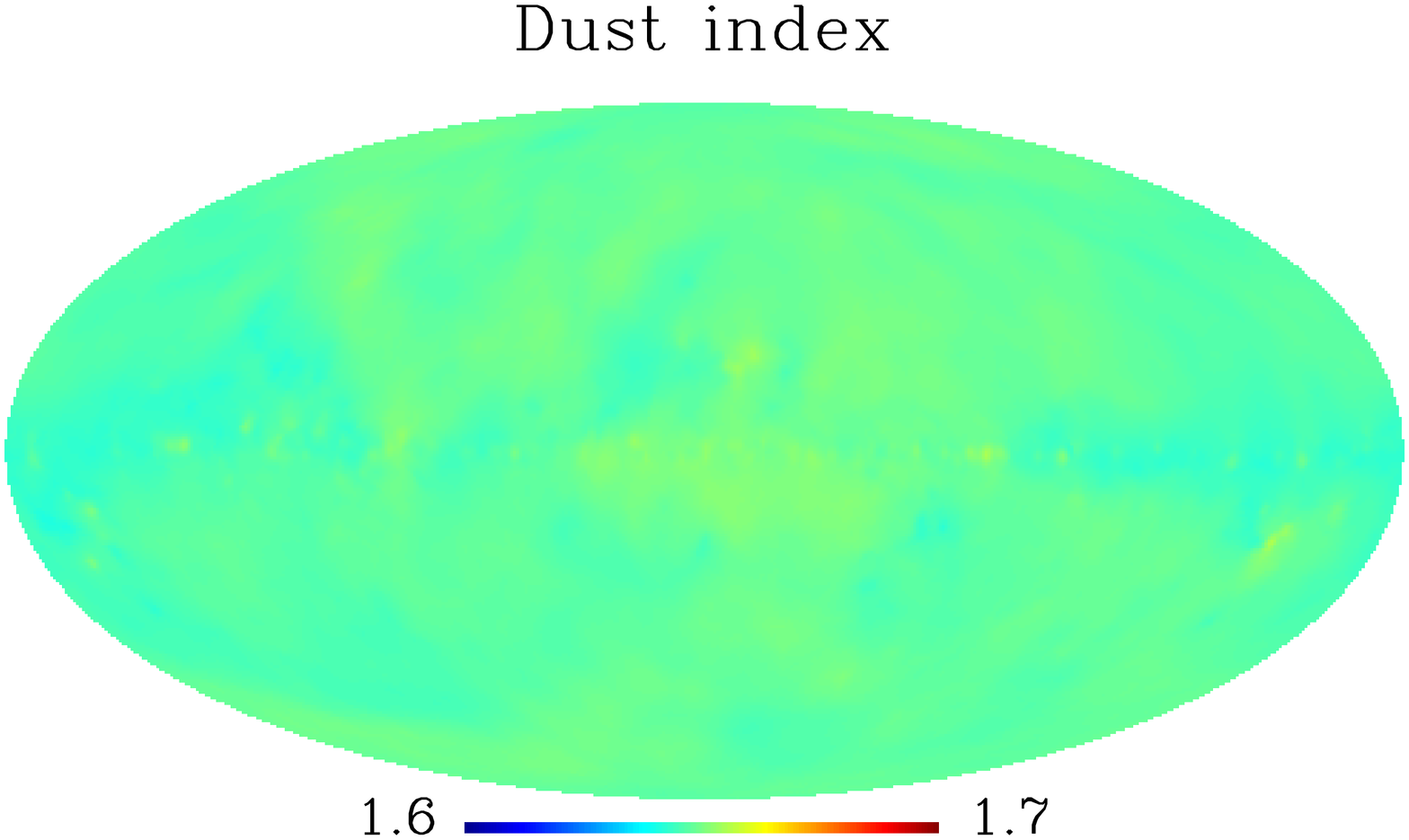,angle=90.,width=7cm,height=4cm}
\caption{Simulated maps for synchrotron (left) and 
dust (right) polarization amplitude in antenna temperature, polarization 
angle, and spectral indices at 90 GHz, 
drawn from the Planck Sky Model and described in the text.
\label{fig:sim_maps}}
}
\end{figure}

The synchrotron component is formed at each frequency $\nu$ using:
\be
Q_{\rm synch}(\nu,\hat n)= (\nu/\nu_s)^{\beta_s(\hat n)}Q_s(\hat n), \quad U_{\rm synch}(\nu,\hat n)= 
(\nu/\nu_s)^{\beta_s(\hat n)}U_s(\hat n)
\ee
For initial studies, 
the maps for $Q_s$, $U_s$ and $\beta_s$ are generated from the 
model of \citet{miville-deschenes/etal:prep}\footnote{As implemented in the PSM v1.6.2.}. They are formed using 
\be
Q_s(\hat n) = f_s g_s(\hat n) \cos(2\gamma_s(\hat n)) I_s(\hat n) \quad U_s(\hat n) = f_s g_s(\hat n) \sin(2\gamma_s(\hat n)) I_s(\hat n),
\label{eqn:synch_map}
\ee
where $I_s$ is the predicted synchrotron intensity at $\nu_s=33$~GHz 
extrapolated from observations at 408 MHz \citep{haslam/etal:1981}. The 
geometric suppression factor $g_s$, the polarization 
angles $\gamma_s$, and the polarization fraction $f_s$ are obtained using a 
magnetic field model constrained using the \wmap\ 23 GHz polarization data 
\citep{page/etal:2007} 
and described in \citet{miville-deschenes/etal:prep}. The geometric suppression
factor accounts for spatially varying 
depolarization due to line-of-sight integration through
the Galactic magnetic field, resulting in a lower observed polarization
fraction than at the point of emission. The factor does not account 
for beam dilution.

The simulated 
spectral index map $\beta_s$ is obtained by first estimating the 
synchrotron intensity at 23 GHz, using the polarization maps 
and the same Galactic magnetic field model described above, assuming 
an emitted polarization fraction of $f_s=0.75$.
A power law is then fit to the 408 MHz Haslam intensity data and this 
estimated 23 GHz intensity map. This ignores any spectral break that one
might expect in that frequency range, but a
scale dependent spectral index can be added to the simulations.
The polarization amplitude, angle, and power law spectral index 
are shown in Figure \ref{fig:sim_maps}. 

A simple parameterization for the synchrotron emission is
\be
\ell(\ell+1)C^{\rm Synch}_\ell/{2\pi} = A_s (\nu/\nu_0)^{2\beta_s} (\ell/\ell_0)^{m_s}, 
\label{eqn:synch}
\ee
although this oversimplifies the true behavior. Estimates for 
these quantities are 
$A_s \sim 0.012$ $\mu {\rm K}^2$, $\beta_s\sim-3.0$, and $m_s\sim -0.6$ 
at $\nu_s= 90$~GHz 
and $\ell_0=10$, based on \wmap\ observations by \citet{page/etal:2007}.
Figure \ref{fig:sim_ang_spec} shows this spectrum as a function of 
angular scale, compared to the primordial CMB signal with tensor-to-scalar 
ratio $r=0.01$. The estimated synchrotron amplitude (given by the square 
root of the spectrum) is about three times bigger than the 
CMB signal for $r=0.01$ at $\ell=100$ over 75\% of the sky at 90 GHz, 
and about ten times bigger than the CMB signal at $\ell=4$. Over the cleanest 3\% of the sky the signal is expected to be comparable to the CMB signal for 
$r=0.01$ .

\begin{figure}[t]
\center{
\includegraphics[width=0.7\hsize]{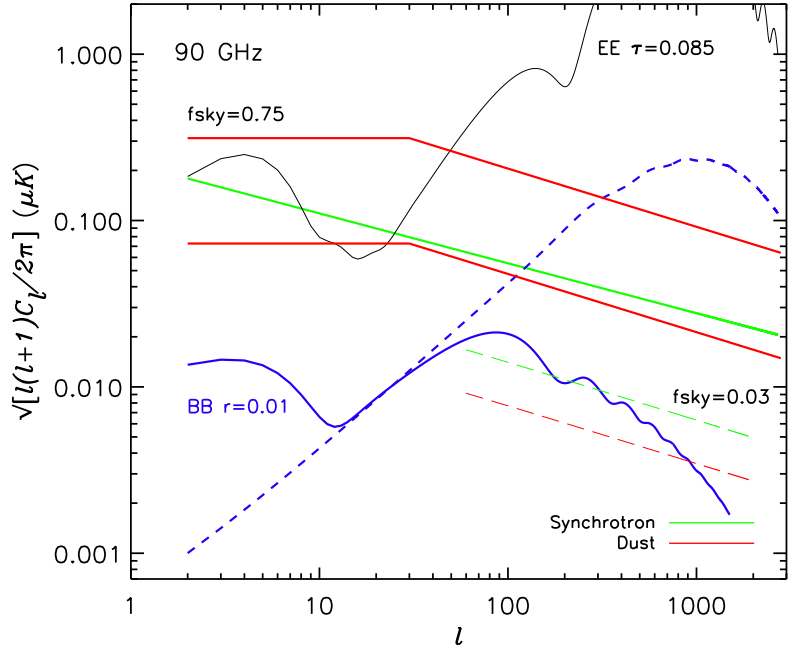}}
\caption{
Compilation of measured and predicted E-mode and B-mode 
amplitudes at 90 GHz for synchrotron and dust emission compared
to the lensed scalar (long dash blue) and tensor (solid blue) CMB signal with $\tau=0.085$ and $r=0.01$. 
Spectra are shown in antenna temperature units for a sky fraction of 
75\%, and 3\%, in the estimated lowest foreground region.  The
synchrotron contribution at large scales is constrained by \wmap\ observations. 
The dust contribution, and the behavior of both components at $\ell \gtsim 30$,
is uncertain. The red lines show two 
realistic levels for the predicted dust spectrum, with  
5\% and 1.5\% average polarization. Only the lower level is shown for the 3\% sky fraction. Recent simulated dust maps by the Planck 
collaboration give a B-mode spectrum that lies within these limits, and has 
a flatter spectrum. 
%These levels are similar to those 
%shown in Figure 4.2 of the Wiess Report \citep{bock/etal:2006}.
\label{fig:sim_ang_spec}
}
\end{figure}

There is far more uncertainty 
in the polarization of the thermal dust component, with about an order of magnitude 
uncertainty in the amplitude, ranging from $\sim$1\% to 10\% polarized. 
To make simulations, the dust component is formed in a similar way 
to the synchrotron, using
\be
Q_{\rm dust}(\nu,\hat n)= (\nu/\nu_d)^{\beta_d(\hat n)}Q_d(\hat n), \quad U_{\rm dust}(\nu,\hat n) 
= (\nu/\nu_d)^{\beta_d(\hat n)}U_d(\hat n).
\label{eqn:dust}
\ee
The amplitudes and spectral indices are taken from the Planck Sky Model, with 
the amplitudes formed using
\be
Q_d(\hat n) = f_d g_d(\hat n) \cos(2\gamma_d(\hat n)) I_d(\hat n), \quad U_d(\hat n) 
= f_d g_d(\hat n) \sin(2\gamma_d(\hat n)) I_d(\hat n).
\label{eqn:dust_map}
\ee
Here $I_d$ is the dust intensity at 94 GHz extrapolated from
\iras\ by \citet{finkbeiner/davis/schlegel:1999}. 
In the maps we adopt as fiducial estimates, the polarization
fraction $f_d$ is set to be 5\%, and the polarization angles and 
geometric suppression factor
are calculated using the Galactic field model described for the synchrotron 
simulations. This corresponds
to the Planck Sky Model v1.6.2, and assumes that the 
dust and synchrotron have the same polarization angle. This is not 
correct along certain lines of sight in the Galaxy 
(see e.g., \citet{page/etal:2007}), but is used as a first approximation. 
The resulting map is only 1-2\% polarized due to depolarization effects. 
In more recent investigations for Planck, a higher polarization 
fraction of $12\%$ at emission is assumed, resulting in an observed 
polarization fraction of $\sim 4\%$.
For the frequency dependence we use power law indices
calculated using FDS Model 8 between 84 and 94~GHz, with an
index $\beta_d=1.65$.  
Figure \ref{fig:sim_maps} shows the polarization amplitude, angle,
and index for these maps at $\nu_d=90$~GHz.
A more realistic model would have multiple polarized dust components 
which can be modeled by adding their contributions to $Q$ and $U$ given
in Eqn. \ref{eqn:dust_map}. A two component model is used in the 
current Planck Sky Model.

Figure \ref{fig:sim_ang_spec} shows the angular power spectrum at 90 GHz for  
the maps outside the P06 sky cut, fit with a power law that 
flattens at large scales. This corresponds to $\sim 1.5\%$ polarization fraction. The power spectrum currently estimated by the Planck collaboration 
(PSM v1.6.4), has a higher polarization (3-4\%) and a 
flatter scale dependence. This model account for more of the turbulent
effects in the Galactic magnetic field that tend to flatten the spectrum 
\citep{prunet/etal:1998,prunet/lazarian:1999}. 
For comparison we also plot 5\% of the FDS dust 
intensity. Parameterizing the dust as
\be
\ell(\ell+1)C^{\rm Dust}_\ell/{2\pi} = A_d (\nu/\nu_0)^{2\beta_d} (\ell/\ell_0)^{m_s}
\ee
fails to correctly describe the large scale behavior, but is a useful starting point. 
Estimates for these quantities over 75\% of the sky 
fall within typical ranges $1.5<\beta_d<2.5$, 
$-0.5<m<0.5$, and $A_d<0.8$ $\mu {\rm K}^2$ for $\ell_0=10$ and $\nu_0=90$~GHz. 
The simulated maps with 1-2\% polarization fraction 
have $A_d=0.004$ $\mu {\rm K}^2$ at 
$\ell=10$, $\beta_d=1.65$, and typically $m=-0.5$. 
For comparison, at 90 GHz, 5\% of the FDS intensity corresponds to $A_d \sim 0.1$ $\mu {\rm K}^2$ at $\ell=10$.  An upper 
limit of $15\%$ is suggested by \citet{draine/fraisse:prep}.
For a smaller patch of sky, taking 
a circle of radius 20 degrees in a region of low 
foreground intensity, the predicted B-mode amplitude is 
significantly lower, with 
$A_d=6 \times 10^{-5}$ $\mu {\rm K}^2$, and $A_s \sim 0.0002$ $\mu {\rm K}^2$  at $\ell=100$ and $\nu=90$~GHz for the simulated maps.

Figure \ref{fig:sim_freq_spec} shows the total predicted power in the 
diffuse foregrounds maps as a function of frequency
for $\ell=80-120$\footnote{The power spectra are 
calculated with PolSpice v2.5.7 (see e.g. \citet{chon/etal:2004}).}. This is based on the PSM model (v1.6.3), with a dust polarization
fraction of only 1-2\% percent. 
%This is about a factor of three lower than the 
%current Planck estimates. 
The CMB amplitude is for a standard $\Lambda$CDM model as derived from 
\wmap\ five-year results \citep{dunkley/etal:prep}, including tensor fluctuations 
with $r=0.01$.
Extragalactic sources are not included. 
The maximum ratio of CMB to foregrounds occurs at $\sim 100$~GHz 
and moves to slightly higher frequencies for cleaner patches of sky.
At 100 GHz and $\ell=100$ 
the foreground signal is $\sim5$ times larger than the B-mode signal 
for $r=0.01$ outside the Galactic plane (in the \wmap\ Kp2 mask), 
but in a clean patch of radius 10 degrees the foreground signal is below the
CMB signal. However, there is still about an 
order of magnitude uncertainty in the total polarized dust 
contribution which should be resolved with observations from Planck, and 
from ground and balloon-based experiments.

\section{Methods for polarized CMB estimation}
\label{sec:methods}

In this section we briefly review and contrast 
methods commonly used to estimate the polarized CMB signal
from raw sky maps. We focus on those that have been applied to current 
polarization data, and that can be used to forecast the 
performance of a future mission. For a description of additional methods 
see e.g., \citet{leach/etal:prep,delabrouille/cardoso:prep}.

\begin{figure}[t]
\begin{center}
\includegraphics[width=12cm,height=16cm]{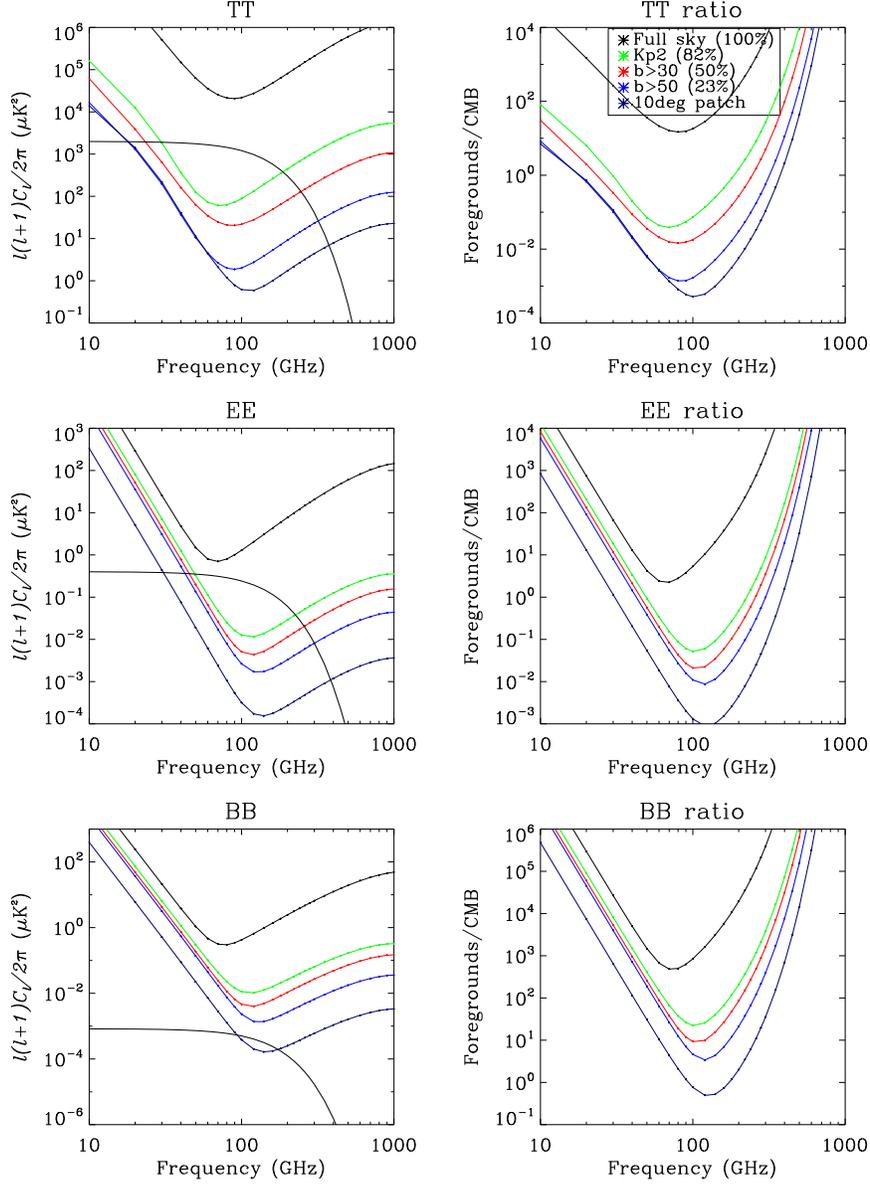}
\end{center}
\caption{
Predicted diffuse foreground power at 
angular scales $\ell=80-120$, as a function of frequency and 
sky coverage, compared to a CMB signal with $r=0.01$. 
From top to bottom are the TT, EE and BB 
power spectra in antenna temperature. 
The CMB is constant in thermodynamic temperature and thus
decreases with frequency in these units. 
{\it Left}: Total power for 
different sky coverage: full-sky, $|b|>10^{\circ}$, $|b|>30^{\circ}$, $|b|>50^{\circ}$, and a clean circular patch of radius $10^{\circ}$ centered on $(l,b)=(240^{\circ},-70^{\circ})$.
{\it Right}: Ratio of the total diffuse foreground power 
to the CMB. The maximum ratio occurs at $\sim 100$~GHz and moves to 
higher frequency for cleaner patches of sky. }
\label{fig:sim_freq_spec}
\end{figure}

\clearpage

 \subsection{Template cleaning}
\label{subsec:template}

A simple foreground cleaning technique
assumes that a map of the microwave sky
$T(p, \nu)$, at some pixel $p$ and frequency $\nu$,
can be described as a superposition of
fixed spatial templates $X(p)$ and noise $n(p, \nu)$,
\begin{equation}
T(p, \nu) = n(p, \nu) + \sum_i \alpha_i(\nu) X_i(p),
\label{template_eq}
\end{equation}
where the coefficients $\alpha_i(\nu)$
describe emission traced by the $i^{\rm th}$ spatial template
at observing frequency $\nu$.
Provided that the noise covariance matrix
(which typically includes both instrument noise and the CMB)
is known,
Eqn. \ref{template_eq} can be solved
for the template coefficients $\alpha_i(\nu)$.
The noise in the foreground-cleaned map,
\begin{equation}
T_c(p, \nu) = T(p, \nu) - \sum_i\alpha_i(\nu) X_i(p)
\label{noise+eq}
\end{equation}
is nearly unaffected by template cleaning
and retains the properties of the noise in the original map.
This is an important consideration
for applications requiring accurate propagation
of noise into, for instance, estimates of cosmological parameters.
The amplitude of the noise in the cleaned map increases
as more parameters are fit;
however,
for most applications 
there are many more pixels than fitted parameters
so the resulting increase is negligible.
A typical sky map with $10^5$ pixels
could simultaneously fit 1000 spatial templates
and suffer only a 1\% increase in the noise amplitude after cleaning.

%--------------------------------------------------------
% Template fig 1: Template sampler
%--------------------------------------------------------
\begin{figure}[t]
\center{
\includegraphics[angle=270,width=1.0\hsize]{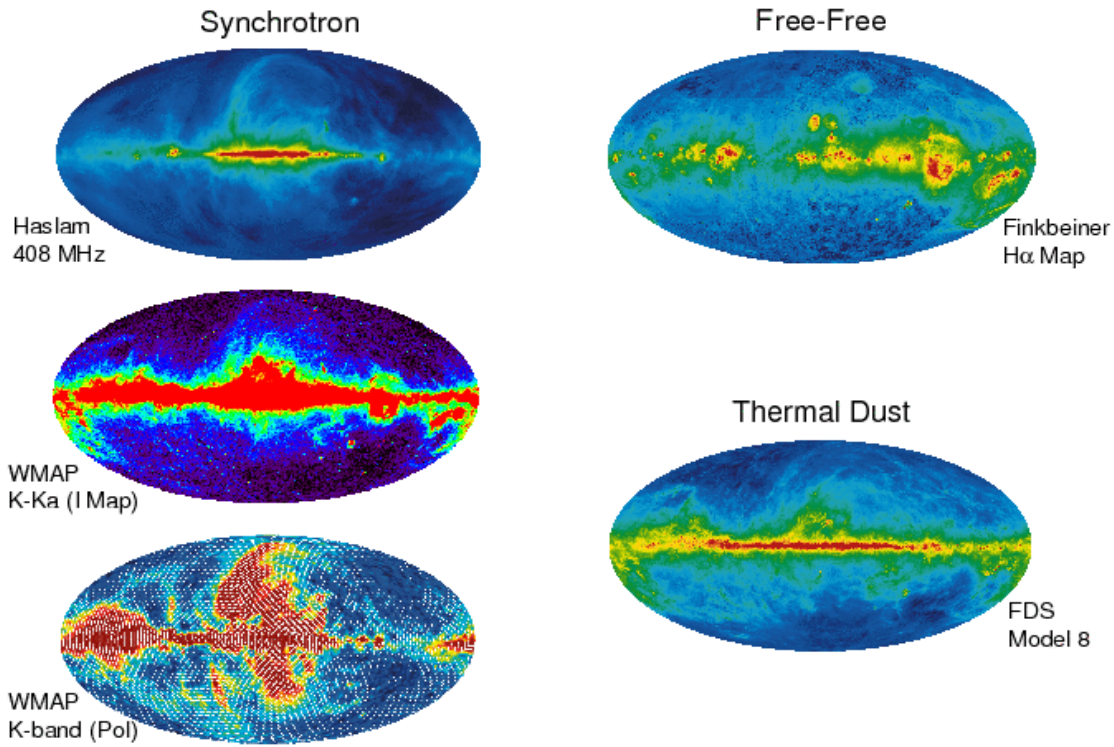}
}
\caption{
A sample of template maps commonly used to clean polarized and unpolarized maps.
\label{template_sampler} 
}
\end{figure}
%---------------------------------------------------------

Template cleaning has a number of other attractive features.
Since each template is fit to all map pixels simultaneously,
the technique makes full use of the 
spatial information in the template map $X_i(p)$.
This is important for the non-stationary, highly non-Gaussian 
emission distribution typical of Galactic foregrounds,
allowing cleaning of the entire map
even if the foreground emission 
is faint compared to the noise in each individual pixel.
The method also allows multiple template maps
to be fit to a single frequency channel,
as opposed to pixel-by-pixel techniques
which generally require at least one frequency channel
per foreground component to be fit.
In addition, template cleaning is insensitive 
to spatial correlations between various foreground components.
Correlations between the template maps
will create non-zero covariance in the fitted parameters $\alpha_i$;
however, this simply complicates the identification of
the associated emission
$T_i = \alpha_i X_i$
with the emission mechanism traced by that template,
but does not bias the total foreground estimate
summed over all templates.

Several cautions apply to template cleaning.
The method assumes that the spatial and frequency dependence
of each emission component can be described using 
separable functions for the spatial and frequency dependence,
$T(p, \nu) = X(p) f(\nu)$.
Many physical emission processes violate this assumption
at least to some extent.
The spectral index of synchrotron emission,
for instance,
is known to vary with position on the sky,
while dust is known to require at least two temperature components
whose ratio also varies across the sky.
Over sufficiently large separations in frequency,
such spatial variation in the frequency scaling
can significantly alter the spatial distribution
of the affected emission component.
In addition, the technique requires a separate template
for each spatially distinct foreground
and is thus ineffective as a `blind' test 
for foregrounds whose spatial distribution is not well matched
to one or more templates.
Template cleaning works best 
when the template map is signal dominated.
Noise in the template map
will be aliased into the cleaned map
and could become significant in clean regions of the sky
if the template noise is sufficiently large.
While this effect can be estimated using Monte Carlo techniques,
some care in template selection and sky cuts should be exercised.

Figure \ref{template_sampler} shows a selection
of template maps widely used for CMB analyses.
Synchrotron emission is often fit
using the unpolarized 408 MHz survey
\citep{haslam/etal:1981}.
Due to concerns over possible spatial variations 
in the synchrotron spectral index,
the \wmap\ team has also constructed a synchrotron-dominated template
using the difference of the two lowest frequency channels
\citep{hinshaw/etal:2007}.
The CMB anisotropy cancels exactly in such a difference map,
leaving a dominant synchrotron signal
with smaller contributions from free-free and dust, as well as 
instrumental noise.
Provided the template fit (Eqn. \ref{template_eq})
includes separate templates for free-free and dust,
the resulting linear combination of templates 
will remove all three foregrounds.
As described in Section \ref{subsec:free}, free-free emission 
from the warm ionized interstellar medium
may be traced using H$\alpha$ emission from the same ionized gas,
subject to non-negligible corrections
for optical extinction by interstellar dust
\citep{finkbeiner:2003,
schlegel/finkbeiner/davis:1998}.
Dust emission is commonly traced
using a model of millimeter-wave dust emission
based on the \cobe\ and \iras\ data
\citep{finkbeiner/davis/schlegel:1999}, described 
in Section \ref{subsec:dust} and shown in Figure \ref{fig:fds}.

Templates for {\it polarized} emission are harder to come by.
Free-free emission is largely unpolarized,
so the \wmap\ 23 GHz polarization map
can be used as a polarized synchrotron template
\citep{page/etal:2007,gold/etal:prep}.
No high signal-to-noise ratio tracer of polarized dust emission
currently exists.
A polarized dust template may be constructed 
using the \citet{finkbeiner/davis/schlegel:1999} unpolarized dust model
convolved with
optical measurements of the dust polarization angle
and a geometric model
of the Galactic magnetic field
along each line of sight
\citep{page/etal:2007}.

%--------------------------------------------------------
% Template Table 1: \wmap\ 5-year template coefficients
%--------------------------------------------------------
\begin{table}[t]
\begin{center}
\begin{tabular}{|c | c c c | c c |}
\hline
Freq  & \multicolumn{3}{c | }{Unpolarized Coefficient}  
      & \multicolumn{2}{c |}{Polarized Coefficient}  \\
(GHz) & Synchrotron & Free-Free & Dust & Synchrotron & Dust \\
\hline
33	& ---  & ---  & ---  & 0.317 & 0.017 \\
41	& 0.24 & 1.00 & 0.20 & 0.177 & 0.015 \\
61	& 0.06 & 0.65 & 0.47 & 0.060 & 0.037 \\
94	& 0.00 & 0.40 & 1.26 & 0.045 & 0.082 \\
\hline
\end{tabular}
\caption{\wmap\ Template Coefficients $\alpha_i(\nu)$
\label{wmap_template_table}}
\end{center}
\end{table}
%--------------------------------------------------------

Template cleaning is widely used for CMB analysis.
As an example, 
Table \ref{wmap_template_table}
shows the coefficients fitted
to the \wmap\ 5-year data
\citep{gold/etal:prep}.
Separate cleaning is performed for the polarized and unpolarized maps.
The unpolarized maps
are fit over the high-latitude sky
using three spatial templates:
the synchrotron-dominated K-Ka difference map,
the \citet{finkbeiner:2003} H$\alpha$ map
corrected for extinction,
and the \citet{finkbeiner/davis/schlegel:1999} 
model 8 dust map
evaluated at 94 GHz.
Since the ``synchrotron'' template 
uses both the K and Ka channels at 23 and 33 GHz,
the template cleaning is applied only to the 3 high-frequency 
unpolarized bands.
The polarized data,
at reduced pixel resolution,
are fit using the \wmap\ 23 GHz map
as a synchrotron tracer, 
plus the \citet{page/etal:2007} template for polarized dust.

Template cleaning is remarkably efficient
at compressing complicated Galactic foregrounds
into a handful of parameters.
Nine parameters describe the unpolarized foreground emission
from some seven million input pixels,
while 8 parameters describe the polarized emission
from nearly 20,000 pixels.
This simple cleaning technique produces results
in agreement with considerably more complicated 
pixel-by-pixel methods which include additional astrophysical information.

Template cleaning can also lead to surprising results.
Figure \ref{anomalous_fig}
shows the template coefficient $\alpha(\nu)$
for a template map of thermal dust emission at 240 $\mu$m
fitted to the \cobedmr\ microwave maps
at 31, 53, and 90 GHz
\citep{kogut/etal:1996a,
kogut/etal:1996b}.
At sub-mm wavelengths 
the coefficients follow the expected behavior for thermal dust emission,
but show an unexpected increase in emission at mm wavelengths.
This ``anomalous emission'', 
correlated with the spatial distribution of thermal dust
but with spectral index $\beta \approx -2.2$
(in antenna temperature)
instead of the values $1.6 < \beta_d < 2$ expected for thermal dust
at mm wavelengths,
has since been confirmed in a number of data sets. This is the same emission
that is discussed in Section \ref{subsec:anom}, and although its
physical origin is still unclear,
it is efficiently removed by template cleaning.

Template cleaning for CMB analysis to date
has operated in two distinct regimes.
Unpolarized foregrounds at high latitude are typically fainter
than the CMB anisotropy,
requiring only a light cleaning of weak foregrounds.
Polarized foregrounds are comparable to or brighter than the CMB polarization
over most of the sky.
Current detections of E-mode polarization,
at amplitudes comparable to the foregrounds,
thus require modest cleaning of bright foregrounds.
A search for primordial B-mode polarization
will push template cleaning into a third regime,
requiring deep cleaning of bright foregrounds.
Detecting a signal with tensor to scalar ratio $r = 0.01$
at multipoles $\ell < 10$
(angular scales $\theta > 20^\circ$)
requires reducing the foreground emission by a factor of roughly 20.

Happily, recent history suggests that such deep cleaning is feasible and should be 
possible with a future mission.
The cosmic infrared background (CIB)
observed at wavelengths $125 < \lambda < 2000 ~\mu$m
results from the integrated infrared emission of galaxies
at redshift $z \sim 3$.
Galactic emission at these wavelengths
is brighter than the CIB
and has a similar spectrum.
Template cleaning of the \cobefiras\ data
using only 3 spatial templates
(the 158 $\mu$m line from ionized carbon,
the 21 cm line from neutral hydrogen,
and the square of the 21 cm line intensity)
successfully reduced a bright, complicated foreground
by a factor of 10
to enable a statistically significant detection
of the CIB intensity and spectrum
\citep{fixsen/etal:1998}.
Furthermore, the ionized carbon template was derived from the \firas\ data 
itself, providing an example of a rich data set being used for internal 
cleaning.
Detecting B-mode polarization in the CMB
would require only modestly deeper cleaning,
and would have the distinct advantage
that the CMB frequency spectrum is 
known {a priori}.

Template cleaning has been demonstrated
to reduce foreground emission by at least a factor of 10.
The ultimate limit for this technique is not known.
Given the large number of existing maps
for various Galactic emission sources
and the small noise penalty to be paid for 
fitting multiple templates,
an interesting scenario would be to 
attempt a simultaneous fit of
polarization data to as many templates as possible,
including (as was done for the CIB)
higher powers of individual templates.
There will be additional templates available in 
the next five years, at low frequencies (5-20~GHz)  
from the C-Band All-Sky Survey (C-BASS), GEM-P \citep{barbosa/etal:2006}, 
and COFE \citep{leonardi/etal:2007}, 
and at higher frequencies from the Planck satellite.

%--------------------------------------------------------
% Template fig 2: anomalous emission
%--------------------------------------------------------
\begin{figure}[t]
\centerline{
\includegraphics[angle=90,width=0.7\hsize]{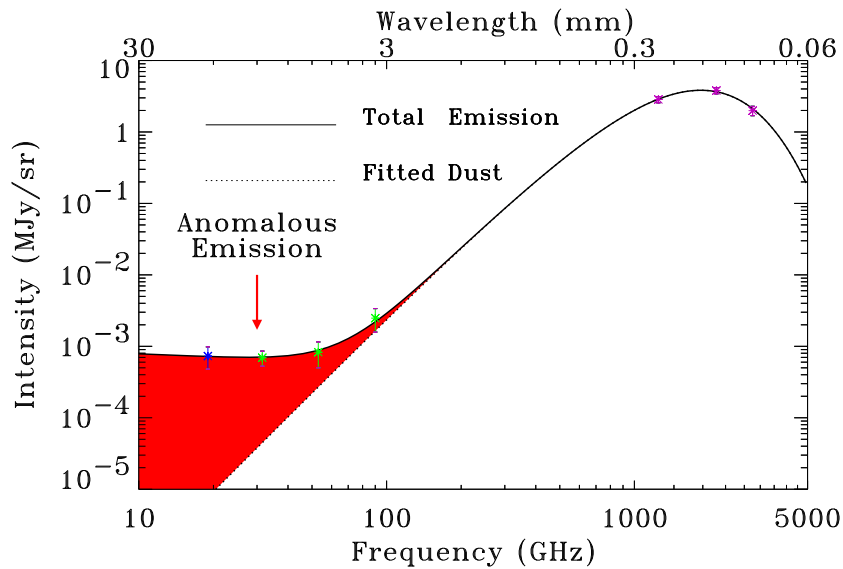}}
\caption{
The frequency dependence of emission correlated with
thermal dust shows an anomalous increase at centimeter wavelengths.
Despite uncertainty in the physical origin of this correlated emission,
it may efficiently be removed using template cleaning.
From \citet{kogut/etal:1996a}.
\label{anomalous_fig} 
}
\end{figure}
%---------------------------------------------------------

\subsection{Parametric methods}
\label{subsec:method_param}

Bayesian parameter estimation methods have begun to be applied to the 
problem of estimating the CMB signal, given observations of the 
sky at multiple frequencies. In contrast to template cleaning, which produces
a cleaned map at each frequency, these methods use the 
multi-frequency information to estimate a single CMB map.
The methods exploit our knowledge of the 
frequency dependence of different components.
The signal is parameterized by a model, 
expressed as the sum of CMB and foreground components at each frequency. 
Current methods are split into two schemes. In the first approach the 
pixel Q and U CMB Stokes parameters are estimated in 
each pixel. In the second approach the CMB angular power spectrum is 
simultaneously estimated with the map parameters.
In this framework, templates comprising the `best-guess' for the 
emission of a particular foreground component can be included as 
priors. One of the main benefits of this method is that errors due to
foreground uncertainty are rigorously propagated
into the estimated CMB maps and power spectra. One of the drawbacks is that 
polarized components that are not in the parameterized model, or that are
incorrectly described by the model, will 
not be properly accounted for. These methods are also typically computationally
demanding.

\subsubsection{Estimation of CMB maps}
\label{subsec:mcmc_map}

\citet{eriksen/etal:2006}, \citet{gold/etal:prep}, 
\citet{dunkley/etal:prepb} and \citet{stompor/etal:prep}
describe a CMB component separation approach 
based on standard Bayesian parameter estimation techniques. A 
parametric model is used for each signal component (including CMB, 
synchrotron, and dust) and the 
probability distribution for the parameters is estimated.

Using Bayes' theorem, the posterior probability is given by
\begin{equation}
P(\theta |\textbf{d}) \propto P(\textbf{d} | \theta) P(\theta) = \mathcal{L} (\theta) P(\theta),
\end{equation}
where $\mathcal{L}(\theta)=P({\textbf d} | \theta)$ is the likelihood of the observed maps, and 
$P(\theta)$ is a prior for model parameters $\theta$. 
The likelihood is given by
\be
-2 \ln {\cal L}=\sum_\nu [\mb d_\nu- \mb S_\nu(\theta)]^T \mb N_\nu^{-1}[\mb d_\nu - \mb S_\nu(\theta)],
\label{eqn:like}
\ee
where $\mb N_\nu$ is the noise covariance at each channel $\nu$, and the data
contains some combination of I, Q, and U in map space.
The model, $S_\nu(\theta)$, is minimally parameterized as the sum of 
CMB, synchrotron emission, and thermal dust emission components: 
\be
\mb S_\nu = \bga_{c,\nu} {\mb A}_c +\bga_{s,\nu} {\mb A}_s + \bga_{d,\nu} {\mb A}_d .
\ee
For the CMB the coefficients are already known: the emission is black-body. For 
dust and synchrotron emission the coefficients are often constrained
to give power law spectral indices, with
$\alpha_\nu(\hat n)=(\nu/\nu_0)^{\beta(\hat n)}$. The amplitudes maps $\mb A_i$ are defined
at pivot frequencies $\nu_0$. The goal is to estimate 
the CMB map $\mb A_c$ and its associated covariance.

The joint distribution for the amplitudes and spectral indices cannot 
be sampled directly, so in 
\citet{eriksen/etal:2006} the MCMC Metropolis-Hastings algorithm 
\citep{metropolis/etal:1953,christensen/etal:2001,lewis/bridle:2002,dunkley/etal:2005} is used to map out the the posterior 
probability distribution for the parameters $\theta$, 
using intensity data ($I$). It has also been applied to 
polarization data (using $I,Q,U$), in studies for the Planck satellite. 
In this case the 
noise is assumed to be diagonal, so each pixel is fit individually.
The full parameter set, including
component amplitudes and spectral indices, is initially 
fit pixel by pixel in low-resolution maps. 
This provides both best-fit values for each parameter and its 
associated uncertainty. 
The second sampling step is performed at higher resolution. 
All non-linear parameters are set at the best-fit values obtained in the 
low-resolution sampling, the parameter fields are smoothed 
spatially, and high-resolution amplitude maps for each component 
are then obtained analytically. 
The residual uncertainty in the measurement of the CMB emission 
in each pixel after foreground signals have been removed is estimated using 
the multiple bands. 
The technique is particularly appropriate given that Galactic emission 
anisotropy power dominates on large scales.
The code `FGFit' provides an implementation of this method and is also a 
useful forecasting tool. A similar 
method has also been implemented by \citet{gold/etal:prep} and applied to the 
5-yr \wmap\ temperature and polarization data.

An alternative method to estimate polarized CMB maps is described in 
\citet{dunkley/etal:prepb}. The principle is the same, but the sampling 
method is modified, with a Metropolis-within-Gibbs MCMC method 
used to map out the joint distribution $p(\theta|d) = p(\mb A, \bga|d)$. 
The Gibbs sampler is a special case of the Metropolis-Hastings MCMC
algorithm. Specifically, suppose the target distribution of interest
is $P(A,B)$, and we want to generate a joint sample $(A,B)$ from this
distribution. This can be achieved by alternately sampling from the
two conditional distributions,
\begin{align}
A^{i+1} &\leftarrow P(A|B^i) \\
B^{i+1} &\leftarrow P(B|A^{i+1}).
\end{align}
Here, the left arrow indicates sampling from the distribution on the
right hand side. 
In this method the parameter distribution is sliced into a conditional 
distribution for the amplitude maps, $p(\mb A|\bgb, \mb d)$ for 
fixed non-linear parameters, and a distribution for 
the spectral indices $p(\bgb|{\bf A}, \mb d)$, for fixed amplitude. The 
first distribution is a Gaussian and can be sampled directly. The second
distribution is sampled using the Metropolis-Hastings algorithm. 
An advantage of this approach is that the noise matrix does not 
have to be assumed diagonal, and spectral indices can for example 
simultaneously be estimated in larger pixels than the amplitudes. 
Priors are imposed on the foreground emission. For example, for 
application to \wmap\ data, 
\citet{dunkley/etal:prepb} assume that 
the spectral indices in Q and U are the same in a given pixel, and that in a 
map with 768 pixels, the spectral indices are only allowed to vary in 
$N_i=48$ larger pixels. In regions of low signal-to-noise priors are imposed on the 
synchrotron and dust spectral indices. A typical choice is a Gaussian 
prior of $\beta_S=-3.0\pm0.3$ on the synchrotron emission index, 
and $\beta_d=2\pm0.3$ (or broader) for the dust emission index.

While the methodology is different for these two methods, 
the principle is the same, and the resulting product is a 
CMB map and covariance matrix estimated 
from the marginalized posterior distribution. 
The likelihood of the estimated maps, 
given a theoretical angular power spectrum, can then be computed using the
method described in \citet{page/etal:2007}.
By varying only the tensor-to-scalar ratio $r$, and calculating the 
likelihood at each value of $r$, this method can be used to estimate 
limits on the tensor-to-scalar ratio including foreground uncertainty.

\subsubsection{Estimation of CMB power spectra}
\label{subsec:mcmc_cl}

Rather than estimating the marginalized maps, and then estimating the CMB 
power spectrum, it is possible to simultaneously estimate both.
Here the main goal is the joint
posterior distribution $P(\mathbf{s}, C_{\ell}, \theta_{\textrm{fg}}
|\mathbf{d})$, where $\mathbf{s}$ is the CMB sky map (denoted $\mb A_c$ in the
previous sub-section), $C_{\ell}$ is
the CMB power spectrum, $\theta_{\textrm{fg}}$ denotes the set of
foreground parameters, and $\mathbf{d}$ represents the data. From this
joint distribution, one may extract any number of marginal
distributions to obtain estimates for each parameter individually, of
which perhaps the most important is the marginal power spectrum
distribution $P(C_{\ell}|\mathbf{d})$. From this one may extract
cosmological parameters using a standard MCMC code such as CosmoMC.

The joint distribution $P(\mathbf{s}, C_{\ell}, \theta_{\textrm{fg}}
|\mathbf{d})$ involves several millions of correlated parameters, and
this poses a serious computational challenge. Brute force evaluation
is impossible. However, in recent years a new Monte Carlo
method based on the Gibbs sampling algorithm has been
developed for precisely this purpose \citep{jewell/levin/anderson:2004,wandelt/larson/lakshminarayanan:2004,eriksen/etal:2004,larson/etal:2007,eriksen/etal:prep}, and
this allows the user to draw samples from the full joint distribution
in a computationally efficient manner. Applications to real-world data
include the analyses presented by \citet{odwyer/etal:2004,eriksen/etal:2007,eriksen/etal:prep}.

In order to use the Gibbs sampler to sample 
from the joint CMB posterior, one must simply
be able to sample from all corresponding conditional distributions,
\begin{align}
\label{eq:C_ell}
C_{\ell}^{i+1} &\leftarrow P(C_{\ell}|\mathbf{s}^{i},
\theta_{\textrm{fg}}^i, \mathbf{d}) \\
\label{eq:s}
\mathbf{s}^{i+1} &\leftarrow P(\mathbf{s}|C_{\ell}^{i+1},
\theta_{\textrm{fg}}^i, \mathbf{d}) \\
\label{eq:theta}
\theta_{\textrm{fg}}^{i+1} &\leftarrow P(\theta_{\textrm{fg}}|C_{\ell}^{i+1},
\mathbf{s}^{i+1}, \mathbf{d}).
\end{align}
Then, after some burn-in period, the set
$(C_{\ell}^{i},\mathbf{s}^{i}, \theta_{\textrm{fg}}^{i})$ will be
drawn from the desired joint distribution. 
Fortunately, sampling from each of these conditionals is
straightforward, as the distribution for $C_{\ell}$ in Eqn.
\ref{eq:C_ell} is an inverse Wishart distribution, the distribution
for $\mathbf{s}$ in Eqn. \ref{eq:s} is a multivariate Gaussian,
and the distribution for $\theta_{\textrm{fg}}$ is in general given by
a $\chi^2$. All of these have well-known sampling algorithms (see,
e.g., Eriksen et al.\ 2008a for a full review).

As in Section \ref{subsec:mcmc_map}, 
the foreground model may be specified quite freely. 
For application to current data, 
the focus has been mainly on models with a small set of
parametric foreground spectra with individual parameters for each
pixel. Typical examples include power-law spectra ($S_{\nu} = A
\nu^{\beta}$; $A$ and $\beta$ free), one- or two-component thermal
dust spectra, SZ spectra and arbitrary tabulated
spectra. Additionally, Gaussian and/or uniform priors may be specified
for each pixel. In the future, 
complementary sets of parameters describing a spatial power spectrum
for each foreground component may be used, a feature which will be useful for
probing the low signal-to-noise regime.

A major advantage of the Gibbs sampling approach is, because of its
conditional nature, its ability to integrate additional sources of
uncertainty without modifying the existing framework. Three specific
examples would be gain calibration uncertainties, beam uncertainties
and noise estimation errors. By adding another conditional sampling
step for each of these to the above Gibbs chain, the corresponding
uncertainties would be seamlessly propagated through to all other
parameters, without requiring any modifications to the main sampling
scheme.

\subsection{Blind component separation}
\label{subsec:blind}

In both the template and parametric methods, the foreground model is 
the sum of specific astrophysical components, and its construction relies on 
knowledge of their spatial behavior and/or frequency dependence.
The goal of blind or semi-blind component separation methods is to
separate the observed sky maps into a CMB component and a set of 
foreground components, with few assumptions about the physical 
behavior of those foreground components. 
Here we describe two methods, Internal Linear Combination (ILC) and 
Independent Component Analysis (ICA). 

\subsubsection{Internal Linear Combination}
\label{subsec:ilc}

The ILC technique in spherical harmonic
space was first described as a way to remove foregrounds from CMB
anisotropies in \citet{tegmark/efstathiou:1996}. It was applied 
successfully to the \wmap\
one, three, and five-year 
temperature data \citep{hinshaw/etal:2003,tegmark/deoliveira-costa/hamilton:2003,deoliveira-costa/tegmark:2006,hinshaw/etal:2007,gold/etal:prep}, 
but it can also
be applied to polarized data, using the E or B mode spherical harmonics 
\citep{amblard/cooray/kaplinghat:2007}. 
The CMB signal is estimated as a 
linear combination of the observed $a^i_{\ell m}$ in each frequency band $i$,
\begin{equation}
a_{\ell m}=\sum_{{\rm freq}=i}w^i_{\ell}a^i_{\ell m} \, ,
\end{equation}
with weights $w_i$ chosen to minimize foreground
contamination. For polarized CMB observations, the
signal at each frequency can be decomposed as
\begin{equation}
a^i_{\ell m} = c_{\ell m} + s^i_{\ell m} + d^i_{\ell m} + n^i_{\ell m} \, ,
\end{equation}
where $c$, $s$, $d$, and $n$ stand for the CMB,
synchrotron, dust, and noise. The number of foregrounds can
be increased, provided that the number of channels is also raised.  
To choose the weights that best extract the CMB signal, the power spectrum 
\begin{equation}
\langle|a_{\ell m}|^2\rangle= {\bf w_\ell}^T {\cal C}_\ell {\bf w_\ell}  \, ,
\label{equ:alm}
\end{equation}
is minimized to optimally remove the foregrounds, where 
${\cal C}^{ij}_\ell = \langle (a^i_{\ell m})^\dagger a^j_{\ell m} \rangle$.
The weights are constrained by $w_\ell^T \cdot {\bf e}=1$, where ${\bf
  e}$ is a column vector of all ones with length equal to the number
of channels. This condition ensures that the CMB signal is retained. 
As derived in \citet{tegmark/deoliveira-costa/hamilton:2003}, the 
minimized weights  are given by
\begin{equation}
{\bf w_\ell} = \frac{ {\cal C}_\ell^{-1} {\bf e}}{e^T {\cal C}_\ell^{-1} e} \, .
\end{equation}
The estimated spherical harmonics are then used for cosmological analysis.
The advantage of the ILC algorithm is that there are no assumptions
made about the frequency dependence of the 
foregrounds.  The algorithm maximizes the
signal to noise of the recovered CMB, and accounts for
differences in beam size between frequencies. The algorithm is computationally
very fast, allowing multiple Monte-Carlo simulations to be
performed to propagate systematics and compute errors.
A disadvantage of the method is that the foregrounds are assumed to 
have a spatially invariant frequency behavior, which is not physically 
realistic. This can be relaxed 
by splitting the full sky maps into several areas. 
Further drawbacks are that the total power spectrum is minimized rather
than just the foreground power, and it is difficult to 
rigorously propagate errors into the estimated CMB maps.

\subsubsection{Independent Component Analysis}
\label{subsec:smica}

As with the ILC method, ICA components are 
linear combinations of the input sky maps, but the weights are chosen using an
alternative method. Two examples are `altICA' and Spectral Matching ICA (SMICA). 

The {\bf altICA} method is based on an algorithm by 
\citet{hyvarinen:1999}, and is described in \citet{maino/etal:2002}\footnote{The altICA algorithm is the result of 
a study supported by NASA LTSA Grant NNG04CG90G.}. The weights given to the sky maps for each
component are estimated by maximizing the 
non-Gaussianity of the component maps. This idea is based on 
the central limit theorem, which states that a variable that is a mixture 
of independent variables is more Gaussian than the 
original ones. By transforming the observed sky maps into a set of
components such that the Gaussianity of the variables is reduced, the
resulting maps should be more independent.  
The algorithm has been implemented as a CMB cleaning procedure, because
the CMB and diffuse foregrounds are expected to be 
statistically independent. However, 
the non-Gaussianity of a given foreground 
component may vary spatially, so this decomposition is not necessarily 
astrophysically motivated.

In this method, the degree of non Gaussianity is quantified by the {\em neg-entropy} statistic, which is more 
robust than many other statistics in the presence of noise. 
The neg-entropy for a map ${\bf y}$ (containing $I,Q$ or $U$, for a given set of weights), is defined as
\be
\label{neg-entropy}
\textrm{neg-entropy}({\bf y})=H({\bf y}_{G})-H({\bf y})\ ,
\ee
where $H({\bf y})=-\int p({\bf y})\log p({\bf y})d{\bf y}$ 
is the entropy associated with the distribution $p({\bf y})$, and
${\bf y}_{G}$ is a Gaussian variable with the same covariance
matrix as ${\bf y}$. 
This method can be extended to include a 
priori knowledge of the frequency scaling of one or more components. 
The method has been applied to 
real and simulated intensity data (including \beast, \cobe\, and \wmap) 
and simulated 
polarization data (see
\citet{maino/etal:2007} and references therein). 
%Monte-Carlo simulations are run for error assessment. 
The performance  
relies on the assumption of
statistical independence for CMB and foregrounds, and operates
most successfully with high resolution data.\\

In the {\bf SMICA}  method 
\citep{delabrouille/cardoso/patanchon:2003,cardoso/etal:prep}, the 
CMB and foreground power spectra are estimated using a 
parametric model of the angular power spectra of the components. For
application to Planck simulations \citep{cardoso/etal:prep},
the angular spectra and cross-spectra, $C_\ell^{ij}$, 
measured at $N_\nu$ frequencies, 
are parameterized as the sum of CMB, foregrounds, and noise:
\be
C_\ell^{ij}=C_\ell^{{\rm CMB},ij} + F_\ell^{ij} + N^{ij}_\ell,
\ee
with $C_\ell^{ij}=1/(2\ell+1)\sum_m(a_{lm}^i)^\dagger a_{lm}^j$ for frequencies $i$ and $j$. 
In practice the spectra are binned in $N_q$ bins in $\ell$, to give spectra 
$C_q^{ij}$. 
The foreground model is parameterized by \(D\) possibly correlated 
angular power spectrum templates, with
\be
F_q= A^TP_q A.
\ee
Here A is a matrix of size $D \times N_\nu$ and $P_q$ is 
a symmetric matrix of size $D \times D$. This gives 
a total of $N_\nu \times D + N_q \times D(D+1)/2$ foreground parameters.
SMICA allows foregrounds to be correlated, which is 
useful for modeling Galactic
components that are naturally
correlated by the geometry of the Galactic plane and the
effect of the Galactic magnetic field.
To capture the complexity of the 
Galactic emission in polarization, at least two templates are needed for 
synchrotron and dust emission. In practice, the number of 
components required to capture the physical behavior 
depends on the frequency range 
and noise levels of the experiment. To study CMBPol 
missions, $D=4$ templates are used for the Galactic emission, 
which captures synchrotron and dust emission, and corrections due 
to variable emission laws. This parameterization of the foregrounds
is considered `semi-blind' as it allows only a limited set of
foreground spectra.

The model for the binned 
CMB angular spectrum, $C_q^{\rm CMB}$, is parameterized by 
cosmological parameters including the tensor-to-scalar ratio. 
Prior knowledge can be introduced in the
modeling, for example by fixing the B-mode power spectrum of the lensed 
CMB to the known value. 
The noise spectra $N_q^{ij}$ 
are considered as additional components in each channel. 
Noise levels are either assumed known, in which case the noise spectra 
are fixed, or else are estimated simultanously with the foreground
and CMB spectra.
This model is fit to the measured angular spectra of multi-frequency sky maps.
Since the spectra of each foreground component are expected to vary spatially,
the fit is done in three spatial regions using apodized masks, 
to cut out the sharp features of the Galactic plane.
The model parameters, $\theta$, are estimated 
by minimizing a mismatch statistic, $\phi$, between the model and the data:
\begin{equation}
  \label{eq:mismatch}
  \phi(\theta) = \sum_{q=1}^{N_q} \frac{w_q}{2} \left[ \rm{tr}(C_q(\theta)^{-1} \hat{C}_q) - \rm{logdet}(C_q(\theta)^{-1} \hat{C}_q) - m \right],
\end{equation}
where \(\mathbf{C}_q(\theta)\) and \(\hat{\mathbf{C}}_q\) are the binned spectra
of the model and data in bin \(q\), and \(w_q\) is 
the effective number of modes in the bin. 
%The factor \(K\) is the Kullback
%divergence between two positive matrices of size $m \times m$ (in this case 
%$m=N_\nu$). 
%The choice of this particular goodness-of-fit measure stems 
%from the maximum likelihood estimator. 
The errors on the CMB spectrum are estimated using the Fisher matrix.

This method has been used for CMB cleaning and power
spectrum estimation for Planck simulations \citep{leach/etal:prep}. 
It has also been used to estimate the temperature 
power spectrum of the CMB for \wmap\
first year data \citep{patanchon/etal:2005} and for Archeops data 
\citep{tristram/etal:2005}.
To estimate the tensor-to-scalar ratio in the presence of 
foreground emission, one can assume
a fixed shape of the CMB B-mode angular power spectrum, and then vary $r$ in the parameter fit
by varying the overall amplitude. 
This method produces estimates of the power spectra of each component. 
Maps of the CMB and of each foreground component can be estimated 
by Wiener filtering the observed multi-frequency maps, using 
the angular power spectrum of the best-fit model as the filter.

\subsection{Discussion}

These three classes of methods have all been applied to observed or simulated
polarization data, and have different advantages and drawbacks. 
The template-cleaning method is simple and has been demonstrated to work 
effectively with current data. We do not yet have 
a good template for the polarized dust, or for low polarized components such 
as free-free, but multi-frequency information from CMBPol should be able 
to provide such templates. 
It is possible to propagate errors due to 
foreground uncertainty into the template-cleaned maps, but 
the framework for doing so is less well-defined than for some other methods. 
The parametric sampling methods are ideal for 
propagating errors due to foreground uncertainty in a fully Bayesian
framework into both CMB maps and estimated power spectra. 
Additional observations can also be included as priors, such that all the 
information we have about the Galactic emission can be used. These
methods could be biased if the physical model does not match the true sky, but 
various models can be tested to establish the most suitable one. 
They can also be numerically intensive. The blind component separation methods
described here are quick to perform, and have the advantage of not assuming a 
particular physical source for the foregrounds. However neither do they
take advantage of the fact that we do have useful knowledge of the emission 
processes, and can make unphysical assumptions 
about the foregrounds, for example including limited spatial variation 
of spectral indices. Cleaning currently observed 
polarization maps from \wmap\ has required methods operating 
in pixel space, so power spectrum methods may need further 
testing with real data. 

At the current time the community has tools ready to apply to polarization
data with the sensitivity of a CMBPol satellite. 
Data from Planck and
ground-based experiments will allow more extensive testing 
and comparisons between these methods, but we are ready now to tackle a 
CMBPol data set. In the event of a gravitational wave 
detection it is clear that having more than one method available will be 
important for cross-checking results.

\section{Numerical forecasts}
\label{sec:forecast}

In this section we attempt to estimate whether a detection of a gravitational
wave signal with $r=0.01$ is possible for a realistic example mission. 
The specifications for this mission are given in Table \ref{table:specs} 
and have arisen from CMBPol studies focused on systematic effects 
and technological capabilities. We also consider 
how to optimize the allocation of detectors with frequency for such a mission.

\begin{table}[t]
\begin{center}
\begin{tabular}{cccc}
\hline
Frequency & FWHM    & $\Delta$T (pol) \\
GHz  & arcmin &  $\mu$K/arcmin \\
\hline
30    & $26$   & $19.2$ \\
45    & $17$   & $8.3$ \\
70    & $11$   & $4.2$ \\
100   & $8$    & $3.2$ \\
150   & $5$    & $3.1$ \\
220   & $3.5$  & $4.8$ \\
340   & $2.3$  & $21.6$ \\
\hline
\end{tabular}
\caption{Specifications for a CMBPol 2m mission, for 
frequencies below 350 GHz. The noise limits $\Delta T$ are polarized 
limits for Q and U.
\label{table:specs}}
\end{center}
\end{table}

\subsection{Constraints on the CMB signal}

We consider two approaches to obtain forecasts on the B-mode CMB signal: 
a Fisher matrix method assuming that the foregrounds can be cleaned to a 
specified level, and a simulation approach, where we apply  
cleaning methods described in Section \ref{sec:methods} 
to simulated maps, to estimate the CMB signal. For simplicity we will focus on 
the possible detection of a primordial signal with tensor-to-scalar ratio 
$r=0.01$. This level is picked as a target that CMBPol should aim to 
reach, based on theoretical arguments discussed in \citet{baumann/etal:prepa}.
In this section all estimates are derived using the experimental specifications
given in Table \ref{table:specs}.

Note that in addition to statistical and systematic
uncertainties estimated in the approaches detailed below, the detection
significance of a B-mode signal is also affected by cosmic variance.
The error bars on the power spectrum, $C_\ell$, with noise spectrum $N_\ell$, are given approximately by 
\be
\sigma(C_\ell) = \sqrt{\frac{2}{(2l+1)f_{sky}}} (C_\ell + N_\ell).
\ee
As $r$ tends to zero, the cosmic variance term also goes to zero. For non-zero
$r$, cosmic variance will therefore affect the confidence interval. 
This means that a simple detection of
B-modes can be confirmed with higher confidence than can a measurement
of a specific small-amplitude signal such as $r=0.01$.

\subsubsection{Fisher matrix approach}

The Fisher matrix approach described in \citet{verde/peiris/jimenez:2006} 
is used for forecasting errors on measured cosmological parameters.  
In this analysis the lensing contamination in the B-mode signal is accounted 
for by assuming that the lensing power spectrum can be accurately 
estimated, but not removed from the maps, 
such that it acts as an additional Gaussian noise term.
To include foregrounds, the 
parameterization given in Section \ref{subsec:sim} is adopted for the total 
foreground signal, with the synchrotron given by Eqn. 
\ref{eqn:synch}. The dust model is given by Eqn. \ref{eqn:dust}  
with $A_d=0.04$ $\mu K^2$ at $\ell=10$, and a scale dependence $m=-0.5$, 
corresponding 
to an observed dust polarization fraction of $\sim$4\%.  At scales $\ell \gtsim 30$ 
this model matches the predicted spectrum from the Planck Sky Model. At 
larger scales the power spectrum of the simulated map flattens, which is 
not captured by this power law behavior\footnote{This 
dust model corresponds to dust `Model A' reported in 
\citet{baumann/etal:prepa}. An additional `Model B' is also 
considered in \citet{baumann/etal:prepa}, with $A_d=0.002$ $\mu K^2$ at $\ell=10$, with scale dependence $m=0.6$. The 
scale dependence of the dust is expected to lie within the $-0.6<m<0.6$ range 
considered in the two models. For simplicity we report only on Model A in this study, as it most closely matches simulated maps.}. 

For a description of the method, see Appendix C in 
\citet{baumann/etal:prepa} and \citet{verde/peiris/jimenez:2006}. 
The method assumes that at each frequency the angular power spectrum of 
the foregrounds can be cleaned to a given residual level. The lowest 
and highest frequencies are discarded, assuming that they will act as foreground
templates. Residual levels of both 30\% and 10\% of the total amplitude (10\% 
and 1\% in power)
are considered, shown in Figure 
\ref{fig:resid} for a 10\% residual at 90 GHz. 
These are chosen to represent a
pessimistic and realistic case. Experience with
previous experiments suggests that reducing the foreground amplitude by a 
factor of 10 (or more) should be possible, corresponding to the realistic 
case. The pessimistic case allows for poorer foreground cleaning, or a 
higher total foreground level than given by the model.

Considering only the large-scale power, with $\ell<15$, a model with $r=0.01$ 
can be detected at almost 7$\sigma$ in the absence of foregrounds, and assuming
all other parameters are known. The error, $\sigma_r=0.0015$, is given in
Table \ref{table:r_limit}, and is unmarginalized, so all other 
cosmological parameters are assumed known. With 
10\% residual foreground amplitude, the primordial signal 
is undetectable according to this method, with $\sigma_r=0.014$. 
However, this should be considered an overly pessimistic 
scenario, as improved cleaning
is likely possible, and more importantly 
the large-scale foreground power is  
likely over-estimated, as the
spectrum of the dust intensity flattens at 
large scales rather than following a pure power 
law. Including smaller scales, prospects are much improved. In the absence of 
foregrounds the $r=0.01$ signal is detected at 20$\sigma$ with 
$\sigma_r=0.00046$\footnote{This limit depends somewhat on assumptions made 
in the forecast. For example, in the absence of lensing noise, 
limits of $\sigma_r=0.0002$ are 
obtained for $f_{sky}=0.87$.}, and 
this is only decreased to 14$\sigma$ with pessimistic residual foregrounds. 
With 10\% residual foregrounds, the estimated degradation of the error on $r$
is small. The analysis, reported in \citet{baumann/etal:prepa}, also
shows that information about the scale dependence of the tensors 
can be extracted by observing a range of scales, with forecasted errors on the
tensor spectral index, $n_t$, of $\sigma(n_t)\sim0.1$.

\begin{figure}[t]
\center{
\includegraphics[width=0.7\hsize]{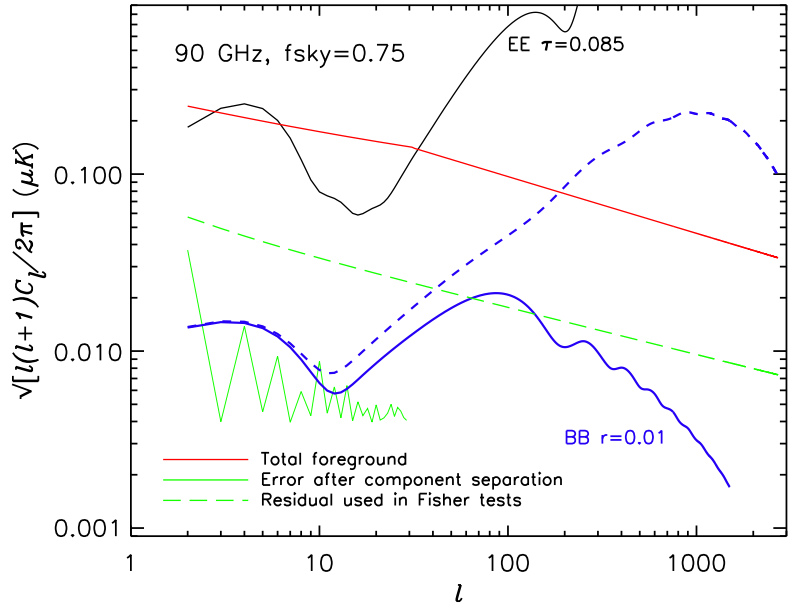}}
\caption{
Comparison of CMB signals (E-mode - solid black; B-mode from inflation - 
solid blue,; total B-mode, including lensing - short dash), and 
an estimate of the residual foreground signal. Foreground spectra (for synchrotron plus dust) 
are given at 90 GHz for the cleanest 75\% of the sky. The red line shows  
the total foreground signal, for simulated maps with mean dust polarization 
fraction 1.5\%. Using a component separation technique on low 
resolution simulated maps gives an instrument noise plus 
foreground residual power spectrum (green), that is about $5\%$ of 
the expected foreground at large scales. 
The green dash line is the power spectrum of foreground residuals assumed 
in the Fisher matrix analysis, where the mean dust polarization fraction is
5\%. This likely over-estimates the residual foreground power at low $\ell$. 
\label{fig:resid} 
}
\end{figure}

%\clearpage

\begin{table}[t]
\begin{center}
\begin{tabular}{ccccc}
\hline
Method & Average dust  & Description  & $\ell<15$ & $\ell<150$ \\
& pol fraction (\%) & & & \\
\hline
Fisher &  0 & No foregrounds  & 0.0015 & 0.00046 \\
Fisher & 5 & 10\% residual  & 0.014 &  0.00052 \\
%Fisher & 5 & 30\% residual  & 0.020 &  0.00070 \\
&&&&\\
Parametric & 1 & Fixed spectral indices  & 0.0015 & 0.0005 \\
Parametric & 1 & Power-law indices   & 0.0025 & ---\\
Parametric & 5 & Power-law indices   & 0.003 & ---\\
&&&&\\
Blind & 4 & SMICA & --- & 0.00055\\ 
&&&&\\
\hline
\end{tabular}
\caption{The forecast $1\sigma$ 
uncertainties on the tensor-to-scalar ratio for fiducial model 
$r=0.01$, assuming other 
cosmological parameters are known, for a set of foreground assumptions. The `Fisher' and `Blind' tests include a residual noise signal from lensing. The `Parametric' tests ignore lensing as a contaminant, a poor approximation at $\ell>15$.
For $\ell<15$,  the errors in the presence of foregrounds from 
the Fisher matrix test are higher than those obtained by 
other methods, as the assumed residual foregrounds are also higher at 
large scales.
\label{table:r_limit}}
\end{center}
\end{table}

These forecasts should be interpreted with care. Some of 
the assumptions made in producing these forecasts, 
for example, whether the lensing treatment is realistic 
(discussed further in \citet{smith/etal:prep}), and the 
assumed polarized dust levels at  $\ell= 100$, need 
to be tested and may affect the resulting error bars. 
The estimated errors also assume that 
there is no effect of leakage of power from E to B modes. 
By using a large fraction of the sky, the errors on the measured 
polarization will vary spatially when foreground uncertainty is 
included, resulting in additional contamination of the B-mode signal from 
E-mode leakage. \citet{amarie/hirata/seljak:2005} approach this problem 
by masking out all regions of sky where the foregrounds exceed some fraction 
of the CMB signal. This leads to larger sky cuts than the 20\% cut
used in this analysis, and larger errors than
obtained here for the $\ell<150$ case. We address this issue further in
the next sub-section.

\subsubsection{Component separation with simulated maps}

An alternative approach is to apply foreground removal methods
to simulated maps, in order to forecast the accuracy to which the CMB signal
is recovered. In this initial study we test the parametric and blind 
component separation methods, as numerical codes already exist to do so. We do 
not consider template-cleaned forecasts here, leaving this approach for 
future investigation.\\ 

\noindent
{\bf Parametric methods:}\\
Simulations at each frequency for the CMBPol mission specified in 
Table \ref{table:specs} are formed using HEALPix with $N_{\rm side}=16$ and 
$N_{\rm side}=128$, using the following model:
\be
\mb m(\nu) =  {\mb A}_{\rm synch} (\nu)+ {\mb A}_{\rm dust} (\nu) + f(\nu){\mb A}_{\rm cmb} + f(\nu) \mb n(\nu),
\ee
wherewhere $f(\nu)$ converts thermodynamic to antenna temperature. The maps
are formed using Eqn. \ref{eqn:synch_map} for the synchrotron, 
and Eqn. \ref{eqn:dust_map} for the dust. Since they
include the synchrotron and dust geometric suppression factors, the average dust polarization is
1\% in the fiducial case (current PSM maps have average 5\% polarization 
fraction; we also consider maps with this increased dust level).
Unlensed CMB maps are generated using the HEALPix synfast routine, for  
$r=0.01$ with all other parameters fixed to the best-fit $\Lambda$CDM 
parameters estimated from the \wmap\ five-year data \cite{komatsu/etal:prep}. 
The $N_{\rm side}=16$ (128) maps are smoothed with a 7 (1.5) 
degree beam. Gaussian noise is added uniformly 
over the sky, and $1/f$ noise is ignored. 
In Figure \ref{fig:resid} we plot the angular power spectrum of the 
total foreground signal in these maps at 90 GHz. Part of the goal 
of performing the component 
separation is to assess the size of the errors on the estimated CMB maps, or 
power spectra, compared 
to the total signal. 

{\it Marginalized CMB maps:} We apply the parametric 
fitting methods described in Section 
\ref{subsec:method_param} to the simulated maps, to estimate the 
foreground-marginalized errors on the observed CMB signal in map space.
For a single random individual pixel 
outside the Galactic plane, the parameters of the model are sampled 
using the MCMC method described in Sec \ref{subsec:mcmc_map}\footnote{Using the FGFit code described in \citet{eriksen/etal:2006}.}, 
and the marginalized error and bias on the CMB pixel amplitude are reported in 
Table \ref{table:single_pixel}, averaged over 5000 simulations.  
The errors are inflated about seven times compared to the 
foreground-free case. This level depends somewhat on the size of pixel in
which the spectral indices are allowed to take unique values, in this case 
two degrees.
We also estimate full marginalized 
CMB maps for this model\footnote{Using the method and code described in \citet{dunkley/etal:prepb}.}. 
The marginalized error map for the Q Stokes parameter is shown in Figure \ref{fig:marg_err}. 
In this case the spectral indices are fit in pixels of side 3.7 degrees, so
errors are slightly reduced compared to the 2 degree pixels used in 
Table \ref{table:single_pixel}. 
The errors should be compared to the 
co-added noise level of 8 nK per pixel, in the absence of foregrounds. 
Increased errors, up to a maximum of 50 nK, 
are due to foregound marginalization.
We then ask how big the marginalized 
errors on the CMB signal are compared to the total foreground signal. 
To do this we take the marginalized error 
maps and compute their angular power 
spectra, masking 25\% of the sky using the \wmap\ P06 mask. This spectrum 
is shown in 
Figure \ref{fig:resid}, and is comparable to the CMB spectrum for $r=0.01$. 
The amplitude is about 5\% of the total foreground signal at 90 GHz. 
This estimate may be over-optimistic if the true sky has more complicated 
spectral behavior and additional low polarization components that were not included in this test. However, in this test we allowed each pixel of side 3.7 degrees to have unique synchrotron and dust spectral index parameters. If we are
only interested in a broad-band detection of the B-mode signal, larger
pixels may be averaged together and could lead to lower residual error levels.

Since there is about an order of magnitude uncertainty in the dust amplitude,
we repeat the tests with varied foreground levels.
For the case of an individual pixel, if we inflate the synchrotron 
amplitude three times, this inflates the 
errors by a further 50\%, but a similar increase in the dust levels, where
we are far less certain, has a  
smaller effect. If the simulation is modified to have a curvature in the 
spectral index, together with a 1\% polarized free-free and 2\% polarized 
spinning dust component with random polarization angles, 
the marginalized error is negligibly different 
from the fiducial case. In this case the model is sufficiently flexible that
the `synchrotron' component absorbs the additional polarized components, 
with an effective spectral index that is the weighted sum of the components. 
In future studies it will be useful to explore simulations that deviate 
further from the model, to test whether there is a significant bias in 
the recovered CMB power when the model fails to match the simulated data.

\begin{table}[t]
\begin{center}
\begin{tabular}{cccc}
\hline
Case & Error (nK) & Case &  Error (nK) \\
\hline
No foregrounds    & $8$  & Increased synchrotron    & $79$ \\
Fiducial model    & $58$ & Increased dust    & $60$ \\
Extended model    & $58$  & Increased dust + synchrotron    & $81$ \\
\hline
\end{tabular}
\caption{Expected per-pixel sensitivities in polarization in 4x4 
degree pixels in nK, including foreground degradation, for the 2m 
CMBPol example mission. The `extended' model has a break in the synchrotron
index, and low levels of polarized free-free and spinning dust emission.
`Increased' means a three-fold increase in amplitude.
\label{table:single_pixel}}
\end{center}
\end{table}

To estimate the CMB power in these simulated maps, we use   
the Gibbs sampling method described in Section \ref{subsec:mcmc_cl}, using 
the Commander numerical code \citep{eriksen/etal:2007}\footnote{We 
cross check results for low resolution maps 
by computing the exact likelihood of 
the marginalized maps, using the method in Section \ref{subsec:mcmc_map}. 
These 
two methods should give the same results for any given set of 
simulated maps, although the implementation is different.}.
Table \ref{table:r_limit} shows 1$\sigma$ errors on $r$, assuming other 
parameters are perfectly measured.
When the foreground 
spectral indices are known perfectly, the tensor-to-scalar ratio
is recovered with $r=0.0106\pm0.0017$ for $\ell<15$, and $r=0.0101\pm0.0005$ 
for $\ell<150$. These error bars are similar to  
those obtained in the Fisher matrix tests in the absence of foregrounds, 
where lensing noise is also included. These errors are too 
optimistic however, as
in reality the synchrotron and dust spectral indices are uncertain and should be
marginalized over.
However, this provides a test of the method and shows that in this simple case 
the recovered signal is not significantly biased. It also indicates that
regions of the sky where the foregrounds are high can still be used to estimate
the CMB power: rather than masking out pixels as in \citet{amarie/hirata/seljak:2005}, the errors are inflated in regions where foreground uncertainty is high, and propagated into the likelihood. 

A more realistic case includes the possible variation of 
synchrotron and dust spectral indices in each pixel. For 
pixels of side 
4 degrees, and scales $\ell<15$, the error bar on $r$ is increased to 
$\sigma_r=0.0025$, indicating only a 4$\sigma$ detection of the primordial 
signal. This is increased to $\sigma_r=0.003$ when the 
dust component in the simulated 
maps is made three times larger (corresponding to $~4\%$ polarization 
fraction). The limits may worsen if 
additional freedom in the foreground model is allowed 
(e.g. polarized free-free emission, 
a two-component dust power law, or a synchrotron spectral curvature), and if the dust is more then 5\% polarized, although 
additional observations may provide priors or templates that would improve 
limits. Observations from the Planck satellite and ground 
based experiments will help to guide such choices for future preparatory 
studies. 
The application of this spectral index marginalization to the 
higher resolution maps, using $\ell<150$, was not calculated in this initial
study, but is an obvious next step.\\

\noindent
{\bf Blind separation methods:}\\
The ILC method in spherical harmonic space described in
Section \ref{subsec:ilc} and the SMICA component separation
method described in Section \ref{subsec:smica}
have also been applied to simulated maps. 

For tests with SMICA, alternative maps to the fiducial simulation 
were used, as part of a Planck preparatory study. 
The results 
can be compared with others considered in this Section, as the maps are 
very similar to our `Increased dust' model, with dust that is
4\% polarized on average. 
The SMICA method assumes perfect knowledge
of all cosmological parameters except $r$, the latter assumed to be measured only
from B-modes. For the mission specifications given in Table \ref{table:specs}, 
limits of $\sigma_r=0.00055$ are found for $\ell<150$ \citep{betoule/etal:prep} 
for a model with $r=0.01$. This includes additional noise from 
the lensing signal, which when ignored leads to a limit of $\sigma_r=0.00046$.
Point sources are assumed to have a negligible impact on limits on $r$.
These are similar to the limits found using the Fisher matrix method. When an 
additional 500~GHz channel is included, estimated limits improve to $\sigma_r=0.0003$, 
indicating that this longer lever arm could provide 
useful additional information. However,
it is possible that extrapolating from these high frequencies can introduce 
errors due to incorrect modeling assumptions.
The numerical experiment with SMICA indicates that even with
some complexity in the foreground modeling, estimation of B modes 
may be performed satisfactorily on a significant fraction of the sky, 
supporting large sky coverage for the CMBPol mission.

Using the ILC method, rather optimistic levels of detection are obtained. 
Confidence levels on a model with $r=0.01$ were not considered, but 
the 99\% confidence level detection of $r$ for fiducial $r=0$ is 0.0002 for 
$\ell<8$, and 0.0004 for $50<\ell<130$ in the 
absence of lensing.\footnote{With lensing as a contaminant, 
this increases to $r<0.001$ for $\ell<8$ and $r<0.008$ for $50<\ell<130$.}
This would imply foreground cleaning to better than $\sim$1\% in amplitude 
at large scales and better than 5\% at intermediate scales, which is likely 
difficult to achieve in practice.   

\begin{figure}[t]
\center{
\epsfig{file=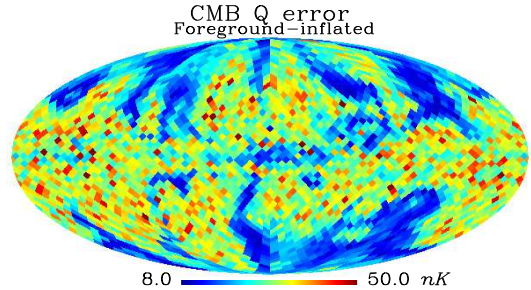,angle=90.,width=7cm,height=4cm}
\caption{Marginalized errors for a CMB Q Stokes map, using the CMBPol 
mission specified in Table \ref{table:specs} and the fiducial 
foreground model. In 
the absence of foregrounds the noise level would be 8~nK in each pixel. The
errors are increased where the signal is most uncertain due to foreground contamination.
 \label{fig:marg_err}}
}
\end{figure}

\subsection{Optimal use of focal plane}

Variables that drive experimental design include frequency range,
bandwidth, and number of channels. To investigate the optimal
configuration one requires either an analytic form for the errors, or
a component separation code that can be run quickly. For this work, we
employ the parametric method described in Section
\ref{subsec:mcmc_map} to perform component separation for a single
pixel, repeated many times with different CMB and noise
realizations.\footnote{FGFIT\_pix was designed exactly for this
purpose, with input files for multiple experimental design and sky
models. The experiment file contains the frequency and noise level for
each channel.} The averaged best-fit value for the CMB amplitude, and
its marginalized error, are estimated. The error is then used as the
statistic to compare the {\it relative}  performance of different
experimental specifications. In principle the number of channels,
positioning in frequency space of the channels, and angular resolution
can be varied.

Here we perform a limited study for a fixed number of ten channels,
chosen to allow for at least three parameters in each pixel for both
synchrotron and thermal dust, and two for additional low-polarization
components. The channels are spaced at logarithmic intervals in
frequency, with the number of detectors at each channel varied such
that the total signal-to-noise ratio (`total' is CMB plus foregrounds)
is kept constant. The maximum and minimum frequencies are then varied,
and the error on the CMB amplitude estimated. The foregrounds are
assumed to have power law spectral indices for the parameterized
model. The preferred minimum frequency is found to be at  $\sim
40\pm5$ GHz, with a poorly constrained maximum at $ \sim 200-300$
GHz. This spans a similar frequency range as proposed in the initial
mission
design, although has more channels. A number of different assumptions
are investigated for the distribution of detectors at each frequency,
using different scalings for sensitivities as a function of frequency.
They give similar results for the optimal frequency range.

To use these tests to guide mission design,  further investigation of
the poorly constrained upper frequency limit would be useful. An
extended parameterization for the foreground model
could also be explored, to test how the optimal range is affected by
the assumption of power law behavior.  The number of channels can be
explored with a similar tool, but was not considered in this initial
study. It is clear that more channels will help to remove the
foregrounds effectively and break degeneracies between spectra of
different emission mechanisms. Sparsely placed frequency channels lead
to uncertainty in the curvature of the foreground frequency spectra.
However, there is always a trade-off between increased sensitivity at
each frequency, and increased number of channels.

If we ignore the effects of lensing there is little motivation to go
to higher resolution than $\approx 1$ degree. However, optimization
for lensing should be considered in order to extract the gravitational
wave signal at smaller scales where the foregrounds are lower than at
the lowest multipoles. See the companion lensing report by
\citet{smith/etal:prep} for more details.

\subsection{Future tests}

The forecasts presented here represent an initial investigation into the
CMBPol mission capabilities. Work remains to be done to 
make concrete forecasts, but we will be in a better
position as more data is obtained from the Planck satellite and 
ground-based experiments. Further preparatory work
for a CMBPol mission could include:

\noindent
{\bf (i) Bias and modeling:} So far we have examined cases where the simulated 
maps are correctly described by the models. In a realistic scenario, the model
will not perfectly capture the observed 
two-dimensional emission. It will be useful to determine how many
parameters we need to fit the true sky `well enough', and whether an incorrect
model leads to significant bias in the estimate of the CMB signal.

\noindent
{\bf (ii) Convergence of estimates:} Different methods exist for forecasting the
constraining abilities of the CMBPol mission.
Consensus of the effect of foregrounds on
the estimated B-mode signal is important in 
the planning of the mission, and will likely improve with better 
observations. However, not all methods will clean the maps equally well, 
so we would not necessarily expect to achieve the same errors on the CMB signal 
from all component separation methods. We plan to 
continue with the program we have begun, to compare the 
estimated residual noise-plus-foreground levels for various methods 
over the $\ell<150$ range, 
accounting for the lensing signal as a source of noise in simulations.
These would include the Fisher methods, parametric and blind separation 
methods described here, as well as template cleaning applied to forecasting.

\noindent
{\bf (iii) Optimizing the experiment:} The community has tools that can be 
applied to experimental optimization. In particular the parametric fitting 
employed in the `FGfit' code allows for easy comparison of the CMB 
errors in a pixel, marginalized over foregrounds, for different 
experimental set-ups. Other methods, including template cleaning and ILC 
component separation, can be used for 
alternative investigations. 
Further exploration with these tools will be useful, 
including tests for the optimal number of channels, 
as well as variation of the complexity of the model. 
One can also test the effect of 
varying the experimental sensitivity, to determine the noise level at which 
the observations become foreground dominated.

\section{Summary}
\label{sec:summary}

Polarized Galactic emission at microwave frequencies 
is dominated by two components, synchrotron 
emission and thermal dust emission. The synchrotron emission 
has been well measured on large angular scales at 23~GHz by \wmap. 
Our knowledge of polarized dust 
emission is relatively poor, particularly 
in the low surface brightness regions out of the Galactic plane 
targeted for CMB observations.
It will soon be better characterized by the Planck satellite, due for 
launch 2009, 
which will measure polarized emission with a spatial resolution $<30$ 
arcmin at 30 - 350 GHz. 
With the assumption of a polarized dust fraction of ~1-10\%, the total
foreground emission is expected to be roughly a factor of eight
brighter, on angular 
scales of two degrees over 75\% of the sky, than a gravitational wave
signal with tensor-to-scalar ratio $r=0.01$. There may also be small
polarization contributions from free-free emission, spinning dust
emission, or other `anomalous' components that have not yet been
detected.

Past experience working on analogous problems suggests that 
the challenge of dealing with 
significant foreground components is not insurmountable, and 
has been addressed successfully with previous experiments such as \firas. 
In this case the CIB foregrounds were reduced 
by a factor of ten to extract the signal, comparable to the cleaning required 
of polarized Galactic foregrounds for the CMB.
Although past performance is no guarantee of future 
success, we have consistently been able to 
clean foregrounds to within a factor of a few times the 
uncertainties of the raw measurements. As a community we have 
a set of tools ready for performing component
separation and estimating the CMB signal from observed sky maps. 
These could be applied today to CMBPol data, and will 
be significantly improved and tested with data from Planck and from 
ground and balloon-based experiments during the next five years.
There is some variation in predicted performance for a specified CMBPol 
mission, so the refinement 
process with upcoming experiments will be useful for  
understanding why some methods perform better than others, and whether this 
depends on details in our simulations. 

Despite the good prospects for cleaning foregrounds, such
prospects rely sensitively on experimental design. Observations over a wide range
of angular scales should be planned in order to mitigate anticipated
residual Galactic contamination on the largest angular scales ($\ell<15$). Multiple 
channels are needed to characterize and remove the foregrounds. Although there
are only two dominant polarized components, they are unlikely to have spectral
indices that can be modeled accurately as power laws. This means 
that at least three channels per component should be included. Since there may 
also be additional low-polarization components, at least ten channels spaced 
logarithmically in the range $\sim 40<\nu<200-300$~GHz would be advantageous, although further studies should be undertaken to explore this in more detail. 
Extrapolating from very low and high frequencies can introduce modeling 
errors, so frequencies below $\sim300$~GHz are more easily used for 
foreground subtraction.

Current methods and simulations indicate that we can clean Galactic foregrounds 
from maps of the polarized sky to at least the $5-10\%$ level. There are 
therefore good prospects that, for a realistic CMBPol mission design, a 
gravitational wave signal with $r=0.01$ could be detected at more 
than $10\sigma$. Successfully disentangling this signal from the emission 
of our own Galaxy will provide rich rewards, as a detection of 
primordial B-modes would open up a new window on 
the earliest moments in the universe.
\\

This research was partly funded by NASA Mission Concept Study award NNX08AT71G S01. 
We thank Scott Dodelson for coordination of the CMBPol 
Theory and Foregrounds workshop and proceedings, and 
acknowledge the organizational work of the Primordial Polarization Program 
Definition Team. We thank David Spergel and 
Alex Lazarian for useful comments.
JD acknowledges support from an RCUK fellowship. GR is 
supported by the US Planck Project, which is funded
by the NASA Science Mission Directorate.
We thank the \wmap\ team for making maps available on LAMBDA, and
acknowledge the use of the Planck Sky Model, developed by the Component 
Separation Working Group (WG2) of the Planck Collaboration. 
We acknowledge use of the HEALPix, PolSpice, CAMB, and CMBFAST packages.

%-------------------------------------------------
% BIBLIOGRAPHY

\addcontentsline{toc}{section}{References}
\small
\setlength{\bibsep}{0.2cm}

%\bibliography{mn-jour,cmbpol,cmbpol_jev,cmbpol_doc}

\begin{thebibliography}{158}
\expandafter\ifx\csname natexlab\endcsname\relax\def\natexlab#1{#1}\fi

\bibitem[{{Aitken}(1996)}]{aitken:1996}
{Aitken}, D.~K. 1996, in ASP Conf.\ Ser.\ 97, Polarimetry of the Interstellar
  Medium, ed. W.~G. Roberge \& D.~C.~B. Whittet, 225

\bibitem[{{Ali-Ha{\"i}moud} {et~al.}(2008){Ali-Ha{\"i}moud}, {Hirata}, \&
  {Dickinson}}]{ali-haimoud/hirata/dickinson:prep}
{Ali-Ha{\"i}moud}, Y., {Hirata}, C.~M., \& {Dickinson}, C. 2008, ArXiv
  e-prints, 0812.2904

\bibitem[{{Amarie} {et~al.}(2005){Amarie}, {Hirata}, \&
  {Seljak}}]{amarie/hirata/seljak:2005}
{Amarie}, M., {Hirata}, C., \& {Seljak}, U. 2005, \prd, 72, 123006,
  arXiv:astro-ph/0508293

\bibitem[{Amblard {et~al.}(2007)Amblard, Cooray, \&
  Kaplinghat}]{amblard/cooray/kaplinghat:2007}
Amblard, A., Cooray, A., \& Kaplinghat, M. 2007, \prd, 75, 083508,
  astro-ph/0610829

\bibitem[{{Barbosa} {et~al.}(2006){Barbosa}, {Fonseca}, {Dos Santos}, {Cupido},
  {Mour{\~a}o}, {Smoot}, \& {Tello}}]{barbosa/etal:2006}
{Barbosa}, D., {Fonseca}, R., {Dos Santos}, D.~M., {Cupido}, L., {Mour{\~a}o},
  A., {Smoot}, G.~F., \& {Tello}, C. 2006, in New Worlds in Astroparticle
  Physics: Proceedings of the Fifth International Workshop, ed. A.~M.
  {Mour{\~a}o}, M.~{Pimenta}, R.~{Potting}, \& P.~M. {S{\'a}}, 233--+

\bibitem[{Barkats {et~al.}(2005)}]{barkats/etal:2005}
Barkats, D., {et~al.} 2005, \apjl, 619, L127, astro-ph/0409380

\bibitem[{{Battistelli} {et~al.}(2006){Battistelli}, {Rebolo},
  {Rubi{\~n}o-Mart{\'{\i}}n}, {Hildebrandt}, {Watson}, {Guti{\'e}rrez}, \&
  {Hoyland}}]{battistelli/etal:2006}
{Battistelli}, E.~S., {Rebolo}, R., {Rubi{\~n}o-Mart{\'{\i}}n}, J.~A.,
  {Hildebrandt}, S.~R., {Watson}, R.~A., {Guti{\'e}rrez}, C., \& {Hoyland},
  R.~J. 2006, \apjl, 645, L141, arXiv:astro-ph/0603379

\bibitem[{{Baumann} {et~al.}(2008{\natexlab{a}}){Baumann}, {Cooray},
  {Dodelson}, {Dunkley}, {Fraisse}, {Jackson}, {Kogut}, {Krauss}, {Smith}, \&
  {Zaldarriaga}}]{baumann/etal:prepb}
{Baumann}, D. {et~al.} 2008{\natexlab{a}}, ArXiv e-prints, 0811.3911

\bibitem[{{Baumann} {et~al.}(2008{\natexlab{b}}){Baumann}, {Jackson},
  {Adshead}, {Amblard}, {Ashoorioon}, {Bartolo}, {Bean}, {Beltran}, {de
  Bernardis}, {Bird}, {Chen}, {Chung}, {Colombo}, {Cooray}, {Creminelli},
  {Dodelson}, {Dunkley}, {Dvorkin}, {Easther}, {Finelli}, {Flauger},
  {Hertzberg}, {Jones-Smith}, {Kachru}, {Kadota}, {Khoury}, {Kinney},
  {Komatsu}, {Krauss}, {Lesgourgues}, {Liddle}, {Liguori}, {Lim}, {Linde},
  {Matarrese}, {Mathur}, {McAllister}, {Melchiorri}, {Nicolis}, {Pagano},
  {Peiris}, {Peloso}, {Pogosian}, {Pierpaoli}, {Riotto}, {Seljak}, {Senatore},
  {Shandera}, {Silverstein}, {Smith}, {Vaudrevange}, {Verde}, {Wandelt},
  {Wands}, {Watson}, {Wyman}, {Yadav}, {Valkenburg}, \&
  {Zaldarriaga}}]{baumann/etal:prepa}
------. 2008{\natexlab{b}}, ArXiv e-prints, 0811.3919

\bibitem[{Beck(2007)}]{beck:2007}
Beck, R. 2007, in EAS Publ.\ Ser.\ 23, Sky Polarisation at far-infrared to
  radio wavelengths: The Galactic Screen before the Cosmic Microwave
  Background, ed. F.~Boulanger \& M.-A. Miville-Desch\^{e}nes (EDP Sciences),
  19

\bibitem[{{Bennett} {et~al.}(2003){Bennett}, {Bay}, {Halpern}, {Hinshaw},
  {Jackson}, {Jarosik}, {Kogut}, {Limon}, {Meyer}, {Page}, {Spergel}, {Tucker},
  {Wilkinson}, {Wollack}, \& {Wright}}]{bennett/etal:2003}
{Bennett}, C.~L. {et~al.} 2003, \apj, 583, 1

\bibitem[{{Berdyugin} {et~al.}(2004){Berdyugin}, {Piirola}, \&
  {Teerikorpi}}]{berdyugin/etal:2004}
{Berdyugin}, A., {Piirola}, V., \& {Teerikorpi}, P. 2004, \aap, 424, 873

\bibitem[{{Betoule} {et~al.}(2008){Betoule}, {Pierpaoli}, {Delabrouille}, {Le
  Jeune}, \& {Cardoso}}]{betoule/etal:prep}
{Betoule}, M., {Pierpaoli}, E., {Delabrouille}, J., {Le Jeune}, M., \&
  {Cardoso}, J.-F. 2008, in preparation for submission to Astronomy and
  Astrophysics

\bibitem[{{Bock} {et~al.}(2006){Bock}, {Church}, {Devlin}, {Hinshaw}, {Lange},
  {Lee}, {Page}, {Partridge}, {Ruhl}, {Tegmark}, {Timbie}, {Weiss}, {Winstein},
  \& {Zaldarriaga}}]{bock/etal:2006}
{Bock}, J. {et~al.} 2006, ArXiv Astrophysics e-prints, arXiv:astro-ph/0604101

\bibitem[{{Bock} {et~al.}(2008){Bock}, {Cooray}, {Hanany}, {Keating}, {Lee},
  {Matsumura}, {Milligan}, {Ponthieu}, {Renbarger}, \& {Tran}}]{bock/etal:prep}
------. 2008, ArXiv e-prints, 0805.4207

\bibitem[{{Bottino} {et~al.}(2008){Bottino}, {Banday}, \&
  {Maino}}]{bottino/banday/maino:2008}
{Bottino}, M., {Banday}, A.~J., \& {Maino}, D. 2008, \mnras, 389, 1190

\bibitem[{{Cardoso} {et~al.}(2008){Cardoso}, {Martin}, {Delabrouille},
  {Betoule}, \& {Patanchon}}]{cardoso/etal:prep}
{Cardoso}, J.-F., {Martin}, M., {Delabrouille}, J., {Betoule}, M., \&
  {Patanchon}, G. 2008, ArXiv e-prints, 0803.1814

\bibitem[{{Casassus} {et~al.}(2006){Casassus}, {Cabrera}, {F{\"o}rster},
  {Pearson}, {Readhead}, \& {Dickinson}}]{casassus/etal:2006}
{Casassus}, S., {Cabrera}, G.~F., {F{\"o}rster}, F., {Pearson}, T.~J.,
  {Readhead}, A.~C.~S., \& {Dickinson}, C. 2006, \apj, 639, 951,
  arXiv:astro-ph/0511283

\bibitem[{Chon {et~al.}(2004)Chon, Challinor, Prunet, Hivon, \&
  Szapudi}]{chon/etal:2004}
Chon, G., Challinor, A., Prunet, S., Hivon, E., \& Szapudi, I. 2004, Mon. Not.
  Roy. Astron. Soc., 350, 914, astro-ph/0303414

\bibitem[{{Christensen} {et~al.}(2001){Christensen}, {Meyer}, {Knox}, \&
  {Luey}}]{christensen/etal:2001}
{Christensen}, N., {Meyer}, R., {Knox}, L., \& {Luey}, B. 2001, Classical and
  Quantum Gravity, 18, 2677

\bibitem[{{Chuss} {et~al.}(2003){Chuss}, {Davidson}, {Dotson}, {Dowell},
  {Hildebrand}, {Novak}, \& {Vaillancourt}}]{chuss/etal:2003}
{Chuss}, D.~T., {Davidson}, J.~A., {Dotson}, J.~L., {Dowell}, C.~D.,
  {Hildebrand}, R.~H., {Novak}, G., \& {Vaillancourt}, J.~E. 2003, \apj, 599,
  1116

\bibitem[{{Crill} {et~al.}(2008){Crill}, {Ade}, {Battistelli}, {Benton},
  {Bihary}, {Bock}, {Bond}, {Brevik}, {Bryan}, {Contaldi}, {Dore}, {Farhang},
  {Fissel}, {Golwala}, {Halpern}, {Hilton}, {Holmes}, {Hristov}, {Irwin},
  {Jones}, {Kuo}, {Lange}, {Lawrie}, {MacTavish}, {Martin}, {Mason}, {Montroy},
  {Netterfield}, {Pascale}, {Riley}, {Ruhl}, {Runyan}, {Trangsrud}, {Tucker},
  {Turner}, {Viero}, \& {Wiebe}}]{crill/etal:2008}
{Crill}, B.~P. {et~al.} 2008, ArXiv e-prints, 0807.1548

\bibitem[{{Curran} \& {Chrysostomou}(2007)}]{curran/chrysostomou:2007}
{Curran}, R.~L., \& {Chrysostomou}, A. 2007, \mnras, 382, 699, arXiv:0709.0256

\bibitem[{{Davies} {et~al.}(2006){Davies}, {Dickinson}, {Banday}, {Jaffe},
  {G{\'o}rski}, \& {Davis}}]{davies/etal:2006}
{Davies}, R.~D., {Dickinson}, C., {Banday}, A.~J., {Jaffe}, T.~R.,
  {G{\'o}rski}, K.~M., \& {Davis}, R.~J. 2006, \mnras, 370, 1125,
  arXiv:astro-ph/0511384

\bibitem[{{Davis} \& {Greenstein}(1951)}]{davis/greenstein:1951}
{Davis}, L.~J., \& {Greenstein}, J.~L. 1951, \apj, 114, 206

\bibitem[{{de Oliveira-Costa} {et~al.}(1997){de Oliveira-Costa}, {Kogut},
  {Devlin}, {Netterfield}, {Page}, \& {Wollack}}]{deoliveira-costa/etal:1997}
{de Oliveira-Costa}, A., {Kogut}, A., {Devlin}, M.~J., {Netterfield}, C.~B.,
  {Page}, L.~A., \& {Wollack}, E.~J. 1997, \apjl, 482, L17+

\bibitem[{{de Oliveira-Costa} \&
  {Tegmark}(2006)}]{deoliveira-costa/tegmark:2006}
{de Oliveira-Costa}, A., \& {Tegmark}, M. 2006, \prd, 74, 023005,
  arXiv:astro-ph/0603369

\bibitem[{de~Oliveira-Costa {et~al.}(2002)}]{deoliveira-costa/etal:2002}
de~Oliveira-Costa, A., {et~al.} 2002, \apj, 567, 363

\bibitem[{{Delabrouille} \& {Cardoso}(2007)}]{delabrouille/cardoso:prep}
{Delabrouille}, J., \& {Cardoso}, J.~. 2007, ArXiv Astrophysics e-prints,
  arXiv:astro-ph/0702198

\bibitem[{{Delabrouille} {et~al.}(2003){Delabrouille}, {Cardoso}, \&
  {Patanchon}}]{delabrouille/cardoso/patanchon:2003}
{Delabrouille}, J., {Cardoso}, J.-F., \& {Patanchon}, G. 2003, \mnras, 346,
  1089, arXiv:astro-ph/0211504

\bibitem[{{Dickinson} {et~al.}(2007){Dickinson}, {Davies}, {Bronfman},
  {Casassus}, {Davis}, {Pearson}, {Readhead}, \&
  {Wilkinson}}]{dickinson/etal:2007}
{Dickinson}, C., {Davies}, R.~D., {Bronfman}, L., {Casassus}, S., {Davis},
  R.~J., {Pearson}, T.~J., {Readhead}, A.~C.~S., \& {Wilkinson}, P.~N. 2007,
  \mnras, 379, 297, arXiv:astro-ph/0702611

\bibitem[{{Dickinson} {et~al.}(2003){Dickinson}, {Davies}, \&
  {Davis}}]{dickinson/etal:2003}
{Dickinson}, C., {Davies}, R.~D., \& {Davis}, R.~J. 2003, \mnras, 341, 369,
  arXiv:astro-ph/0302024

\bibitem[{{Dobler} {et~al.}(2008){Dobler}, {Draine}, \&
  {Finkbeiner}}]{dobler/draine/finkbeiner:prep}
{Dobler}, G., {Draine}, B.~T., \& {Finkbeiner}, D.~P. 2008, ArXiv e-prints,
  0811.1040

\bibitem[{{Dobler} \&
  {Finkbeiner}(2008{\natexlab{a}})}]{dobler/finkbeiner:2008a}
{Dobler}, G., \& {Finkbeiner}, D.~P. 2008{\natexlab{a}}, \apj, 680, 1222,
  0712.1038

\bibitem[{{Dobler} \&
  {Finkbeiner}(2008{\natexlab{b}})}]{dobler/finkbeiner:2008b}
------. 2008{\natexlab{b}}, \apj, 680, 1235, 0712.2238

\bibitem[{{Dolginov}(1972)}]{dolginov:1972}
{Dolginov}, A.~Z. 1972, \apss, 18, 337

\bibitem[{{Dolginov} \& {Mytrophanov}(1976)}]{dolginov/mytrophanov:1976}
{Dolginov}, A.~Z., \& {Mytrophanov}, I.~G. 1976, \apss, 43, 257

\bibitem[{Dotson {et~al.}(2000)Dotson, Davidson, Dowell, Schleuning, \&
  Hildebrand}]{dotson/etal:2000}
Dotson, J.~L., Davidson, J., Dowell, C.~D., Schleuning, D.~A., \& Hildebrand,
  R.~H. 2000, \apjs, 128, 335

\bibitem[{Dotson {et~al.}(2009)Dotson, Davidson, Dowell, Hildebrand, Kirby, \&
  Vaillancourt}]{dotson/etal:2009}
Dotson, J.~L., Davidson, J.~A., Dowell, C.~D., Hildebrand, R.~H., Kirby, L., \&
  Vaillancourt, J.~E. 2009, {ApJS, submitted}

\bibitem[{{Draine} \& {Fraisse}(2008)}]{draine/fraisse:prep}
{Draine}, B.~T., \& {Fraisse}, A.~A. 2008, ArXiv e-prints, 0809.2094

\bibitem[{{Draine} \& {Lazarian}(1998{\natexlab{a}})}]{draine/lazarian:1998a}
{Draine}, B.~T., \& {Lazarian}, A. 1998{\natexlab{a}}, \apjl, 494, L19

\bibitem[{{Draine} \& {Lazarian}(1998{\natexlab{b}})}]{draine/lazarian:1998b}
------. 1998{\natexlab{b}}, \apj, 508, 157

\bibitem[{{Draine} \& {Lazarian}(1999)}]{draine/lazarian:1999}
------. 1999, \apj, 512, 740

\bibitem[{{Draine} \& {Li}(2001)}]{draine/li:2001}
{Draine}, B.~T., \& {Li}, A. 2001, \apj, 551, 807, arXiv:astro-ph/0011318

\bibitem[{Draine \& Weingartner(1996)}]{draine/weingartner:1996}
Draine, B.~T., \& Weingartner, J.~C. 1996, \apj, 470, 551

\bibitem[{Draine \& Weingartner(1997)}]{draine/weingartner:1997}
------. 1997, \apj, 480, 633

\bibitem[{{Dunkley} {et~al.}(2005){Dunkley}, {Bucher}, {Ferreira}, {Moodley},
  \& {Skordis}}]{dunkley/etal:2005}
{Dunkley}, J., {Bucher}, M., {Ferreira}, P.~G., {Moodley}, K., \& {Skordis}, C.
  2005, \mnras, 356, 925, arXiv:astro-ph/0405462

\bibitem[{{Dunkley} {et~al.}(2008{\natexlab{a}}){Dunkley}, {Komatsu}, {Nolta},
  {Spergel}, {Larson}, {Hinshaw}, {Page}, {Bennett}, {Gold}, {Jarosik},
  {Weiland}, {Halpern}, {Hill}, {Kogut}, {Limon}, {Meyer}, {Tucker}, {Wollack},
  \& {Wright}}]{dunkley/etal:prep}
{Dunkley}, J. {et~al.} 2008{\natexlab{a}}, ArXiv e-prints, 803, 0803.0586

\bibitem[{{Dunkley} {et~al.}(2008{\natexlab{b}}){Dunkley}, {Spergel},
  {Komatsu}, {Hinshaw}, {Larson}, {Nolta}, {Odegard}, {Page}, {Bennett},
  {Gold}, {Hill}, {Jarosik}, {Weiland}, {Halpern}, {Kogut}, {Limon}, {Meyer},
  {Tucker}, {Wollack}, \& {Wright}}]{dunkley/etal:prepb}
------. 2008{\natexlab{b}}, ArXiv e-prints, 0811.4280

\bibitem[{{Eriksen} {et~al.}(2006){Eriksen}, {Dickinson}, {Lawrence},
  {Baccigalupi}, {Banday}, {G{\'o}rski}, {Hansen}, {Lilje}, {Pierpaoli},
  {Seiffert}, {Smith}, \& {Vanderlinde}}]{eriksen/etal:2006}
{Eriksen}, H.~K. {et~al.} 2006, \apj, 641, 665, arXiv:astro-ph/0508268

\bibitem[{{Eriksen} {et~al.}(2004){Eriksen}, {Hansen}, {Banday}, {G{\' o}rski},
  \& {Lilje}}]{eriksen/etal:2004}
{Eriksen}, H.~K., {Hansen}, F.~K., {Banday}, A.~J., {G{\' o}rski}, K.~M., \&
  {Lilje}, P.~B. 2004, \apj, 605, 14

\bibitem[{{Eriksen} {et~al.}(2007{\natexlab{a}}){Eriksen}, {Huey}, {Saha},
  {Hansen}, {Dick}, {Banday}, {G{\'o}rski}, {Jain}, {Jewell}, {Knox}, {Larson},
  {O'Dwyer}, {Souradeep}, \& {Wandelt}}]{eriksen/etal:2007}
{Eriksen}, H.~K. {et~al.} 2007{\natexlab{a}}, \apj, 656, 641,
  arXiv:astro-ph/0606088

\bibitem[{{Eriksen} {et~al.}(2007{\natexlab{b}}){Eriksen}, {Jewell},
  {Dickinson}, {Banday}, {Gorski}, \& {Lawrence}}]{eriksen/etal:prep}
{Eriksen}, H.~K., {Jewell}, J.~B., {Dickinson}, C., {Banday}, A.~J., {Gorski},
  K.~M., \& {Lawrence}, C.~R. 2007{\natexlab{b}}, ArXiv e-prints, 709,
  0709.1058

\bibitem[{{Finkbeiner}(2003)}]{finkbeiner:2003}
{Finkbeiner}, D.~P. 2003, \apjs, 146, 407, accepted (astro-ph/0301558)

\bibitem[{{Finkbeiner}(2004)}]{finkbeiner:2004}
------. 2004, \apj, 614, 186

\bibitem[{Finkbeiner {et~al.}(1999)Finkbeiner, Davis, \&
  Schlegel}]{finkbeiner/davis/schlegel:1999}
Finkbeiner, D.~P., Davis, M., \& Schlegel, D.~J. 1999, \apj, 524, 867,
  astro-ph/9905128

\bibitem[{{Finkbeiner} {et~al.}(2004){Finkbeiner}, {Langston}, \&
  {Minter}}]{finkbeiner/langston/minter:2004}
{Finkbeiner}, D.~P., {Langston}, G.~I., \& {Minter}, A.~H. 2004, \apj, 617, 350

\bibitem[{{Fixsen} {et~al.}(1998){Fixsen}, {Dwek}, {Mather}, {Bennett}, \&
  {Shafer}}]{fixsen/etal:1998}
{Fixsen}, D.~J., {Dwek}, E., {Mather}, J.~C., {Bennett}, C.~L., \& {Shafer},
  R.~A. 1998, \apj, 508, 123, arXiv:astro-ph/9803021

\bibitem[{{Fosalba} {et~al.}(2002){Fosalba}, {Lazarian}, {Prunet}, \&
  {Tauber}}]{fosalba/etal:2002}
{Fosalba}, P., {Lazarian}, A., {Prunet}, S., \& {Tauber}, J.~A. 2002, \apj,
  564, 762

\bibitem[{{Fraisse} {et~al.}(2008){Fraisse}, {Brown}, {Dobler}, {Dotson},
  {Draine}, {Frisch}, {Haverkorn}, {Hirata}, {Jansson}, {Lazarian},
  {Magalh{\~a}es}, {Waelkens}, \& {Wolleben}}]{fraisse/etal:prep}
{Fraisse}, A.~A. {et~al.} 2008, ArXiv e-prints, 0811.3920

\bibitem[{{Gold} {et~al.}(2008)}]{gold/etal:prep}
{Gold}, B., {et~al.} 2008, \apjs

\bibitem[{{Gold}(1952)}]{gold:1952}
{Gold}, T. 1952, \mnras, 112, 215

\bibitem[{{Halverson} {et~al.}(2002){Halverson}, {Leitch}, {Pryke}, {Kovac},
  {Carlstrom}, {Holzapfel}, {Dragovan}, {Cartwright}, {Mason}, {Padin},
  {Pearson}, {Readhead}, \& {Shepherd}}]{halverson/etal:2002}
{Halverson}, N.~W. {et~al.} 2002, \apj, 568, 38

\bibitem[{{Han}(2007)}]{han:2007}
{Han}, J.~L. 2007, in IAU Symposium, Vol. 242, IAU Symposium, 55--63

\bibitem[{{Han} {et~al.}(2006){Han}, Manchester, Lyne, Qiao, \& van
  Straten}]{han/etal:2006}
{Han}, J.-L., Manchester, R.~N., Lyne, A.~G., Qiao, G.~J., \& van Straten, W.
  2006, \apj, accepted (astro-ph/0601357)

\bibitem[{{Han} \& {Zhang}(2007)}]{han/zhang:2007}
{Han}, J.~L., \& {Zhang}, J.~S. 2007, \aap, 464, 609, arXiv:astro-ph/0611213

\bibitem[{{Haslam} {et~al.}(1981){Haslam}, {Klein}, {Salter}, {Stoffel},
  {Wilson}, {Cleary}, {Cooke}, \& {Thomasson}}]{haslam/etal:1981}
{Haslam}, C.~G.~T., {Klein}, U., {Salter}, C.~J., {Stoffel}, H., {Wilson},
  W.~E., {Cleary}, M.~N., {Cooke}, D.~J., \& {Thomasson}, P. 1981, \aap, 100,
  209

\bibitem[{{Heiles}(1996)}]{heiles:1996}
{Heiles}, C. 1996, \apj, 462, 316

\bibitem[{{Heiles}(2000)}]{heiles:2000}
------. 2000, \aj, 119, 923

\bibitem[{Hildebrand \& Kirby(2004)}]{hildebrand/kirby:2004}
Hildebrand, R., \& Kirby, L. 2004, in ASP Conf.\ Ser. 309, Astrophysics of
  Dust, ed. A.~N. Witt, G.~C. Clayton, \& B.~T. Draine (San Francisco: ASP),
  515

\bibitem[{Hildebrand {et~al.}(1999)Hildebrand, Dotson, Dowell, Schleuning, \&
  Vaillancourt}]{hildebrand/etal:1999}
Hildebrand, R.~H., Dotson, J.~L., Dowell, C.~D., Schleuning, D.~A., \&
  Vaillancourt, J.~E. 1999, \apj, 516, 834

\bibitem[{{Hildebrand} \& {Dragovan}(1995)}]{hildebrand/dragovan:1995}
{Hildebrand}, R.~H., \& {Dragovan}, M. 1995, \apj, 450, 663

\bibitem[{{Hildebrandt} {et~al.}(2007){Hildebrandt}, {Rebolo},
  {Rubi{\~n}o-Mart{\'{\i}}n}, {Watson}, {Guti{\'e}rrez}, {Hoyland}, \&
  {Battistelli}}]{hildebrandt/etal:2007}
{Hildebrandt}, S.~R., {Rebolo}, R., {Rubi{\~n}o-Mart{\'{\i}}n}, J.~A.,
  {Watson}, R.~A., {Guti{\'e}rrez}, C.~M., {Hoyland}, R.~J., \& {Battistelli},
  E.~S. 2007, \mnras, 382, 594, 0706.1873

\bibitem[{{Hinshaw} {et~al.}(2007){Hinshaw}, {Nolta}, {Bennett}, {Bean},
  {Dor{\'e}}, {Greason}, {Halpern}, {Hill}, {Jarosik}, {Kogut}, {Komatsu},
  {Limon}, {Odegard}, {Meyer}, {Page}, {Peiris}, {Spergel}, {Tucker}, {Verde},
  {Weiland}, {Wollack}, \& {Wright}}]{hinshaw/etal:2007}
{Hinshaw}, G. {et~al.} 2007, \apjs, 170, 288, arXiv:astro-ph/0603451

\bibitem[{{Hinshaw} {et~al.}(2003){Hinshaw}, {Spergel}, {Verde}, {Hill},
  {Meyer}, {Barnes}, {Bennett}, {Halpern}, {Jarosik}, {Kogut}, {Komatsu},
  {Limon}, {Page}, {Tucker}, {Weiland}, {Wollack}, \&
  {Wright}}]{hinshaw/etal:2003}
------. 2003, \apjs, 148, 135

\bibitem[{{Hinshaw} {et~al.}(2008){Hinshaw}, {Weiland}, {Hill}, {Odegard},
  {Larson}, {Bennett}, {Dunkley}, {Gold}, {Greason}, {Jarosik}, {Komatsu},
  {Nolta}, {Page}, {Spergel}, {Wollack}, {Halpern}, {Kogut}, {Limon}, {Meyer},
  {Tucker}, \& {Wright}}]{hinshaw/etal:prep}
------. 2008, ArXiv e-prints, 803, 0803.0732

\bibitem[{{Hoang} \& {Lazarian}(2008)}]{Hoang/Lazarian:2008}
{Hoang}, T., \& {Lazarian}, A. 2008, \mnras, 388, 117, 0707.3645

\bibitem[{Hyvarinen(1999)}]{hyvarinen:1999}
Hyvarinen, A. 1999, in Neural Processing Letters, 1--5

\bibitem[{{Jewell} {et~al.}(2004){Jewell}, {Levin}, \&
  {Anderson}}]{jewell/levin/anderson:2004}
{Jewell}, J., {Levin}, S., \& {Anderson}, C.~H. 2004, \apj, 609, 1,
  arXiv:astro-ph/0209560

\bibitem[{{Keating} {et~al.}(1998){Keating}, {Timbie}, {Polnarev}, \&
  {Steinberger}}]{keating/etal:1998}
{Keating}, B.~G., {Timbie}, P.~T., {Polnarev}, A., \& {Steinberger}, J. 1998,
  \apj, 495, 580

\bibitem[{{Kogut} {et~al.}(1996{\natexlab{a}}){Kogut}, {Banday}, {Bennett},
  {G{\'o}rski}, {Hinshaw}, \& {Reach}}]{kogut/etal:1996a}
{Kogut}, A., {Banday}, A.~J., {Bennett}, C.~L., {G{\'o}rski}, K.~M., {Hinshaw},
  G., \& {Reach}, W.~T. 1996{\natexlab{a}}, \apj, 460, 1

\bibitem[{{Kogut} {et~al.}(1996{\natexlab{b}}){Kogut}, {Banday}, {Bennett},
  {G{\'o}rski}, {Hinshaw}, {Smoot}, \& {Wright}}]{kogut/etal:1996b}
{Kogut}, A., {Banday}, A.~J., {Bennett}, C.~L., {G{\'o}rski}, K.~M., {Hinshaw},
  G., {Smoot}, G.~F., \& {Wright}, E.~I. 1996{\natexlab{b}}, \apjl, 464, L5

\bibitem[{{Kogut} {et~al.}(2007){Kogut}, {Dunkley}, {Bennett}, {Dor{\'e}},
  {Gold}, {Halpern}, {Hinshaw}, {Jarosik}, {Komatsu}, {Nolta}, {Odegard},
  {Page}, {Spergel}, {Tucker}, {Weiland}, {Wollack}, \&
  {Wright}}]{kogut/etal:2007}
{Kogut}, A. {et~al.} 2007, \apj, 665, 355, arXiv:0704.3991

\bibitem[{{Komatsu} {et~al.}(2008){Komatsu}, {Dunkley}, {Nolta}, {Bennett},
  {Gold}, {Hinshaw}, {Jarosik}, {Larson}, {Limon}, {Page}, {Spergel},
  {Halpern}, {Hill}, {Kogut}, {Meyer}, {Tucker}, {Weiland}, {Wollack}, \&
  {Wright}}]{komatsu/etal:prep}
{Komatsu}, E. {et~al.} 2008, ArXiv e-prints, 803, 0803.0547

\bibitem[{{Kovac} {et~al.}(2002){Kovac}, {Leitch}, {Pryke}, {Carlstrom},
  {Halverson}, \& {Holzapfel}}]{kovac/etal:2002}
{Kovac}, J.~M., {Leitch}, E.~M., {Pryke}, C., {Carlstrom}, J.~E., {Halverson},
  N.~W., \& {Holzapfel}, W.~L. 2002, \nat, 420, 772

\bibitem[{{La Porta} {et~al.}(2008){La Porta}, {Burigana}, {Reich}, \&
  {Reich}}]{laporta/etal:2008}
{La Porta}, L., {Burigana}, C., {Reich}, W., \& {Reich}, P. 2008, \aap, 479,
  641, 0801.0547

\bibitem[{{Larson} {et~al.}(2007){Larson}, {Eriksen}, {Wandelt}, {G{\'o}rski},
  {Huey}, {Jewell}, \& {O'Dwyer}}]{larson/etal:2007}
{Larson}, D.~L., {Eriksen}, H.~K., {Wandelt}, B.~D., {G{\'o}rski}, K.~M.,
  {Huey}, G., {Jewell}, J.~B., \& {O'Dwyer}, I.~J. 2007, \apj, 656, 653,
  arXiv:astro-ph/0608007

\bibitem[{{Lazarian}(2003)}]{lazarian:2003}
{Lazarian}, A. 2003, J.\ Quant.\ Spectros.\ Radiat.\ Transfer, 79, 881,
  arXiv:astro-ph/0208487

\bibitem[{{Lazarian}(2007)}]{lazarian:2007}
------. 2007, J.\ Quant.\ Spectros.\ Radiat.\ Transfer, 106, 225,
  arXiv:astro-ph/0208487

\bibitem[{{Lazarian} \& {Draine}(2000)}]{lazarian/draine:2000}
{Lazarian}, A., \& {Draine}, B.~T. 2000, \apjl, 536, L15

\bibitem[{{Lazarian} \& {Finkbeiner}(2003)}]{lazarian/finkbeiner:2003}
{Lazarian}, A., \& {Finkbeiner}, D. 2003, New Astronomy Review, 47, 1107

\bibitem[{{Lazarian} \& {Hoang}(2007)}]{lazarian/hoang:2007}
{Lazarian}, A., \& {Hoang}, T. 2007, \mnras, 378, 910, 0707.0886

\bibitem[{{Leach} {et~al.}(2008){Leach}, {Cardoso}, {Baccigalupi}, {Barreiro},
  {Betoule}, {Bobin}, {Bonaldi}, {de Zotti}, {Delabrouille}, {Dickinson},
  {Eriksen}, {Gonz{\'a}lez-Nuevo}, {Hansen}, {Herranz}, {LeJeune},
  {L{\'o}pez-Caniego}, {Martinez-Gonz{\'a}lez}, {Massardi}, {Melin},
  {Miville-Desch{\^e}nes}, {Patanchon}, {Prunet}, {Ricciardi}, {Salerno},
  {Sanz}, {Starck}, {Stivoli}, {Stolyarov}, {Stompor}, \&
  {Vielva}}]{leach/etal:prep}
{Leach}, S.~M. {et~al.} 2008, ArXiv e-prints, 0805.0269

\bibitem[{Lee {et~al.}(2001)}]{lee/etal:2001}
Lee, A.~T., {et~al.} 2001, \apjl, 561, L1, astro-ph/0104459

\bibitem[{{Leitch} {et~al.}(2002){Leitch}, {Kovac}, {Pryke}, {Carlstrom},
  {Halverson}, {Holzapfel}, {Dragovan}, {Reddall}, \&
  {Sandberg}}]{leitch/etal:2002}
{Leitch}, E.~M. {et~al.} 2002, \nat, 420, 763

\bibitem[{{Leitch} {et~al.}(1997){Leitch}, {Readhead}, {Pearson}, \&
  {Myers}}]{leitch/etal:1997}
{Leitch}, E.~M., {Readhead}, A.~C.~S., {Pearson}, T.~J., \& {Myers}, S.~T.
  1997, \apjl, 486, L23

\bibitem[{{Leonardi} {et~al.}(2007){Leonardi}, {Williams}, {Bersanelli},
  {Ferreira}, {Lubin}, {Meinhold}, {O'Neill}, {Stebor}, {Villa}, {Villela}, \&
  {Wuensche}}]{leonardi/etal:2007}
{Leonardi}, R. {et~al.} 2007, ArXiv e-prints, 0704.0810

\bibitem[{{Lewis} \& {Bridle}(2002)}]{lewis/bridle:2002}
{Lewis}, A., \& {Bridle}, S. 2002, \prd, 66, 103511

\bibitem[{{Magalh{\~a}es} {et~al.}(2005){Magalh{\~a}es}, {Pereyra},
  {Melgarejo}, {de Matos}, {Carciofi}, {Benedito}, {Valentim}, {Vidotto}, {da
  Silva}, {de Souza}, {Faria}, \& {Gabriel}}]{magalhaes/etal:2005}
{Magalh{\~a}es}, A.~M. {et~al.} 2005, in Astronomical Society of the Pacific
  Conference Series, Vol. 343, Astronomical Polarimetry: Current Status and
  Future Directions, ed. A.~{Adamson}, C.~{Aspin}, C.~{Davis}, \&
  T.~{Fujiyoshi}, 305--+

\bibitem[{{Maino} {et~al.}(2007){Maino}, {Donzelli}, {Banday}, {Stivoli}, \&
  {Baccigalupi}}]{maino/etal:2007}
{Maino}, D., {Donzelli}, S., {Banday}, A.~J., {Stivoli}, F., \& {Baccigalupi},
  C. 2007, \mnras, 374, 1207

\bibitem[{{Maino} {et~al.}(2002){Maino}, {Farusi}, {Baccigalupi}, {Perrotta},
  {Banday}, {Bedini}, {Burigana}, {De Zotti}, {G{\'o}rski}, \&
  {Salerno}}]{maino/etal:2002}
{Maino}, D. {et~al.} 2002, \mnras, 334, 53, arXiv:astro-ph/0108362

\bibitem[{Martin(2007)}]{martin:2007}
Martin, P.~G. 2007, in EAS Publ.\ Ser.\ 23, Sky Polarisation at far-infrared to
  radio wavelengths: The Galactic Screen before the Cosmic Microwave
  Background, ed. F.~Boulanger \& M.-A. Miville-Desch\^{e}nes (EDP Sciences),
  165

\bibitem[{{Mason} {et~al.}(2008){Mason}, {Robishaw}, \&
  {Finkbeiner}}]{mason/robishaw/finkbeiner:2008}
{Mason}, B., {Robishaw}, T., \& {Finkbeiner}, D. 2008, in Astronomical Society
  of the Pacific Conference Series, Vol. 395, Frontiers of Astrophysics: A
  Celebration of NRAO's 50th Anniversary, ed. A.~H. {Bridle}, J.~J. {Condon},
  \& G.~C. {Hunt}, 373--+

\bibitem[{{Mathis} {et~al.}(1977){Mathis}, {Rumpl}, \&
  {Nordsieck}}]{mathis/etal:1977}
{Mathis}, J.~S., {Rumpl}, W., \& {Nordsieck}, K.~H. 1977, \apj, 217, 425

\bibitem[{Matthews {et~al.}(2008)Matthews, McPhee, Fissel, \&
  Curran}]{matthews/etal:2008}
Matthews, B.~C., McPhee, C., Fissel, L., \& Curran, R. 2008, \apjs, submitted

\bibitem[{{Matthews} {et~al.}(2001){Matthews}, {Wilson}, \&
  {Fiege}}]{matthews/etal:2001}
{Matthews}, B.~C., {Wilson}, C.~D., \& {Fiege}, J.~D. 2001, \apj, 562, 400

\bibitem[{{Metropolis} {et~al.}(1953){Metropolis}, {Rosenbluth}, \&
  {Rosenbluth}}]{metropolis/etal:1953}
{Metropolis}, N., {Rosenbluth}, A.~W., \& {Rosenbluth}, M.~N.~and~{Teller},
  A.~H. 1953, J.~Chem.~Phys., 21, 1087

\bibitem[{Miller {et~al.}(1999)}]{miller/etal:1999}
Miller, A.~D., {et~al.} 1999, \apjl, 524, L1, astro-ph/9906421

\bibitem[{{Miville-Deschenes} {et~al.}(2008){Miville-Deschenes}, {Ysard},
  {Lavabre}, {Ponthieu}, {Macias-Perez}, {Aumont}, \&
  {Bernard}}]{miville-deschenes/etal:prep}
{Miville-Deschenes}, M.~., {Ysard}, N., {Lavabre}, A., {Ponthieu}, N.,
  {Macias-Perez}, J.~F., {Aumont}, J., \& {Bernard}, J.~P. 2008, ArXiv
  e-prints, 802, 0802.3345

\bibitem[{{Montroy} {et~al.}(2006){Montroy}, {Ade}, {Bock}, {Bond}, {Borrill},
  {Boscaleri}, {Cabella}, {Contaldi}, {Crill}, {de Bernardis}, {De Gasperis},
  {de Oliveira-Costa}, {De Troia}, {di Stefano}, {Hivon}, {Jaffe}, {Kisner},
  {Jones}, {Lange}, {Masi}, {Mauskopf}, {MacTavish}, {Melchiorri}, {Natoli},
  {Netterfield}, {Pascale}, {Piacentini}, {Pogosyan}, {Polenta}, {Prunet},
  {Ricciardi}, {Romeo}, {Ruhl}, {Santini}, {Tegmark}, {Veneziani}, \&
  {Vittorio}}]{montroy/etal:2006}
{Montroy}, T.~E. {et~al.} 2006, \apj, 647, 813, arXiv:astro-ph/0507514

\bibitem[{Netterfield {et~al.}(2002)}]{netterfield/etal:2002}
Netterfield, C.~B., {et~al.} 2002, \apj, 571, 604, astro-ph/0104460

\bibitem[{{Neugebauer} {et~al.}(1984){Neugebauer}, {Habing}, {van Duinen},
  {Aumann}, {Baud}, {Beichman}, {Beintema}, {Boggess}, {Clegg}, {de Jong},
  {Emerson}, {Gautier}, {Gillett}, {Harris}, {Hauser}, {Houck}, {Jennings},
  {Low}, {Marsden}, {Miley}, {Olnon}, {Pottasch}, {Raimond}, {Rowan-Robinson},
  {Soifer}, {Walker}, {Wesselius}, \& {Young}}]{neugebauer/etal:1984}
{Neugebauer}, G. {et~al.} 1984, \apjl, 278, L1

\bibitem[{O'Dwyer {et~al.}(2004)}]{odwyer/etal:2004}
O'Dwyer, I.~J., {et~al.} 2004, Astrophys. J., 617, L99, astro-ph/0407027

\bibitem[{{Oxley} {et~al.}(2004){Oxley}, {Ade}, {Baccigalupi}, {deBernardis},
  {Cho}, {Devlin}, {Hanany}, {Johnson}, {Jones}, {Lee}, {Matsumura}, {Miller},
  {Milligan}, {Renbarger}, {Spieler}, {Stompor}, {Tucker}, \&
  {Zaldarriaga}}]{oxley/etal:2004}
{Oxley}, P. {et~al.} 2004, in Presented at the Society of Photo-Optical
  Instrumentation Engineers (SPIE) Conference, Vol. 5543, Society of
  Photo-Optical Instrumentation Engineers (SPIE) Conference Series, ed.
  M.~{Strojnik}, 320--331

\bibitem[{{P.~Ade} {et~al.}(2007){P.~Ade}, {Bock}, {Bowden}, {Brown}, {Cahill},
  {Carlstrom}, {Castro}, {Church}, {Culverhouse}, {Friedman}, {Ganga}, {Gear},
  {Hinderks}, {Kovac}, {Lange}, {Leitch}, {Melhuish}, {Murphy}, {Orlando},
  {Schwarz}, {O'Sullivan}, {Piccirillo}, {Pryke}, {Rajguru}, {Rusholme},
  {Taylor}, {Thompson}, {Wu}, \& {Zemcov}}]{ade/etal:2007}
{P.~Ade} {et~al.} 2007, ArXiv e-prints, 705, 0705.2359

\bibitem[{{Page} {et~al.}(2007){Page}, {Hinshaw}, {Komatsu}, {Nolta},
  {Spergel}, {Bennett}, {Barnes}, {Bean}, {Dor{\'e}}, {Dunkley}, {Halpern},
  {Hill}, {Jarosik}, {Kogut}, {Limon}, {Meyer}, {Odegard}, {Peiris}, {Tucker},
  {Verde}, {Weiland}, {Wollack}, \& {Wright}}]{page/etal:2007}
{Page}, L. {et~al.} 2007, \apjs, 170, 335, arXiv:astro-ph/0603450

\bibitem[{{Patanchon} {et~al.}(2005){Patanchon}, {Cardoso}, {Delabrouille}, \&
  {Vielva}}]{patanchon/etal:2005}
{Patanchon}, G., {Cardoso}, J.-F., {Delabrouille}, J., \& {Vielva}, P. 2005,
  \mnras, 364, 1185, arXiv:astro-ph/0410280

\bibitem[{{Pearson} {et~al.}(2003){Pearson}, {Mason}, {Readhead}, {Shepherd},
  {Sievers}, {Udomprasert}, {Cartwright}, {Farmer}, {Padin}, {Myers}, {Bond},
  {Contaldi}, {Pen}, {Prunet}, {Pogosyan}, {Carlstrom}, {Kovac}, {Leitch},
  {Pryke}, {Halverson}, {Holzapfel}, {Altamirano}, {Bronfman}, {Casassus},
  {May}, \& {Joy}}]{pearson/etal:2003}
{Pearson}, T.~J. {et~al.} 2003, \apj, 591, 556

\bibitem[{{Pereyra} \& {Magalh{\~a}es}(2004)}]{pereyra/magalhaes:2004}
{Pereyra}, A., \& {Magalh{\~a}es}, A.~M. 2004, \apj, 603, 584

\bibitem[{{Pereyra} \& {Magalh{\~a}es}(2007)}]{pereyra/magalhaes:2007}
------. 2007, \apj, 662, 1014, arXiv:astro-ph/0702550

\bibitem[{Ponthieu {et~al.}(2005)}]{ponthieu/etal:2005}
Ponthieu, N., {et~al.} 2005, Astron. \& Astro,, 607, 655, astro-ph/0501427

\bibitem[{{Prunet} \& {Lazarian}(1999)}]{prunet/lazarian:1999}
{Prunet}, S., \& {Lazarian}, A. 1999, in Microwave Foregrounds. Sloan Summit,
  Institute for Advanced Study, Princeton, New Jersey 14-15 November 1998. Eds:
  A. Rde Oliveira-Costa and M. Tegmark., Vol. 181 (Astronomical Society of the
  Pacific), 113, astro-ph/9902314

\bibitem[{{Prunet} {et~al.}(1998){Prunet}, {Sethi}, {Bouchet}, \&
  {Miville-Deschenes}}]{prunet/etal:1998}
{Prunet}, S., {Sethi}, S.~K., {Bouchet}, F.~R., \& {Miville-Deschenes}, M.-A.
  1998, \aap, 339, 187

\bibitem[{{Pryke} {et~al.}(2008){Pryke}, {Ade}, {Bock}, {Bowden}, {Brown},
  {Cahill}, {Castro}, {Church}, {Culverhouse}, {Friedman}, {Ganga}, {Gear},
  {Gupta}, {Hinderks}, {Kovac}, {Lange}, {Leitch}, {Melhuish}, {Memari},
  {Murphy}, {Orlando}, {Schwarz}, {O'Sullivan}, {Piccirillo}, {Rajguru},
  {Rusholme}, {Taylor}, {Thompson}, {Turner}, {Wu}, \&
  {Zemcov}}]{pryke/etal:prep}
{Pryke}, C. {et~al.} 2008, ArXiv e-prints, 0805.1944

\bibitem[{{Readhead} {et~al.}(2004){Readhead}, {Mason}, {Contaldi}, {Pearson},
  {Bond}, {Myers}, {Padin}, {Sievers}, {Cartwright}, {Shepherd}, {Pogosyan},
  {Prunet}, {Altamirano}, {Bustos}, {Bronfman}, {Casassus}, {Holzapfel}, {May},
  {Pen}, {Torres}, \& {Udomprasert}}]{readhead/etal:2004}
{Readhead}, A.~C.~S. {et~al.} 2004, \apj, 609, 498

\bibitem[{Roberge(2004)}]{roberge:2004}
Roberge, W.~G. 2004, in ASP Conf.\ Ser. 309, Astrophysics of Dust, ed. A.~N.
  Witt, G.~C. Clayton, \& B.~T. Draine, 467

\bibitem[{{Rybicki} \& {Lightman}(1979)}]{rybicki/lightman:1979}
{Rybicki}, G.~B., \& {Lightman}, A. 1979, {Radiative Processes in Astrophysics
  } (Wiley \& Sons: New York)

\bibitem[{{Samtleben}(2008)}]{samtleben:2008}
{Samtleben}, D. 2008, ArXiv e-prints, 0806.4334

\bibitem[{{Savage} \& {Sembach}(1996)}]{savage/sembach:1996}
{Savage}, B.~D., \& {Sembach}, K.~R. 1996, \araa, 34, 279

\bibitem[{{Schlegel} {et~al.}(1998){Schlegel}, {Finkbeiner}, \&
  {Davis}}]{schlegel/finkbeiner/davis:1998}
{Schlegel}, D.~J., {Finkbeiner}, D.~P., \& {Davis}, M. 1998, \apj, 500, 525,
  astro-ph/9710327

\bibitem[{Schleuning(1998)}]{schleuning:1998}
Schleuning, D.~A. 1998, \apj, 493, 811

\bibitem[{Schleuning {et~al.}(2000)Schleuning, Vaillancourt, Hildebrand,
  Dowell, Novak, Dotson, \& Davidson}]{schleuning/etal:2000}
Schleuning, D.~A., Vaillancourt, J.~E., Hildebrand, R.~H., Dowell, C.~D.,
  Novak, G., Dotson, J.~L., \& Davidson, J.~A. 2000, \apj, 535, 913

\bibitem[{{Scott} {et~al.}(2003){Scott}, {Carreira}, {Cleary}, {Davies},
  {Davis}, {Dickinson}, {Grainge}, {Guti{\' e}rrez}, {Hobson}, {Jones},
  {Kneissl}, {Lasenby}, {Maisinger}, {Pooley}, {Rebolo}, {Rubi{\~ n}o-Martin},
  {Sosa Molina}, {Rusholme}, {Saunders}, {Savage}, {Slosar}, {Taylor},
  {Titterington}, {Waldram}, {Watson}, \& {Wilkinson}}]{scott/etal:2003}
{Scott}, P.~F. {et~al.} 2003, \mnras, 341, 1076, astro-ph/0205380

\bibitem[{{Sembach} \& {Savage}(1996)}]{sembach/savage:1996}
{Sembach}, K.~R., \& {Savage}, B.~D. 1996, \apj, 457, 211

\bibitem[{{Smith} {et~al.}(2008){Smith}, {Cooray}, {Das}, {Dor{\'e}}, {Hanson},
  {Hirata}, {Kaplinghat}, {Keating}, {LoVerde}, {Miller}, {Rocha}, {Shimon}, \&
  {Zahn}}]{smith/etal:prep}
{Smith}, K.~M. {et~al.} 2008, ArXiv e-prints, 0811.3916

\bibitem[{{Spergel} {et~al.}(2007){Spergel}, {Bean}, {Dor{\'e}}, {Nolta},
  {Bennett}, {Dunkley}, {Hinshaw}, {Jarosik}, {Komatsu}, {Page}, {Peiris},
  {Verde}, {Halpern}, {Hill}, {Kogut}, {Limon}, {Meyer}, {Odegard}, {Tucker},
  {Weiland}, {Wollack}, \& {Wright}}]{spergel/etal:2007}
{Spergel}, D.~N. {et~al.} 2007, \apjs, 170, 377, arXiv:astro-ph/0603449

\bibitem[{{Spergel} {et~al.}(2003){Spergel}, {Verde}, {Peiris}, {Komatsu},
  {Nolta}, {Bennett}, {Halpern}, {Hinshaw}, {Jarosik}, {Kogut}, {Limon},
  {Meyer}, {Page}, {Tucker}, {Weiland}, {Wollack}, \&
  {Wright}}]{spergel/etal:2003}
------. 2003, \apjs, 148, 175

\bibitem[{{Stompor} {et~al.}(2008){Stompor}, {Leach}, {Stivoli}, \&
  {Baccigalupi}}]{stompor/etal:prep}
{Stompor}, R., {Leach}, S., {Stivoli}, F., \& {Baccigalupi}, C. 2008, ArXiv
  e-prints, 0804.2645

\bibitem[{{Strong} {et~al.}(2007){Strong}, {Moskalenko}, \&
  {Ptuskin}}]{strong/moskalenko/ptuskin:2007}
{Strong}, A.~W., {Moskalenko}, I.~V., \& {Ptuskin}, V.~S. 2007, Annual Review
  of Nuclear and Particle Science, 57, 285, arXiv:astro-ph/0701517

\bibitem[{{Takahashi} {et~al.}(2008){Takahashi}, {Barkats}, {Battle},
  {Bierman}, {Bock}, {Chiang}, {Dowell}, {Hivon}, {Holzapfel}, {Hristov},
  {Jones}, {Kaufman}, {Keating}, {Kovac}, {Kuo}, {Lange}, {Leitch}, {Mason},
  {Matsumura}, {Nguyen}, {Ponthieu}, {Rocha}, {Yoon}, {Ade}, \&
  {Duband}}]{takahashi/etal:2008}
{Takahashi}, Y.~D. {et~al.} 2008, in Proc.\ SPIE 7020, Millimeter and
  Submillimeter Detectors and Instrumentation, ed. W.~D. Duncan, W.~S. Holland,
  S.~Withington, \& J.~Zmuidzinas, 7020--1D

\bibitem[{{Taylor}(2006)}]{taylor:2006}
{Taylor}, A.~C. 2006, New Astronomy Review, 50, 993, arXiv:astro-ph/0610716

\bibitem[{{Taylor} {et~al.}(2003){Taylor}, {Gibson}, {Peracaula}, {Martin},
  {Landecker}, {Brunt}, {Dewdney}, {Dougherty}, {Gray}, {Higgs}, {Kerton},
  {Knee}, {Kothes}, {Purton}, {Uyaniker}, {Wallace}, {Willis}, \&
  {Durand}}]{taylor/etal:2003}
{Taylor}, A.~R. {et~al.} 2003, \aj, 125, 3145

\bibitem[{{Tegmark} {et~al.}(2003){Tegmark}, {de Oliveira-Costa}, \&
  {Hamilton}}]{tegmark/deoliveira-costa/hamilton:2003}
{Tegmark}, M., {de Oliveira-Costa}, A., \& {Hamilton}, A.~J. 2003, \prd, 68,
  123523

\bibitem[{Tegmark \& Efstathiou(1996)}]{tegmark/efstathiou:1996}
Tegmark, M., \& Efstathiou, G. 1996, MNRAS, 281, 1297, astro-ph/9507009

\bibitem[{{The Planck Collaboration}(2006)}]{planck:2006}
{The Planck Collaboration}. 2006, ArXiv Astrophysics e-prints,
  arXiv:astro-ph/0604069

\bibitem[{{Tristram} {et~al.}(2005){Tristram}, {Patanchon},
  {Mac{\'{\i}}as-P{\'e}rez}, {Ade}, {Amblard}, {Ansari}, {Aubourg},
  {Beno{\^i}t}, {Bernard}, {Blanchard}, {Bock}, {Bouchet}, {Bourrachot},
  {Camus}, {Cardoso}, {Couchot}, {de Bernardis}, {Delabrouille}, {D{\'e}sert},
  {Douspis}, {Dumoulin}, {Filliatre}, {Fosalba}, {Giard}, {Giraud-H{\'e}raud},
  {Gispert}, {Guglielmi}, {Hamilton}, {Hanany}, {Henrot-Versill{\'e}},
  {Kaplan}, {Lagache}, {Lamarre}, {Lange}, {Madet}, {Maffei}, {Magneville},
  {Masi}, {Mayet}, {Nati}, {Perdereau}, {Plaszczynski}, {Piat}, {Ponthieu},
  {Prunet}, {Renault}, {Rosset}, {Santos}, {Vibert}, \&
  {Yvon}}]{tristram/etal:2005}
{Tristram}, M. {et~al.} 2005, \aap, 436, 785, arXiv:astro-ph/0411633

\bibitem[{{Uyaniker} {et~al.}(2003){Uyaniker}, {Landecker}, {Gray}, \&
  {Kothes}}]{uyaniker/etal:2003}
{Uyaniker}, B., {Landecker}, T.~L., {Gray}, A.~D., \& {Kothes}, R. 2003, \apj,
  585, 785, arXiv:astro-ph/0211436

\bibitem[{Vaillan\-court(2002)}]{vaillancourt:2002}
Vaillan\-court, J.~E. 2002, \apjs, 142, 53

\bibitem[{Vaillan\-court(2007)}]{vaillancourt:2007}
Vaillan\-court, J.~E. 2007, in EAS Publ.\ Ser.\ 23, Sky Polarisation at
  far-infrared to radio wavelengths: The Galactic Screen before the Cosmic
  Microwave Background, ed. F.~Boulanger \& M.-A. Miville-Desch\^{e}nes (EDP
  Sciences), 147

\bibitem[{{Verde} {et~al.}(2006){Verde}, {Peiris}, \&
  {Jimenez}}]{verde/peiris/jimenez:2006}
{Verde}, L., {Peiris}, H., \& {Jimenez}, R. 2006, JCAP, 019, astro-ph/0506036

\bibitem[{{Wandelt} {et~al.}(2004){Wandelt}, {Larson}, \&
  {Lakshminarayanan}}]{wandelt/larson/lakshminarayanan:2004}
{Wandelt}, B.~D., {Larson}, D.~L., \& {Lakshminarayanan}, A. 2004, \prd, 70,
  083511, arXiv:astro-ph/0310080

\bibitem[{{Watson} {et~al.}(2005){Watson}, {Rebolo},
  {Rubi{\~n}o-Mart{\'{\i}}n}, {Hildebrandt}, {Guti{\'e}rrez},
  {Fern{\'a}ndez-Cerezo}, {Hoyland}, \& {Battistelli}}]{watson/etal:2005}
{Watson}, R.~A., {Rebolo}, R., {Rubi{\~n}o-Mart{\'{\i}}n}, J.~A.,
  {Hildebrandt}, S., {Guti{\'e}rrez}, C.~M., {Fern{\'a}ndez-Cerezo}, S.,
  {Hoyland}, R.~J., \& {Battistelli}, E.~S. 2005, \apjl, 624, L89

\bibitem[{Whittet(2003)}]{whittet:2003}
Whittet, D. C.~B. 2003, Dust in the Galactic Environment, 2nd edn., Series on
  Astronomy and Astrophysics (Philadelphia: Institute of Physics Publishing)

\bibitem[{Whittet(2004)}]{whittet:2004}
Whittet, D. C.~B. 2004, in ASP Conf.\ Ser. 309, Astrophysics of Dust, ed. A.~N.
  Witt, G.~C. Clayton, \& B.~T. Draine, 65

\bibitem[{{Whittet} {et~al.}(2001){Whittet}, {Gerakines}, {Hough}, \&
  {Shenoy}}]{whittet/etal:2001}
{Whittet}, D.~C.~B., {Gerakines}, P.~A., {Hough}, J.~H., \& {Shenoy}, S.~S.
  2001, \apj, 547, 872

\bibitem[{{Wolleben} {et~al.}(2006){Wolleben}, {Landecker}, {Reich}, \&
  {Wielebinski}}]{wolleben/etal:2006}
{Wolleben}, M., {Landecker}, T.~L., {Reich}, W., \& {Wielebinski}, R. 2006,
  \aap, 448, 411, arXiv:astro-ph/0510456

\bibitem[{{Zaldarriaga} {et~al.}(2008){Zaldarriaga}, {Colombo}, {Komatsu},
  {Lidz}, {Mortonson}, {Oh}, {Pierpaoli}, {Verde}, \&
  {Zahn}}]{zaldarriaga/etal:prep}
{Zaldarriaga}, M. {et~al.} 2008, ArXiv e-prints, 0811.3918

\bibitem[{{Zweibel} \& {Heiles}(1997)}]{zweibel/heiles:1997}
{Zweibel}, E.~G., \& {Heiles}, C. 1997, \nat, 385, 131

\end{thebibliography}
%\bibliographystyle{hapj}

\end{document}